\documentclass[english,onecolumn]{IEEEtran}
\usepackage[T1]{fontenc}
\usepackage[latin9]{inputenc}
\usepackage{amsmath}
\usepackage{amsthm}
\usepackage{amssymb}
\usepackage{graphicx}
\usepackage{esint}

\makeatletter
\theoremstyle{plain}
\newtheorem{thm}{\protect\theoremname}
\theoremstyle{definition}
\newtheorem{defn}{\protect\definitionname}
\theoremstyle{plain}
\newtheorem{lem}{\protect\lemmaname}
\theoremstyle{plain}
\newtheorem{prop}{\protect\propositionname}
\theoremstyle{remark}
\newtheorem{rem}{\protect\remarkname}

\usepackage{mathrsfs}

\makeatother

\usepackage{babel}
\providecommand{\definitionname}{Definition}
\providecommand{\lemmaname}{Lemma}
\providecommand{\propositionname}{Proposition}
\providecommand{\remarkname}{Remark}
\providecommand{\theoremname}{Theorem}

\begin{document}

\title{A Unified Framework for One-shot Achievability via the Poisson Matching
Lemma}

\author{Cheuk Ting Li and Venkat Anantharam\\
	EECS, UC Berkeley, Berkeley, CA, USA\\
	Email: ctli@berkeley.edu, ananth@eecs.berkeley.edu}
\maketitle
\begin{abstract}
We introduce a fundamental lemma called the Poisson matching lemma,
and apply it to prove one-shot achievability results for various settings,
namely channels with state information at the encoder, lossy source
coding with side information at the decoder, joint source-channel
coding, broadcast channels, distributed lossy source coding, multiple
access channels, channel resolvability and wiretap channels. Our one-shot
bounds improve upon the best known one-shot bounds in most of the
aforementioned settings (except multiple access channels, channel
resolvability and wiretap channels, where we recover bounds comparable
to the best known bounds), with shorter proofs in some settings even
when compared to the conventional asymptotic approach using typicality.
The Poisson matching lemma replaces both the packing and covering
lemmas, greatly simplifying the error analysis. This paper extends
the work of Li and El Gamal on Poisson functional representation,
which mainly considered variable-length source coding settings, whereas
this paper studies fixed-length settings, and is not limited to source
coding, showing that the Poisson functional representation is a viable
alternative to typicality for most problems in network information
theory.
\end{abstract}

\section{Introduction}

\allowdisplaybreaks

The Poisson functional representation was introduced by Li and El
Gamal \cite{sfrl_trans} to prove the strong functional representation
lemma: for any pair of random variables $(X,Y)$, there exists a random
variable $Z$ independent of $X$ such that $Y$ is a function of
$(X,Z)$, and $H(Y|Z)\le I(X;Y)+\log(I(X;Y)+1)+4$. The lemma is applied
to show various one-shot variable-length lossy source coding results,
and a simple proof of the asymptotic achievability in the Gelfand-Pinsker
theorem \cite{gelfand1980coding}.

In this paper, we introduce the Poisson matching lemma, which gives
a bound on the probability of mismatch between the Poisson functional
representations applied on different distributions, and use it to
prove one-shot achievability results for various settings, namely
channels with state information at the encoder, lossy source coding
with side information at the decoder, joint source-channel coding,
broadcast channels, distributed lossy source coding, multiple access
channels, channel resolvability and wiretap channels. The Poisson
matching lemma can replace both the packing and covering lemmas (and
generalizations such as the mutual covering lemma) in asymptotic typicality-based
proofs. The one-shot bounds in this paper subsume the corresponding
asymptotic achievability results by straightforward applications of
the law of large numbers.

Various non-asymptotic alternatives to typicality have been proposed,
e.g. one-shot packing and covering lemmas \cite{verdu2012nonasymp,liu2015oneshotmutual},
stochastic likelihood coder \cite{yassaee2013oneshot}, likelihood
encoder \cite{song2016likelihood} and random binning \cite{yassaee2013binning}.
However, these non-asymptotic approaches generally require more complex
proofs than their asymptotic counterparts, whereas proofs using the
Poisson matching lemma can be even simpler than asymptotic proofs.

Our approach is better than the conventional asymptotic
approach using typicality (and previous one-shot results, e.g. \cite{verdu2012nonasymp,yassaee2013oneshot}),
in the following ways:
\begin{enumerate}
\item We can give one-shot bounds stronger than the best known one-shot
bounds in many settings discussed in this paper, with the exception
of channel coding, multiple access channels, channel resolvability
and wiretap channels, which are included for demonstration purposes,
where we recover bounds comparable to the best known bounds.
\item Our proofs work for random variables in general Polish spaces.
\item To the best of our knowledge, for the achievability in the Gelfand-Pinsker
theorem \cite{gelfand1980coding} (for channels with state information
at the encoder) and the Wyner-Ziv theorem \cite{wyner1976ratedistort,wyner1978rate}
(for lossy source coding with side information at the decoder), our
proofs are significantly shorter than all previous proofs (another
short proof of the achievability in the Gelfand-Pinsker theorem is
given in \cite{sfrl_trans}, though it is asymptotic). Using our approach,
we can also greatly shorten the proof of the achievability of the
dispersion in joint source-channel coding \cite{kostina2013joint}.
\item Our proofs only use the Poisson matching lemma introduced in this
paper, which replaces both the packing and covering lemmas in proofs
using typicality. The Poisson matching lemma can also be used to prove
a soft covering lemma. Hence the Poisson matching lemma can be the
only tool needed to prove a wide range of results in network information
theory.
\item Our analyses usually involve fewer (or no) uses of sub-codebooks and
binning. As a result, we can reduce the number of error events and
give sharper second-order bounds. For example:
\begin{enumerate}
\item Conventional proofs of the Gelfand-Pinsker theorem involve one sub-codebook,
giving an additional error event, whereas we do not use any sub-codebook.
\item Conventional proofs of the Wyner-Ziv theorem and the Berger-Tung inner
bound \cite{berger1978multiterminal,tung1978multiterminal} (for distributed
lossy source coding) use binning, giving additional error events,
whereas we do not require binning.
\item Conventional proofs of Marton's inner bound \cite{marton1979broadcast}
(for broadcast channels) involve two sub-codebooks, whereas we use
only one.
\end{enumerate}
\item In our approach, the encoders and decoders are characterized using
a common framework (the Poisson functional representation), which
is noteworthy since the roles of an encoder and a decoder in an operational
setting are very different, and their constructions usually have little
in common in conventional approaches.
\end{enumerate}
\medskip{}

\subsection*{Notation}

Throughout this paper, we assume that $\log$ is to base 2 and the
entropy $H$ is in bits. We write $\mathrm{exp}_{a}(b)$ for $a^{b}$.

The set of positive integers is denoted as $\mathbb{N}=\{1,2,\ldots\}$.
We use the notation: $X_{a}^{b}:=(X_{a},\ldots,X_{b})$, $X^{n}:=X_{1}^{n}$
and $[a:b]:=[a,b]\cap\mathbb{Z}$. The conditional information density
is denoted as 
\[
\iota_{X;Y|Z}(x;y|z):=\log\frac{dP_{XY|Z=z}}{d(P_{X|Z=z}\times P_{Y|Z=z})}(x,y).
\]
We consider $\iota_{X;Y|Z}(x;y|z)$ to be defined only if $P_{XY|Z=z}\ll P_{X|Z=z}\times P_{Y|Z=z}$.

For discrete $X$, we write the probability mass function as $p_{X}$.
For continuous $X$, we write the probability density function as
$f_{X}$. For a general random variable $X$ in a measurable space,
we write its distribution as $P_{X}$. The uniform distribution over
a finite set $S$ is denoted as $\mathrm{Unif}(S)$. The joint distribution
of $X_{1},\ldots,X_{n}\stackrel{iid}{\sim}P_{X}$ is written as $P_{X}^{\otimes n}$.
The degenerate distribution $\mathbf{P}\{X=a\}=1$ is denoted as
$\delta_{a}$. The conditional independence of $X$ and $Z$ given
$Y$ is denoted as $X\leftrightarrow Y\leftrightarrow Z$.

The Q-function and its inverse are denoted as $\mathcal{Q}(x)$ and
$\mathcal{Q}^{-1}(\epsilon)$ respectively. For $V\in\mathbb{R}^{n\times n}$
positive semidefinite, define $\mathcal{Q}^{-1}(V,\epsilon)=\{x\in\mathbb{R}^{n}:\,\mathbf{P}\{X\le x\}\ge1-\epsilon\}$
where $X\sim N(0,V)$ and $X\le x$ denotes entrywise comparison.

We assume that every random variable mentioned in this paper lies
in a Polish space with its Borel $\sigma$-algebra, and all functions
mentioned (e.g. distortion measures, the function $x(u,s)$ in Theorem
\ref{thm:channel_state}) are measurable. The Lebesgue measure over
$\mathbb{R}$ is denoted as $\lambda$. The Lebesgue measure restricted
to the set $S\subseteq\mathbb{R}$ is denoted as $\lambda_{S}$. For
two measures $\mu,\nu$ over $\mathcal{X}$ (a Polish space with its
Borel $\sigma$-algebra) such that $\nu$ is absolutely continuous
with respect to $\mu$ (denoted as $\nu\ll\mu$), the Radon-Nikodym
derivative is written as 
\[
\frac{d\nu}{d\mu}:\,\mathcal{X}\to[0,\infty).
\]
If $\nu_{1},\nu_{2}\ll\mu$ (but $\nu_{1}\ll\nu_{2}$ may not hold),
we write
\begin{equation}
\frac{d\nu_{1}}{d\nu_{2}}(x)=\frac{d\nu_{1}}{d\mu}(x)\left(\frac{d\nu_{2}}{d\mu}(x)\right)^{-1}\in[0,\infty],\label{eq:rnderiv}
\end{equation}
which is $0$ if $(d\nu_{1}/d\mu)(x)=0$, and is $\infty$ if $(d\nu_{1}/d\mu)(x)>0$
and $(d\nu_{2}/d\mu)(x)=0$.

The total variation distance between two distributions $P,Q$ over
$\mathcal{X}$ is denoted as $\Vert P-Q\Vert_{\mathrm{TV}}=\sup_{A\subseteq\mathcal{X}\,\mathrm{measurable}}|P(A)-Q(A)|$.\medskip{}

\section{Poisson Matching Lemma}

We first state the definition of Poisson functional representation
in \cite{sfrl_trans}, with a different notation that allows the proofs
to be written in a simpler and more intuitive manner.
\begin{defn}
[Poisson functional representation]Let $\{\bar{U}_{i},T_{i}\}_{i\in\mathbb{N}}$
be the points of a Poisson process with intensity measure $\mu\times\lambda_{\mathbb{R}_{\ge0}}$
on $\mathcal{U}\times\mathbb{R}_{\ge0}$ (where $\mathcal{U}$ is
a Polish space with its Borel $\sigma$-algebra, and $\mu$ is $\sigma$-finite).
For $P\ll\mu$ a probability measure over $\mathcal{U}$, define
\[
\tilde{U}_{P}\left(\{\bar{U}_{i},T_{i}\}_{i\in\mathbb{N}}\right):=\bar{U}_{K_{P}(\{\bar{U}_{i},T_{i}\}_{i\in\mathbb{N}})},
\]
where
\[
K_{P}\left(\{\bar{U}_{i},T_{i}\}_{i\in\mathbb{N}}\right):=\underset{i:\,\frac{dP}{d\mu}(\bar{U}_{i})>0}{\arg\min}T_{i}\left(\frac{dP}{d\mu}(\bar{U}_{i})\right)^{-1},
\]
with arbitrary tie-breaking (a tie occurs with probability 0). We
omit $\{\bar{U}_{i},T_{i}\}_{i\in\mathbb{N}}$ and only write $\tilde{U}_{P}$
if the Poisson process is clear from the context. If the Poisson process
is $\{\bar{X}_{i},T_{i}\}_{i\in\mathbb{N}}$ instead of $\{\bar{U}_{i},T_{i}\}_{i\in\mathbb{N}}$,
then the Poisson functional representation is likewise denoted as
$\tilde{X}_{P}$. If $\bar{U}_{i}=(\bar{X}_{i},\bar{Y}_{i})$ is multivariate,
and $P$ is a distribution over $\mathcal{X}\times\mathcal{Y}$, the
Poisson functional representation is denoted as $(\tilde{X},\tilde{Y})_{P}$.
We write its components as $(\tilde{X},\tilde{Y})_{P}=(\tilde{X}_{P},\tilde{Y}_{P})$.

Note that while $dP/d\mu$ is only uniquely defined up to a $\mu$-null
set, changing the value of $dP/d\mu$ on a $\mu$-null set will only
affect the values of $\tilde{U}_{P}$ on a null set with respect to
the distribution of $\{\bar{U}_{i},T_{i}\}_{i\in\mathbb{N}}$, since
the probability that there exists $\bar{U}_{i}$ on that $\mu$-null
set is zero. Therefore $\tilde{U}_{P}$ is uniquely defined up to
a null set.

\end{defn}
By the mapping theorem \cite{kingman1992poisson,last2017lectures}
(also see Appendix A of \cite{sfrl_trans}), we have $\tilde{U}_{P}\sim P$.
This is termed Poisson functional representation in \cite{sfrl_trans}
since it can be regarded as a construction for the functional representation
lemma \cite{elgamal2011network}. Consider the distribution $P_{U,X}$.
Let $\{\bar{U}_{i},T_{i}\}_{i\in\mathbb{N}}$ be the points of a Poisson
process with intensity measure $P_{U}\times\lambda_{\mathbb{R}_{\ge0}}$,
$X\sim P_{X}$ independent of the process, and $U:=\tilde{U}_{P_{U|X}(\cdot|X)}$.
Then $(U,X)\sim P_{U,X}$. Hence we can express $U$ as a function
of $X$ and $\{\bar{U}_{i},T_{i}\}$ (which is independent of $X$).
This fact will be used repeatedly throughout the proofs in this paper.

For two different distributions $P$ and $Q$, $\tilde{U}_{P}$ and
$\tilde{U}_{Q}$ are coupled in such a way that $\tilde{U}_{P}=\tilde{U}_{Q}$
occurs with a probability that can be bounded in terms of $dP/dQ$.
We now present the core lemma of this paper. The proof is given in
Appendix \ref{subsec:pf_phidiv}.
\begin{lem}
[Poisson matching lemma]\label{lem:phidiv}Let $\{\bar{U}_{i},T_{i}\}_{i\in\mathbb{N}}$
be the points of a Poisson process with intensity measure $\mu\times\lambda_{\mathbb{R}_{\ge0}}$,
and $P,Q$ be probability measures on $\mathcal{U}$ with $P,Q\ll\mu$.
Then we have the following almost surely:
\begin{equation}
\mathbf{P}\left\{ \left.\tilde{U}_{Q}\neq\tilde{U}_{P}\,\right|\,\tilde{U}_{P}\right\} \le1-\left(1+\frac{dP}{dQ}(\tilde{U}_{P})\right)^{-1},\label{eq:phidiv}
\end{equation}
where we write $(dP/dQ)(u)=(dP/d\mu)(u)/((dQ/d\mu)(u))$ as in \eqref{eq:rnderiv}
(we do not require $P\ll Q$). The right hand side of \eqref{eq:phidiv}
is considered to be 1 if $(dP/d\mu)(\tilde{U}_{P})>0$ and $(dQ/d\mu)(\tilde{U}_{P})=0$.
\end{lem}
The exact expression for the left hand side of \eqref{eq:phidiv}
is in \eqref{eq:prob_exact}.

We usually do not apply the Poisson matching lemma on fixed $P,Q$,
but rather on conditional distributions. The following conditional
version of the Poisson matching lemma follows directly from applying
the lemma on $(P,Q)\leftarrow(P_{U|X}(\cdot|X),Q_{U|Y}(\cdot|Y))$.
The proof is given in Appendix \ref{subsec:phidiv_cond} for the sake
of completeness.
\begin{lem}
[Conditional Poisson matching lemma]\label{lem:cond_phidiv}Fix a
distribution $P_{X,U,Y}$ and a probability kernel $Q_{U|Y}$ (that
is not necessarily $P_{U|Y}$) satisfying $P_{U|X}(\cdot|X),Q_{U|Y}(\cdot|Y)\ll\mu$
almost surely. Let $X\sim P_{X}$, and $\{\bar{U}_{i},T_{i}\}_{i\in\mathbb{N}}$
be the points of a Poisson process with intensity measure $\mu\times\lambda_{\mathbb{R}_{\ge0}}$
independent of $X$. Let $U=\tilde{U}_{P_{U|X}(\cdot|X)}$ and $Y|(X,U,\{\bar{U}_{i},T_{i}\}_{i})\sim P_{Y|X,U}(\cdot|X,U)$
(note that $(X,U,Y)\sim P_{X,U,Y}$ and $Y\leftrightarrow(X,U)\leftrightarrow\{\bar{U}_{i},T_{i}\}_{i}$).
Then we have the following almost surely: 
\[
\mathbf{P}\left\{ \left.\tilde{U}_{Q_{U|Y}(\cdot|Y)}\neq U\,\right|\,X,U,Y\right\} \le1-\left(1+\frac{dP_{U|X}(\cdot|X)}{dQ_{U|Y}(\cdot|Y)}(U)\right)^{-1}.
\]
\medskip{}
\end{lem}
The condition that $P_{U|X}(\cdot|X),Q_{U|Y}(\cdot|Y)\ll\mu$ almost
surely is satisfied, for example, when $\mu=P_{U}$, $Q_{U|Y}=P_{U|Y}$,
$P_{UX}\ll P_{U}\times P_{X}$ and $P_{UY}\ll P_{U}\times P_{Y}$.
Note that since $X{\perp\!\!\!\perp}\{\bar{U}_{i},T_{i}\}_{i}$, we
have $\tilde{U}_{P_{U|X}(\cdot|X)}|X\sim P_{U|X}$, whereas $Y$ may
not be independent of $\{\bar{U}_{i},T_{i}\}_{i}$, so $\tilde{U}_{Q_{U|Y}(\cdot|Y)}$
may not follow the conditional distribution $Q_{U|Y}$.

\section{One-shot Channel Coding\label{sec:channel}}

To demonstrate the application of the Poisson matching lemma, we apply
it to recover a bound for one-shot channel coding in \cite{yassaee2013oneshot}
(with a slight penalty of having $\mathsf{L}$ instead of $\mathsf{L}-1$).
Upon observing $M\sim\mathrm{Unif}[1:\mathsf{L}]$, the encoder produces
$X$, which is sent through the channel $P_{Y|X}$. The decoder observes
$Y$ and recovers $\hat{M}$ with error probability $P_{e}=\mathbf{P}\{M\neq\hat{M}\}$.
\begin{prop}
\label{prop:channel}Fix any $P_{X}$. There exists a code for the
channel $P_{Y|X}$, with message $M\sim\mathrm{Unif}[1:\mathsf{L}]$,
with average error probability
\[
P_{e}\le\mathbf{E}\left[1-\left(1+\mathsf{L}2^{-\iota_{X;Y}(X;Y)}\right)^{-1}\right]
\]
if $P_{XY}\ll P_{X}\times P_{Y}$.
\end{prop}
\begin{IEEEproof}
Let $\{(\bar{X}_{i},\bar{M}_{i}),T_{i}\}_{i\in\mathbb{N}}$ be the
points of a Poisson process with intensity measure $P_{X}\times P_{M}\times\lambda_{\mathbb{R}_{\ge0}}$
(where $P_{M}$ is $\mathrm{Unif}[1:\mathsf{L}]$) independent of
$M$. The encoding function is $m\mapsto\tilde{X}_{P_{X}\times\delta_{m}}$
(i.e., $X=\tilde{X}_{P_{X}\times\delta_{M}}$), and the decoding function
is $y\mapsto\tilde{M}_{P_{X|Y}(\cdot|y)\times P_{M}}$ (i.e., $\hat{M}=\tilde{M}_{P_{X|Y}(\cdot|Y)\times P_{M}}$).
Note that the encoding and decoding functions also depend on the common
randomness $\{(\bar{X}_{i},\bar{M}_{i}),T_{i}\}_{i\in\mathbb{N}}$,
which will be fixed later. We have $(M,X,Y)\sim P_{M}\times P_{X}P_{Y|X}$.
\begin{align*}
 & \mathbf{P}\left\{ M\neq\tilde{M}_{P_{X|Y}(\cdot|Y)\times P_{M}}\right\} \\
 & \le\mathbf{P}\left\{ (X,M)\neq(\tilde{X},\tilde{M})_{P_{X|Y}(\cdot|Y)\times P_{M}}\right\} \\
 & =\mathbf{E}\left[\mathbf{P}\left\{ \left.(X,M)\neq(\tilde{X},\tilde{M})_{P_{X|Y}(\cdot|Y)\times P_{M}}\,\right|\,M,X,Y\right\} \right]\\
 & \stackrel{(a)}{\le}\mathbf{E}\left[1-\left(1+\frac{dP_{X}\times\delta_{M}}{dP_{X|Y}(\cdot|Y)\times P_{M}}(X,M)\right)^{-1}\right]\\
 & =\mathbf{E}\left[1-(1+\mathsf{L}2^{-\iota_{X;Y}(X;Y)})^{-1}\right],
\end{align*}
where (a) is by the conditional Poisson matching lemma (Lemma \ref{lem:cond_phidiv})
on $(X,U,Y,Q_{U|Y})\leftarrow(M,(X,M),Y,P_{X|Y}\times P_{M})$ (note
that $P_{X,M|M}=P_{X}\times\delta_{M}$). Therefore there exists a
fixed $\{(\bar{x}_{i},\bar{m}_{i}),t_{i}\}_{i\in\mathbb{N}}$ such
that conditioned on $\{(\bar{X}_{i},\bar{M}_{i}),T_{i}\}_{i\in\mathbb{N}}=\{(\bar{x}_{i},\bar{m}_{i}),t_{i}\}_{i\in\mathbb{N}}$,
the average probability of error is bounded by $\mathbf{E}\left[1-(1+\mathsf{L}2^{-\iota_{X;Y}(X;Y)})^{-1}\right]$.
\end{IEEEproof}
\bigskip{}

Compared to the scheme in \cite{yassaee2013oneshot}, we use the Poisson
process $\{(\bar{X}_{i},\bar{M}_{i}),T_{i}\}$ to create a codebook,
instead of the conventional i.i.d. random codebook in \cite{yassaee2013oneshot}.
While the codewords for different $m$'s are still i.i.d., we attach
a bias $T_{i}$ to each codeword. Our scheme does not use a stochastic
decoder as in \cite{yassaee2013oneshot}, but rather a biased maximum
likelihood decoder $\tilde{M}_{P_{X|Y}(\cdot|y)\times P_{M}}=\bar{M}_{K}$
where $K=\arg\max_{i}T_{i}^{-1}(dP_{X|Y}(\cdot|y)/dP_{X})(\bar{X}_{i})$.
In the following sections, we will demonstrate how our approach can
lead to simpler proofs and sharper bounds compared to \cite{yassaee2013oneshot}.

Using the generalized Poisson matching lemma that will be introduced
in Section \ref{sec:gen_pml}, we can prove the following bound. The
proof is in Appendix \ref{subsec:channel2}.
\begin{thm}
\label{thm:channel2}Fix any $P_{X}$. There exists a code for the
channel $P_{Y|X}$, with message $M\sim\mathrm{Unif}[1:\mathsf{L}]$,
with average error probability
\[
P_{e}\le\mathbf{E}\left[1-\left(1-\min\left\{ 2^{-\iota_{X;Y}(X;Y)},\,1\right\} \right)^{(\mathsf{L}+1)/2}\right]
\]
if $P_{XY}\ll P_{X}\times P_{Y}$.\medskip{}
\end{thm}
Compare this to the dependence testing bound \cite{polyanskiy2010channel}:
\[
P_{e}\le\mathbf{E}\left[\min\left\{ \frac{\mathsf{L}-1}{2}\cdot2^{-\iota_{X;Y}(X;Y)},\,1\right\} \right].
\]
Theorem \ref{thm:channel2} is at least as strong (with a slight penalty
of having $(\mathsf{L}+1)/2$ instead of $(\mathsf{L}-1)/2$) since
\begin{align*}
 & \mathbf{E}\left[1-\left(1-\min\left\{ 2^{-\iota_{X;Y}(X;Y)},\,1\right\} \right)^{(\mathsf{L}+1)/2}\right]\\
 & \le\mathbf{E}\left[\min\left\{ \frac{\mathsf{L}+1}{2}\cdot2^{-\iota_{X;Y}(X;Y)},\,1\right\} \right].
\end{align*}

\begin{rem}
Apart from the dependence testing bound \cite{polyanskiy2010channel},
there are other one-shot bounds for channel coding such as the random-coding
union (RCU) bound and the $\kappa\beta$ bound in \cite{polyanskiy2010channel},
which are tighter in certain situations (e.g. the RCU bound is suitable
for error exponent analysis). The technique introduced in this paper
is suitable for first and second order analysis, but does not seem
to give tight error exponent bounds.
\end{rem}

\section{One-shot Coding for Channels with State Information at the Encoder\label{sec:channel_state}}

The one-shot coding setting for a channel with state information at
the encoder is described as follows. Upon observing $M\sim\mathrm{Unif}[1:\mathsf{L}]$
and $S\sim P_{S}$, the encoder produces $X$, which is sent through
the channel $P_{Y|X,S}$ with state $S$. The decoder observes $Y$
and recovers $\hat{M}$ with error probability $P_{e}=\mathbf{P}\{M\neq\hat{M}\}$.

We show a one-shot version of the Gelfand-Pinsker theorem \cite{gelfand1980coding}.
This is the first one-shot bound attaining the best known second order
result in \cite{scarlett2015dispersions} (which considers a finite-blocklength,
not one-shot scenario). Our bound is stronger than the one-shot bounds
in \cite{verdu2012nonasymp,yassaee2013oneshot,watanabe2015nonasymp}
(in the second order), and significantly simpler to state and prove
than all the aforementioned results. Unlike previous approaches, our
proof does not require sub-codebooks.
\begin{thm}
\label{thm:channel_state}Fix any $P_{U|S}$ and function $x:\mathcal{U}\times\mathcal{S}\to\mathcal{X}$.
There exists a code for the channel $P_{Y|X,S}$ with state distribution
$P_{S}$ with message $M\sim\mathrm{Unif}[1:\mathsf{L}]$, with error
probability
\[
P_{e}\le\mathbf{E}\left[1-(1+\mathsf{L}2^{\iota_{U;S}(U;S)-\iota_{U;Y}(U;Y)})^{-1}\right]
\]
if $P_{US}\ll P_{U}\times P_{S}$ and $P_{UY}\ll P_{U}\times P_{Y}$,
where $(S,U,X,Y)\sim P_{S}P_{U|S}\delta_{x(U,S)}P_{Y|X,S}$.
\end{thm}
\begin{IEEEproof}
Let $\{(\bar{U}_{i},\bar{M}_{i}),T_{i}\}_{i\in\mathbb{N}}$ be the
points of a Poisson process with intensity measure $P_{U}\times P_{M}\times\lambda_{\mathbb{R}_{\ge0}}$
independent of $M,S$. The encoding function is $(m,s)\mapsto x(\tilde{U}_{P_{U|S}(\cdot|s)\times\delta_{m}},s)$
(let $U=\tilde{U}_{P_{U|S}(\cdot|S)\times\delta_{M}}$, $X=x(U,S)$),
and the decoding function is $y\mapsto\tilde{M}_{P_{U|Y}(\cdot|y)\times P_{M}}$
(i.e., $\hat{M}=\tilde{M}_{P_{U|Y}(\cdot|Y)\times P_{M}}$). Note
that $(M,S,U,X,Y)\sim P_{M}\times P_{S}P_{U|S}\delta_{x(U,S)}P_{Y|X,S}$.
We have
\begin{align*}
 & \mathbf{P}\{M\neq\tilde{M}_{P_{U|Y}(\cdot|Y)\times P_{M}}\}\\
 & \le\mathbf{P}\{(U,M)\neq(\tilde{U},\tilde{M})_{P_{U|Y}(\cdot|Y)\times P_{M}}\}\\
 & =\mathbf{E}\left[\mathbf{P}\left\{ \left.(U,M)\neq(\tilde{U},\tilde{M})_{P_{U|Y}(\cdot|Y)\times P_{M}}\,\right|\,M,S,U,Y\right\} \right]\\
 & \stackrel{(a)}{\le}\mathbf{E}\left[1-\left(1+\frac{dP_{U|S}(\cdot|S)\times\delta_{M}}{dP_{U|Y}(\cdot|Y)\times P_{M}}(U,M)\right)^{-1}\right]\\
 & =\mathbf{E}\left[1-(1+\mathsf{L}2^{\iota_{U;S}(U;S)-\iota_{U;Y}(U;Y)})^{-1}\right].
\end{align*}
where (a) is by the conditional Poisson matching lemma on $((M,S),\,(U,M),\,Y,\,P_{U|Y}\times P_{M})$
(note that $P_{U,M|M,S}=P_{U|S}\times\delta_{M}$). Therefore there
exists a fixed $\{(\bar{u}_{i},\bar{m}_{i}),t_{i}\}_{i\in\mathbb{N}}$
attaining the desired bound.
\end{IEEEproof}
\bigskip{}

Compared to Theorem 3 in \cite{verdu2012nonasymp}:
\[
P_{e}\le\mathbf{P}\{\iota_{U;S}(U;S)>\log\mathsf{J}-\gamma\}+\mathbf{P}\{\iota_{U;Y}(U;Y)\le\log\mathsf{L}\mathsf{J}+\gamma\}+2^{-\gamma}+e^{-2^{\gamma}}
\]
for any $\gamma>0$, $\mathsf{J}\in\mathbb{N}$, our result is strictly
stronger since
\begin{align*}
 & \mathbf{E}\left[1-\left(1+\mathsf{L}2^{\iota_{U;S}(U;S)-\iota_{U;Y}(U;Y)}\right)^{-1}\right]\\
 & \le\mathbf{P}\{\iota_{U;S}(U;S)>\log\mathsf{J}-\gamma\}+\mathbf{P}\{\iota_{U;Y}(U;Y)\le\log\mathsf{L}\mathsf{J}+\gamma\}\\
 & \;\;\;\;+\mathbf{E}\left[1-\left(1+\mathsf{L}2^{\iota_{U;S}(U;S)-\iota_{U;Y}(U;Y)}\right)^{-1}\,|\,\iota_{U;S}(U;S)\le\log\mathsf{J}-\gamma,\,\iota_{U;Y}(U;Y)>\log\mathsf{L}\mathsf{J}+\gamma\right]\\
 & \le\mathbf{P}\{\iota_{U;S}(U;S)>\log\mathsf{J}-\gamma\}+\mathbf{P}\{\iota_{U;Y}(U;Y)\le\log\mathsf{L}\mathsf{J}+\gamma\}+2^{-2\gamma}\\
 & <\mathbf{P}\{\iota_{U;S}(U;S)>\log\mathsf{J}-\gamma\}+\mathbf{P}\{\iota_{U;Y}(U;Y)\le\log\mathsf{L}\mathsf{J}+\gamma\}+2^{-\gamma}+e^{-2^{\gamma}}.
\end{align*}
This is due to the fact that the Poisson matching lemma simultaneously
replaces both the covering and the packing lemma, resulting in only
one error event.

Next, we prove a second-order result. Fix $\epsilon>0$. Let $C:=I(U;Y)-I(U;S)$,
$V:=\mathrm{Var}[\iota_{U;S}(U;S)-\iota_{U;Y}(U;Y)]$. We apply Theorem
\ref{thm:channel_state} on $n$ uses of the memoryless channel with
i.i.d. state sequence $S^{n}=(S_{1},\ldots,S_{n})$, and
\[
\mathsf{L}:=\left\lfloor \mathrm{exp}_{2}\left(nC-\sqrt{nV}\mathcal{Q}^{-1}\left(\epsilon-\frac{\alpha}{\sqrt{n}}\right)-\frac{1}{2}\log n\right)\right\rfloor ,
\]
where $\alpha$ is a constant that depends on $P_{S,U,Y}$. For $n>\alpha^{2}\epsilon^{-2}$,
by the Berry-Esseen theorem \cite{berry1941accuracy,esseen1942liapunov,feller1971introduction},
we have
\begin{align*}
P_{e} & \le\mathbf{E}\left[\min\left\{ 2^{\log\mathsf{L}+\iota_{U^{n};S^{n}}(U^{n};S^{n})-\iota_{U^{n};Y^{n}}(U^{n};Y^{n})},\,1\right\} \right]\\
 & \le\frac{1}{\sqrt{n}}+\mathbf{P}\left\{ 2^{\log\mathsf{L}+\iota_{U^{n};S^{n}}(U^{n};S^{n})-\iota_{U^{n};Y^{n}}(U^{n};Y^{n})}>\frac{1}{\sqrt{n}}\right\} \\
 & \le\frac{1}{\sqrt{n}}+\mathbf{P}\left\{ \frac{1}{\sqrt{n}}\sum_{i=1}^{n}\left(\iota_{U;Y}(U_{i};Y_{i})-\iota_{U;S}(U_{i};S_{i})-C\right)<-\sqrt{V}\mathcal{Q}^{-1}\left(\epsilon-\frac{\alpha}{\sqrt{n}}\right)\right\} \\
 & \le\frac{1}{\sqrt{n}}+\epsilon-\frac{\alpha}{\sqrt{n}}+\frac{\alpha-1}{\sqrt{n}}\\
 & \le\epsilon
\end{align*}
if we let $\alpha-1$ be the constant given by the Berry-Esseen theorem.
This coincides with the best known second order result in \cite{scarlett2015dispersions},
which is stronger than the second order results implied by \cite{verdu2012nonasymp,yassaee2013oneshot,watanabe2015nonasymp}.
We bound $\iota_{U;S}(U;S)-\iota_{U;Y}(U;Y)$ as a single quantity,
instead of bounding the two terms separately as in \cite{verdu2012nonasymp,yassaee2013oneshot,watanabe2015nonasymp},
resulting in a sharper second order bound.

\section{One-shot Lossy Source Coding with Side Information at the Decoder}

The one-shot lossy source coding setting with side information at
the decoder is described as follows. Upon observing $X\sim P_{X}$,
the encoder produces $M\in[1:\mathsf{L}]$. The decoder observes $M$
and $Y\sim P_{Y|X}$ and recovers $\hat{Z}\in\mathcal{Z}$ with probability
of excess distortion $P_{e}=\mathbf{P}\{\mathsf{d}(X,\hat{Z})>\mathsf{D}\}$,
where $\mathsf{d}:\mathcal{X}\times\mathcal{Z}\to\mathbb{R}_{\ge0}$
is a distortion measure.

We show a one-shot version of the Wyner-Ziv theorem \cite{wyner1976ratedistort,wyner1978rate}.
Our bound is stronger than those in \cite{verdu2012nonasymp,watanabe2015nonasymp},
and significantly simpler to state and prove. Unlike previous approaches,
our proof does not require binning.
\begin{thm}
\label{thm:lscsi}Fix any $P_{U|X}$ and function $z:\mathcal{U}\times\mathcal{Y}\to\mathcal{Z}$.
There exists a code for lossy source coding with source distribution
$P_{X}$, side information at the decoder given by $P_{Y|X}$, and
message size $\mathsf{L}$, with probability of excess distortion
\[
P_{e}\le\mathbf{E}\left[1-\mathbf{1}\{\mathsf{d}(X,Z)\le\mathsf{D}\}(1+\mathsf{L}^{-1}2^{\iota_{U;X}(U;X)-\iota_{U;Y}(U;Y)})^{-1}\right]
\]
if $P_{UX}\ll P_{U}\times P_{X}$ and $P_{UY}\ll P_{U}\times P_{Y}$,
where $(X,Y,U,Z)\sim P_{X}P_{Y|X}P_{U|X}\delta_{z(U,Y)}$.
\end{thm}
\begin{IEEEproof}
Let $\{(\bar{U}_{i},\bar{M}_{i}),T_{i}\}_{i\in\mathbb{N}}$ be the
points of a Poisson process with intensity measure $P_{U}\times P_{M}\times\lambda_{\mathbb{R}_{\ge0}}$
independent of $X$, where $P_{M}$ is $\mathrm{Unif}[1:\mathsf{L}]$.
The encoding function is $x\mapsto\tilde{M}_{P_{U|X}(\cdot|x)\times P_{M}}$
(i.e., $M=\tilde{M}_{P_{U|X}(\cdot|X)\times P_{M}}$), and the decoding
function is $(m,y)\mapsto z(\tilde{U}_{P_{U|Y}(\cdot|y)\times\delta_{m}},y)$
(let $\hat{U}=\tilde{U}_{P_{U|Y}(\cdot|Y)\times\delta_{M}}$, $\hat{Z}=z(\hat{U},Y)$).
Also define $U=\tilde{U}_{P_{U|X}(\cdot|X)\times P_{M}}$, $Z=z(U,Y)$.
Note that $(M,X,Y,U,Z)\sim P_{M}\times P_{X}P_{Y|X}P_{U|X}\delta_{z(U,Y)}$.
We have
\begin{align*}
 & \mathbf{P}\{\mathsf{d}(X,\hat{Z})>\mathsf{D}\}\\
 & \le1-\mathbf{P}\{\mathsf{d}(X,Z)\le\mathsf{D}\,\mathrm{and}\,U=\hat{U}\}\\
 & \le\mathbf{E}\left[1-\mathbf{1}\{\mathsf{d}(X,Z)\le\mathsf{D}\}\mathbf{P}\{(U,M)=(\tilde{U},\tilde{M})_{P_{U|Y}(\cdot|Y)\times\delta_{M}}\,|\,M,X,Y,U\}\right]\\
 & \stackrel{(a)}{\le}\mathbf{E}\left[1-\mathbf{1}\{\mathsf{d}(X,Z)\le\mathsf{D}\}\left(1+\frac{dP_{U|X}(\cdot|X)\times P_{M}}{dP_{U|Y}(\cdot|Y)\times\delta_{M}}(U,M)\right)^{-1}\right]\\
 & \le\mathbf{E}\left[1-\mathbf{1}\{\mathsf{d}(X,Z)\le\mathsf{D}\}(1+\mathsf{L}^{-1}2^{\iota_{U;X}(U;X)-\iota_{U;Y}(U;Y)})^{-1}\right].
\end{align*}
where (a) is by the conditional Poisson matching lemma on $(X,\,(U,M),\,(M,Y),\,P_{U|Y}\times\delta_{M})$
(note that $P_{U,M|X}=P_{U|X}\times P_{M}$). Therefore there exists
a fixed $\{(\bar{u}_{i},\bar{m}_{i}),t_{i}\}_{i\in\mathbb{N}}$ attaining
the desired bound.
\end{IEEEproof}
\bigskip{}

This reduces to lossy source coding (without side information) when
$Y=\emptyset$. Note that the encoder is designed in the same way
with or without side information. An encoder for lossy source coding
is sufficient to achieve the bound in Theorem \ref{thm:lscsi} even
when side information is present. Binning is not required at the encoder.

Similar to the case in Section \ref{sec:channel_state}, it can be
checked that our bound is stronger than that in Theorem 2 in \cite{verdu2012nonasymp}.
Compared to Corollary 9 in \cite{watanabe2015nonasymp}:
\begin{align}
P_{e} & \le\mathbf{P}\{\iota_{U;X}(U;X)>\gamma_{\mathrm{c}}\;\mathrm{or}\;\iota_{U;Y}(U;Y)<\gamma_{\mathrm{p}}\;\mathrm{or}\;\mathsf{d}(X,Z)>\mathsf{D}\}+\frac{\mathsf{J}}{2^{\gamma_{\mathrm{p}}}\mathsf{L}}+\frac{1}{2}\sqrt{\frac{2^{\gamma_{\mathrm{c}}}}{\mathsf{J}}}\label{eq:lscsi_watanabe}
\end{align}
for any $\gamma_{\mathrm{p}},\gamma_{\mathrm{c}}>0$, $\mathsf{J}\in\mathbb{N}$,
our result is stronger since
\begin{align*}
 & \mathbf{E}\left[1-\mathbf{1}\{\mathsf{d}(X,Z)\le\mathsf{D}\}(1+\mathsf{L}^{-1}2^{\iota_{U;X}(U;X)-\iota_{U;Y}(U;Y)})^{-1}\right]\\
 & \le\mathbf{P}\{\iota_{U;X}(U;X)>\gamma_{\mathrm{c}}\;\mathrm{or}\;\iota_{U;Y}(U;Y)<\gamma_{\mathrm{p}}\;\mathrm{or}\;\mathsf{d}(X,Z)>\mathsf{D}\}+\mathsf{L}^{-1}2^{\gamma_{\mathrm{c}}-\gamma_{\mathrm{p}}}\\
 & \le\mathbf{P}\{\iota_{U;X}(U;X)>\gamma_{\mathrm{c}}\;\mathrm{or}\;\iota_{U;Y}(U;Y)<\gamma_{\mathrm{p}}\;\mathrm{or}\;\mathsf{d}(X,Z)>\mathsf{D}\}+\frac{\mathsf{J}}{2^{\gamma_{\mathrm{p}}}\mathsf{L}}+\frac{1}{2}\sqrt{\frac{2^{\gamma_{\mathrm{c}}}}{\mathsf{J}}},
\end{align*}
where the last inequality is due to 
\[
a+b\ge(a+b)^{3}=27\left(\frac{a+2(b/2)}{3}\right)^{3}\ge27a(b/2)^{2}\ge4ab^{2}
\]
by the AM-GM inequality for $a,b\ge0$, $a+b\le1$ (since the right
hand side of \eqref{eq:lscsi_watanabe} $\le1$ for it to be meaningful).
We bound $\iota_{U;X}(U;X)-\iota_{U;Y}(U;Y)$ as a single quantity,
instead of bounding the two terms separately, resulting in a sharper
bound.

\medskip{}

\section{One-shot Joint Source-Channel Coding}

The one-shot joint source-channel coding setting is described as follows.
Upon observing the source symbol $W\sim P_{W}$, the encoder produces
$X\in\mathcal{X}$, which is sent through the channel $P_{Y|X}$.
The decoder observes $Y$ and recovers $\hat{Z}\in\mathcal{Z}$ with
probability of excess distortion $P_{e}=\mathbf{P}\{\mathsf{d}(W,\hat{Z})>\mathsf{D}\}$,
where $\mathsf{d}:\mathcal{W}\times\mathcal{Z}\to\mathbb{R}_{\ge0}$
is a distortion measure.

We show a one-shot joint source-channel coding result that achieves
the optimal dispersion in \cite{kostina2013joint}.
\begin{thm}
\label{thm:jscc}Fix any $P_{X}$ and $P_{Z}$. There exists a code
for the source distribution $P_{W}$ and channel $P_{Y|X}$, with
probability of excess distortion
\[
P_{e}\le\mathbf{E}\left[\left(1+P_{Z}(\mathcal{B}_{\mathsf{D}}(W))2^{\iota_{X;Y}(X;Y)}\right)^{-1}\right]
\]
if $P_{XY}\ll P_{X}\times P_{Y}$, where $(W,X,Y)\sim P_{W}\times P_{X}P_{Y|X}$,
and $\mathcal{B}_{\mathsf{D}}(w):=\{z:\,\mathsf{d}(w,z)\le\mathsf{D}\}$.
\end{thm}
\begin{IEEEproof}
Let $\{(\bar{X}_{i},\bar{Z}_{i}),T_{i}\}_{i\in\mathbb{N}}$ be the
points of a Poisson process with intensity measure $P_{X}\times P_{Z}\times\lambda_{\mathbb{R}_{\ge0}}$
independent of $W$. Let $\rho(w):=P_{Z}(\mathcal{B}_{\mathsf{D}}(w))$.
Let $P_{\check{Z}|W}$ be defined as
\[
P_{\check{Z}|W}(A|w):=\begin{cases}
P_{Z}(A\cap\mathcal{B}_{\mathsf{D}}(w))/\rho(w) & \mathrm{if}\,\rho(w)>0\\
P_{Z}(A) & \mathrm{if}\,\rho(w)=0.
\end{cases}
\]
The encoding function is $w\mapsto\tilde{X}_{P_{X}\times P_{\check{Z}|W}(\cdot|w)}$
(i.e., $X=\tilde{X}_{P_{X}\times P_{\check{Z}|W}(\cdot|W)}$). The
decoding function is $y\mapsto\tilde{Z}_{P_{X|Y}(\cdot|y)\times P_{Z}}$
(i.e., $\hat{Z}=\tilde{Z}_{P_{X|Y}(\cdot|Y)\times P_{Z}}$). Also
define $\check{Z}=\tilde{Z}_{P_{X}\times P_{\check{Z}|W}(\cdot|W)}$.
We have $(X,Y,W,\check{Z})\sim P_{X}P_{Y|X}\times P_{W}P_{\check{Z}|W}$.
\begin{align*}
 & \mathbf{P}\{\mathsf{d}(W,\hat{Z})>\mathsf{D}\}\\
 & \le\mathbf{P}\{\rho(W)=0\}+\mathbf{P}\{\rho(W)>0\;\mathrm{and}\;\check{Z}\neq\hat{Z}\}\\
 & \le\mathbf{P}\{\rho(W)=0\}+\mathbf{E}\left[\mathbf{1}\{\rho(W)>0\}\mathbf{P}\{(X,\check{Z})\neq(\tilde{X},\tilde{Z})_{P_{X|Y}(\cdot|Y)\times P_{Z}}\,|\,X,Y,W,\check{Z}\}\right]\\
 & \stackrel{(a)}{\le}\mathbf{P}\{\rho(W)=0\}+\mathbf{E}\left[\mathbf{1}\{\rho(W)>0\}\left(1-\left(1+\frac{dP_{X}\times P_{\check{Z}|W}(\cdot|W)}{dP_{X|Y}(\cdot|Y)\times P_{Z}}(X,\check{Z})\right)^{-1}\right)\right]\\
 & =\mathbf{P}\{\rho(W)=0\}+\mathbf{E}\left[\mathbf{1}\{\rho(W)>0\}\left(1-\left(1+(\rho(W))^{-1}2^{-\iota_{X;Y}(X;Y)}\right)^{-1}\right)\right]\\
 & =\mathbf{E}\left[\left(1+\rho(W)2^{\iota_{X;Y}(X;Y)}\right)^{-1}\right],
\end{align*}
where (a) is by the conditional Poisson matching lemma on $(W,\,(X,\check{Z}),\,Y,\,P_{X|Y}\times P_{Z})$
(note that $P_{X,\check{Z}|W}=P_{X}\times P_{\check{Z}|W}$). Therefore
there exists a fixed $\{(\bar{x}_{i},\bar{z}_{i}),t_{i}\}_{i\in\mathbb{N}}$
attaining the desired bound.
\end{IEEEproof}
\bigskip{}

Compare this to Theorem 7 in \cite{kostina2013joint}:
\begin{equation}
P_{e}\le\mathbf{E}\left[\min\left\{ J2^{-\iota_{X;Y}(X;Y)},\,1\right\} \right]+\mathbf{E}\left[(1-P_{Z}(\mathcal{B}_{\mathsf{D}}(W)))^{J}\right]\label{eq:jscc_kostina}
\end{equation}
for any $P_{J|W}$, $J\in\mathbb{N}$. While neither of the bounds
implies the other, our bound is at least within a factor of 2 from
\eqref{eq:jscc_kostina}, since
\begin{align*}
 & \mathbf{E}\left[\left(1+P_{Z}(\mathcal{B}_{\mathsf{D}}(W))2^{\iota_{X;Y}(X;Y)}\right)^{-1}\right]\\
 & \le\mathbf{E}\left[\left(1+(2J)^{-1}2^{\iota_{X;Y}(X;Y)}\right)^{-1}\right]+\mathbf{P}\left\{ (2J)^{-1}\ge P_{Z}(\mathcal{B}_{\mathsf{D}}(W))\right\} \\
 & \le\mathbf{E}\left[\min\left\{ 2J2^{-\iota_{X;Y}(X;Y)},\,1\right\} \right]+2\mathbf{E}\left[\max\{1-JP_{Z}(\mathcal{B}_{\mathsf{D}}(W)),\,0\}\right]\\
 & \le2\mathbf{E}\left[\min\left\{ J2^{-\iota_{X;Y}(X;Y)},\,1\right\} \right]+2\mathbf{E}\left[(1-P_{Z}(\mathcal{B}_{\mathsf{D}}(W)))^{J}\right].
\end{align*}
However, \eqref{eq:jscc_kostina} does not imply a bound that is within
a constant factor from our bound. Theorem 8 in \cite{kostina2013joint}
is obtained by substituting $J=\lfloor\gamma/P_{Z}(\mathcal{B}_{\mathsf{D}}(W))\rfloor$
in \eqref{eq:jscc_kostina}:
\[
P_{e}\le\mathbf{E}\left[\min\left\{ \gamma P_{Z}(\mathcal{B}_{\mathsf{D}}(W))^{-1}2^{-\iota_{X;Y}(X;Y)},\,1\right\} \right]+e^{1-\gamma},
\]
which is strictly weaker than our bound with an unbounded multiplicative
gap $\gamma$ (that tends to $\infty$ when the bound tends to 0).
Hence our bound is stronger than Theorem 7 and 8 in \cite{kostina2013joint}
(ignoring constant multiplicative gaps). Also our proof is significantly
shorter than that of Theorem 7 in \cite{kostina2013joint}. 

Please refer to Appendix \ref{subsec:second_jscc} for the proof that
Theorem \ref{thm:jscc} achieves the optimal dispersion.

\medskip{}

\section{Poisson Matching Lemma Beyond the First Index\label{sec:gen_pml}}

The Poisson functional representation concerns the point with the
smallest $T_{i}((dP/d\mu)(\bar{U}_{i}))^{-1}$. We can generalize
it to obtain a sequence ordered in ascending order of $T_{i}((dP/d\mu)(\bar{U}_{i}))^{-1}$.
\begin{defn}
[Mapped Poisson process] Let $\{\bar{U}_{i},T_{i}\}_{i\in\mathbb{N}}$
be the points of a Poisson process with intensity measure $\mu\times\lambda_{\mathbb{R}_{\ge0}}$
on $\mathcal{U}\times\mathbb{R}_{\ge0}$ (where $\mathcal{U}$ is
a Polish space with its Borel $\sigma$-algebra, and $\mu$ is $\sigma$-finite).
For $P\ll\mu$ a probability measure over $\mathcal{U}$, let $i_{P,1},i_{P,2},\ldots\in\mathbb{N}$
be a sequence of distinct integers such that $\bigcup_{j=1}^{\infty}\{i_{P,j}\}=\{i:\,(dP/d\mu)(\bar{U}_{i})>0\}$
and $\{T_{i_{P,j}}((dP/d\mu)(\bar{U}_{i_{P,j}}))^{-1}\}_{j\in\mathbb{N}}$
is sorted in ascending order with arbitrary tie-breaking (a tie occurs
with probability 0). For $j\in\mathbb{N},\,u\in\mathcal{U}$, define
the \emph{mapped Poisson process with respect to $P$ }as 
\begin{equation}
\left\{ \tilde{U}_{P}\left(\{\bar{U}_{i},T_{i}\}_{i\in\mathbb{N}},\,j\right),\,\tilde{T}_{P}\left(\{\bar{U}_{i},T_{i}\}_{i\in\mathbb{N}},\,j\right)\right\} _{j\in\mathbb{N}},\label{eq:mappedpp}
\end{equation}
where
\[
\tilde{T}_{P}\left(\{\bar{U}_{i},T_{i}\}_{i\in\mathbb{N}},\,j\right):=T_{i_{P,j}}\left(\frac{dP}{d\mu}(\bar{U}_{i_{P,j}})\right)^{-1},
\]
\[
\tilde{U}_{P}\left(\{\bar{U}_{i},T_{i}\}_{i\in\mathbb{N}},\,j\right):=\bar{U}_{i_{P,j}}.
\]
For $P,Q\ll\mu$ probability measures over $\mathcal{U}$, define
$i_{P,1},i_{P,2},\ldots\in\mathbb{N}$ and $i_{Q,1},i_{Q,2},\ldots\in\mathbb{N}$
as above. Define
\[
\Upsilon_{P\Vert Q}\left(\{\bar{U}_{i},T_{i}\}_{i\in\mathbb{N}},\,j\right):=\min\{k\in\mathbb{N}:\,i_{Q,k}=i_{P,j}\},
\]
where the minimum is $\infty$ if such $k$ does not exist. We omit
$\{\bar{U}_{i},T_{i}\}_{i\in\mathbb{N}}$ and only write $\tilde{U}_{P}(j)$,
$\tilde{T}_{P}(j)$, $\Upsilon_{P\Vert Q}(j)$ if the Poisson process
is clear from the context. Note that, with probability 1, we have
either $\tilde{U}_{Q}(\Upsilon_{P\Vert Q}(j))=\tilde{U}_{P}(j)$ or
$\Upsilon_{P\Vert Q}(j)=\infty$. Also, for any $j,k\in\mathbb{N}$,
$\Upsilon_{P\Vert Q}(j)=k\Leftrightarrow\Upsilon_{Q\Vert P}(k)=j$.
Loosely speaking, $\Upsilon_{P\Vert Q}(j)$ can be regarded as ``$\tilde{U}_{Q}^{-1}(\tilde{U}_{P}(j))$''
(if there are no atoms in $\mu$), i.e., finding the $j$-th point
in the mapped Poisson process w.r.t. $P$, then finding its index
in the mapped Poisson process w.r.t. $Q$.

While $dP/d\mu$ is only uniquely defined up to a $\mu$-null set,
changing the value of $dP/d\mu$ on a $\mu$-null set will only affect
the values of $\{\tilde{U}_{P}(j),\,\tilde{T}_{P}(j)\}_{j\in\mathbb{N}}$
on a null set with respect to the distribution of $\{\bar{U}_{i},T_{i}\}_{i\in\mathbb{N}}$,
since the probability that there exists $\bar{U}_{i}$ in that $\mu$-null
set is zero. Therefore $\{\tilde{U}_{P}(j),\,\tilde{T}_{P}(j)\}_{j\in\mathbb{N}}$
is uniquely defined up to a null set. The same is true for $\Upsilon_{P\Vert Q}(j)$.

\end{defn}
By the mapping theorem \cite{kingman1992poisson,last2017lectures}
(also see Appendix A of \cite{sfrl_trans}), 
\[
\{\bar{U}_{i_{P,j}},\,T_{i_{P,j}}((dP/d\mu)(U_{i_{P,j}}))^{-1}\}_{j\in\mathbb{N}}=\{\tilde{U}_{P}(j),\tilde{T}_{P}(j)\}_{j\in\mathbb{N}}
\]
is a Poisson process with intensity measure $P\times\lambda_{\mathbb{R}_{\ge0}}$.
Hence
\[
\tilde{U}_{P}(1),\tilde{U}_{P}(2),\ldots\stackrel{iid}{\sim}P.
\]

We present a generalized Poisson matching lemma concerning the indices
beyond the first. The proof is given in Appendix \ref{subsec:pf_phidiv}.
\begin{lem}
[Generalized Poisson matching lemma]\label{lem:phidiv_gen}Let $\{\bar{U}_{i},T_{i}\}_{i\in\mathbb{N}}$
be the points of a Poisson process with intensity measure $\mu\times\lambda_{\mathbb{R}_{\ge0}}$
on $\mathcal{U}\times\mathbb{R}_{\ge0}$, and $P,Q$ be probability
measures over $\mathcal{U}$ with $P,Q\ll\mu$. Fix any $j\in\mathbb{N}$.
Then we have the following almost surely:
\[
\mathbf{E}\left[\left.\Upsilon_{P\Vert Q}(j)\,\right|\,\tilde{U}_{P}(j)\right]\le j\frac{dP}{dQ}(\tilde{U}_{P}(j))+1,
\]
where we write $(dP/dQ)(u)=(dP/d\mu)(u)/((dQ/d\mu)(u))$ as in \eqref{eq:rnderiv}
(we do not require $P\ll Q$). As a result, we have the following
almost surely: for all $k\in\mathbb{N}$,
\begin{align*}
\mathbf{P}\left\{ \left.\tilde{U}_{P}(j)\notin\{\tilde{U}_{Q}(i)\}_{i\in[1:k]}\,\right|\,\tilde{U}_{P}(j)\right\}  & \le\mathbf{P}\left\{ \left.\Upsilon_{P\Vert Q}(j)>k\,\right|\,\tilde{U}_{P}(j)\right\} \\
 & \le\min\left\{ \frac{j}{k}\frac{dP}{dQ}(\tilde{U}_{P}(j)),\,1\right\} .
\end{align*}
For $k=1$, this can be slightly strengthened to
\[
\mathbf{P}\left\{ \left.\Upsilon_{P\Vert Q}(j)>1\,\right|\,\tilde{U}_{P}(j)\right\} \le1-\left(1-\min\left\{ \frac{dP}{dQ}(\tilde{U}_{P}(j)),\,1\right\} \right)^{j}.
\]
For $j=1$, this can be slightly strengthened to: for all $k\in\mathbb{N}$,
\begin{align*}
\mathbf{P}\left\{ \left.\Upsilon_{P\Vert Q}(1)>k\,\right|\,\tilde{U}_{P}(1)\right\}  & \le\Bigl(1-\Bigl(1+\frac{dP}{dQ}(\tilde{U}_{P}(1))\Bigr)^{-1}\Bigr)^{k}\\
 & \le1-\Bigl(1+k^{-1}\frac{dP}{dQ}(\tilde{U}_{P}(1))\Bigr)^{-1}.
\end{align*}
\end{lem}
\medskip{}

The exact distribution of $\Upsilon_{P\Vert Q}(j)$ is given in \eqref{eq:dist_exact}.

Similar to Lemma \ref{lem:cond_phidiv}, we can state a conditional
version of the generalized Poisson matching lemma. The proof follows
the same logic as Lemma \ref{lem:cond_phidiv} and is omitted.
\begin{lem}
[Conditional generalized Poisson matching lemma]\label{lem:cond_phidiv_gen}Fix
a distribution $P_{X,J,U,Y}$ and a probability kernel $Q_{U|Y}$,
satisfying $J\in\mathbb{N}$ and $P_{U|X,J}(\cdot|X,J),Q_{U|Y}(\cdot|Y)\ll\mu$
almost surely. Let $(X,J)\sim P_{X,J}$, and $\{\bar{U}_{i},T_{i}\}_{i\in\mathbb{N}}$
be the points of a Poisson process with intensity measure $\mu\times\lambda_{\mathbb{R}_{\ge0}}$
independent of $(X,J)$. Let $U=\tilde{U}_{P_{U|X,J}(\cdot|X,J)}(J)$
and $Y|(X,J,U,\{\bar{U}_{i},T_{i}\}_{i})\sim P_{Y|X,J,U}(\cdot|X,J,U)$
(note that $(X,J,U,Y)\sim P_{X,J,U,Y}$ and $Y\leftrightarrow(X,J,U)\leftrightarrow\{\bar{U}_{i},T_{i}\}_{i}$).
Then we have the following almost surely:
\[
\mathbf{E}\left[\left.\Upsilon_{P_{U|X,J}(\cdot|X,J)\Vert Q_{U|Y}(\cdot|Y)}(J)\,\right|\,X,J,U,Y\right]\le J\frac{dP_{U|X,J}(\cdot|X,J)}{dQ_{U|Y}(\cdot|Y)}(U)+1,
\]
and for all $k\in\mathbb{N}$,
\[
\mathbf{P}\left\{ \left.\Upsilon_{P_{U|X,J}(\cdot|X,J)\Vert Q_{U|Y}(\cdot|Y)}(J)>k\,\right|\,X,J,U,Y\right\} \le\min\left\{ \frac{J}{k}\frac{dP_{U|X,J}(\cdot|X,J)}{dQ_{U|Y}(\cdot|Y)}(U),\,1\right\} ,
\]
and
\[
\mathbf{P}\left\{ \left.\Upsilon_{P_{U|X,J}(\cdot|X,J)\Vert Q_{U|Y}(\cdot|Y)}(J)>1\,\right|\,X,J,U,Y\right\} \le1-\left(1-\min\left\{ \frac{dP_{U|X,J}(\cdot|X,J)}{dQ_{U|Y}(\cdot|Y)}(U),\,1\right\} \right)^{J}.
\]
If $J=1$ almost surely, then we also have the following almost surely:
for all $k\in\mathbb{N}$,
\begin{align*}
\mathbf{P}\left\{ \left.\Upsilon_{P_{U|X}(\cdot|X)\Vert Q_{U|Y}(\cdot|Y)}(1)>k\,\right|\,X,U,Y\right\}  & \le\Bigl(1-\Bigl(1+\frac{dP_{U|X}(\cdot|X)}{dQ_{U|Y}(\cdot|Y)}(U)\Bigr)^{-1}\Bigr)^{k}\\
 & \le1-\Bigl(1+k^{-1}\frac{dP_{U|X}(\cdot|X)}{dQ_{U|Y}(\cdot|Y)}(U)\Bigr)^{-1}.
\end{align*}

\end{lem}
\medskip{}

\begin{rem}
We can use the generalized Poisson matching lemma to extend Proposition
\ref{prop:channel} to the list decoding setting with fixed list size
$\mathsf{J}$. The decoder outputs the list $\{\tilde{M}_{P_{X|Y}(\cdot|Y)\times P_{M}}(j)\}_{j\in[1:\mathsf{J}]}$.
The error event becomes $(X,M)\notin\{(\tilde{X},\tilde{M})_{P_{X|Y}(\cdot|Y)\times P_{M}}(j)\}_{j\in[1:\mathsf{J}]}$.
The probability of error is bounded by $\mathbf{E}\left[(1-(1+\mathsf{L}2^{-\iota_{X;Y}(X;Y)})^{-1})^{\mathsf{J}}\right]$.
\end{rem}

\section{One-shot Coding for Broadcast Channels and Mutual Covering}

The one-shot coding setting for the broadcast channel with common
message is described as follows. Upon observing three independent
messages $M_{j}\sim\mathrm{Unif}[1:\mathsf{L}_{j}]$, $j=0,1,2$,
the encoder produces $X$, which is sent through the broadcast channel
$P_{Y_{1},Y_{2}|X}$. Decoder $j$ observes $Y_{j}$ and recovers
$\hat{M}_{0j}$ and $\hat{M}_{j}$ ($j=1,2$). The error probability
is $P_{e}=\mathbf{P}\{(M_{0},M_{0},M_{1},M_{2})\neq(\hat{M}_{01},\hat{M}_{02},\hat{M}_{1},\hat{M}_{2})\}$.

We show a one-shot version of the inner bound in \cite[Theorem 5]{liang2007broadcast}
(which is shown to be equivalent to \cite[Theorem 1]{gelfand1980capacity}
in \cite{liang2011equivalence}). The proof is given in Appendix \ref{subsec:pf_bc_cm}.
\begin{thm}
\label{thm:bc_cm}Fix any $P_{U_{0},U_{1},U_{2}}$ and function $x:\mathcal{U}_{0}\times\mathcal{U}_{1}\times\mathcal{U}_{2}\to\mathcal{X}$.
For any $\mathsf{J},\mathsf{K}_{1},\mathsf{K}_{2}\in\mathbb{N}$,
there exists a code for the broadcast channel $P_{Y_{1},Y_{2}|X}$
for independent messages $M_{j}\sim\mathrm{Unif}[1:\mathsf{L}_{j}]$,
$j=0,1,2$, with the error probability bounded by
\begin{align*}
P_{e} & \le\mathbf{E}\biggl[\min\biggl\{\tilde{\mathsf{L}}_{0}\tilde{\mathsf{L}}_{1}\mathsf{J}A2^{-\iota_{U_{0},U_{1};Y_{1}}(U_{0},U_{1};Y_{1})}+\tilde{\mathsf{L}}_{1}\mathsf{J}A2^{-\iota_{U_{1};Y_{1}|U_{0}}(U_{1};Y_{1}|U_{0})}\\
 & \;\;\;\;\;\;+\tilde{\mathsf{L}}_{0}\tilde{\mathsf{L}}_{2}\mathsf{J}^{-1}B2^{\iota_{U_{1};U_{2}|U_{0}}(U_{1};U_{2}|U_{0})-\iota_{U_{0},U_{2};Y_{2}}(U_{0},U_{2};Y_{2})}+\tilde{\mathsf{L}}_{0}\tilde{\mathsf{L}}_{2}(1-\mathsf{J}^{-1})B2^{-\iota_{U_{0},U_{2};Y_{2}}(U_{0},U_{2};Y_{2})}\\
 & \;\;\;\;\;\;+\tilde{\mathsf{L}}_{2}\mathsf{J}^{-1}B2^{\iota_{U_{1};U_{2}|U_{0}}(U_{1};U_{2}|U_{0})-\iota_{U_{2},Y_{2}|U_{0}}(U_{2};Y_{2}|U_{0})}+\tilde{\mathsf{L}}_{2}(1-\mathsf{J}^{-1})B2^{-\iota_{U_{2},Y_{2}|U_{0}}(U_{2};Y_{2}|U_{0})},\,1\biggr\}\biggr]
\end{align*}
if all the information density terms are defined almost surely, where
\begin{align*}
\tilde{\mathsf{L}}_{0} & :=\mathsf{L}_{0}\mathsf{K}_{1}\mathsf{K}_{2},\\
\tilde{\mathsf{L}}_{a} & :=\lceil\mathsf{L}_{a}/\mathsf{K}_{a}\rceil\;\mathrm{for}\,a=1,2,\\
A & :=(\log(\tilde{\mathsf{L}}_{1}^{-1}\mathsf{J}^{-1}2^{\iota_{U_{1};Y_{1}|U_{0}}(U_{1};Y_{1}|U_{0})}+1)+1)^{2},\\
B & :=\bigl(\log((\tilde{\mathsf{L}}_{2}\mathsf{J}^{-1}2^{\iota_{U_{1};U_{2}|U_{0}}(U_{1};U_{2}|U_{0})-\iota_{U_{2},Y_{2}|U_{0}}(U_{2};Y_{2}|U_{0})}\\
 & \;\;\;\;+\tilde{\mathsf{L}}_{2}(1-\mathsf{J}^{-1})2^{-\iota_{U_{2},Y_{2}|U_{0}}(U_{2};Y_{2}|U_{0})})^{-1}+1)+1\bigr)^{2}.
\end{align*}
As a result, for $\gamma>0$,
\begin{align}
P_{e} & \le\mathbf{P}\biggl\{\log\tilde{\mathsf{L}}_{1}\mathsf{J}>\iota_{U_{1};Y_{1}|U_{0}}(U_{1};Y_{1}|U_{0})-\gamma\;\;\mathrm{or}\;\;\log\tilde{\mathsf{L}}_{2}>\iota_{U_{2},Y_{2}|U_{0}}(U_{2};Y_{2}|U_{0})-\gamma\nonumber \\
 & \;\;\;\;\mathrm{or}\;\;\log\tilde{\mathsf{L}}_{2}\mathsf{J}^{-1}>\iota_{U_{2},Y_{2}|U_{0}}(U_{2};Y_{2}|U_{0})-\iota_{U_{1};U_{2}|U_{0}}(U_{1};U_{2}|U_{0})-\gamma\nonumber \\
 & \;\;\;\;\mathrm{or}\;\;\log\tilde{\mathsf{L}}_{0}\tilde{\mathsf{L}}_{1}\mathsf{J}>\iota_{U_{0},U_{1};Y_{1}}(U_{0},U_{1};Y_{1})-\gamma\;\;\mathrm{or}\;\;\log\tilde{\mathsf{L}}_{0}\tilde{\mathsf{L}}_{2}>\iota_{U_{0},U_{2};Y_{2}}(U_{0},U_{2};Y_{2})-\gamma\nonumber \\
 & \;\;\;\;\mathrm{or}\;\;\log\tilde{\mathsf{L}}_{0}\tilde{\mathsf{L}}_{2}\mathsf{J}^{-1}>\iota_{U_{0},U_{2};Y_{2}}(U_{0},U_{2};Y_{2})-\iota_{U_{1};U_{2}|U_{0}}(U_{1};U_{2}|U_{0})-\gamma\biggr\}\nonumber \\
 & +2^{-\gamma}\left(8\mathbf{E}\left[(\iota_{U_{1};Y_{1}|U_{0}}(U_{1};Y_{1}|U_{0}))^{2}+(\iota_{U_{2},Y_{2}|U_{0}}(U_{2};Y_{2}|U_{0}))^{2}\right]+12\gamma^{2}+84\right).\label{eq:bc_pe2}
\end{align}
\end{thm}
The logarithmic terms $A$ and $B$ (or the last term in \eqref{eq:bc_pe2})
result in an $O(n^{-1}\log n)$ penalty on the rate in the finite
blocklength regime, and do not affect the second order result. Ignoring
the last term in \eqref{eq:bc_pe2}, the error event in \eqref{eq:bc_pe2}
is a strict subset of those in \cite[eqn (32)]{yassaee2013oneshot}
 and \cite[eqn (49)]{liu2015oneshotmutual}. This is because the error
event in \cite{yassaee2013oneshot} is a superset of \eqref{eq:bc_pe2}
by Fourier-Motzkin elimination on $\mathsf{J}_{2}$ in the error event
in \cite{yassaee2013oneshot}, but the reverse is not true since Fourier-Motzkin
elimination only guarantees the existence of a random variable for
$\mathsf{J}_{2}$ (that depends on the information density terms)
satisfying the bounds, but $\mathsf{J}_{2}$ must be a constant since
it is a parameter of the code construction in \cite{yassaee2013oneshot}.

Theorem \ref{thm:bc_cm} gives the following second order bound. Consider
$n$ independent channel uses. Let $\mathsf{L}_{a}=2^{nR_{a}}$ for
$a=0,1,2$. By the multi-dimensional Berry-Esseen theorem \cite{bentkus2003dependence}
(using the notation in \cite{yassaee2013oneshot}), we have $P_{e}\le\epsilon$
if there exists $\bar{R},\hat{R}_{1},\hat{R}_{2}\ge0$ such that
\[
\left[\begin{array}{c}
\tilde{R}_{1}+\bar{R}\\
\tilde{R}_{2}\\
\tilde{R}_{2}-\bar{R}\\
\tilde{R}_{0}+\tilde{R}_{1}+\bar{R}\\
\tilde{R}_{0}+\tilde{R}_{2}\\
\tilde{R}_{0}+\tilde{R}_{2}-\bar{R}
\end{array}\right]\in\mathbf{E}[I]-\frac{1}{\sqrt{n}}\mathcal{Q}^{-1}\left(\mathrm{Cov}[I],\,\epsilon-\frac{\beta}{\sqrt{n}}\right)-\frac{\beta\log n}{n}
\]
if $n>\beta^{2}\epsilon^{-2}$, where $\beta$ is a constant that
depends on $P_{U_{0},U_{1},U_{2},Y_{1},Y_{2}}$, and $\tilde{R}_{0}=R_{0}+\hat{R}_{1}+\hat{R}_{2}$,
$\tilde{R}_{a}=R_{a}-\hat{R}_{a}$ for $a=1,2$, and
\[
I=\left[\begin{array}{c}
\iota_{U_{1};Y_{1}|U_{0}}(U_{1};Y_{1}|U_{0})\\
\iota_{U_{2},Y_{2}|U_{0}}(U_{2};Y_{2}|U_{0})\\
\iota_{U_{2},Y_{2}|U_{0}}(U_{2};Y_{2}|U_{0})-\iota_{U_{1};U_{2}|U_{0}}(U_{1};U_{2}|U_{0})\\
\iota_{U_{0},U_{1};Y_{1}}(U_{0},U_{1};Y_{1})\\
\iota_{U_{0},U_{2};Y_{2}}(U_{0},U_{2};Y_{2})\\
\iota_{U_{0},U_{2};Y_{2}}(U_{0},U_{2};Y_{2})-\iota_{U_{1};U_{2}|U_{0}}(U_{1};U_{2}|U_{0})
\end{array}\right].
\]

To demonstrate the use of the generalized Poisson matching lemma in
place of the mutual covering lemma, we prove a one-shot version of
Marton's inner bound without common message \cite{marton1979broadcast}
(i.e., $\mathsf{L}_{0}=1$). Our bound is stronger than that in \cite{verdu2012nonasymp}
in the sense that our bound implies \cite{verdu2012nonasymp} (with
a slight penalty of having $2^{1-\gamma}+2^{-2\gamma}$ instead of
$2^{1-\gamma}+e^{-2^{\gamma}}$), but \cite{verdu2012nonasymp} does
not imply our bound. We also note that a finite-blocklength bound
is given in \cite{yassaee2013binning}. Nevertheless, the analysis
in \cite{yassaee2013binning} only works for discrete auxiliary random
variables $U_{1},U_{2}$, and does not appear to yield a one-shot
bound due to the use of typical sequences.

\begin{figure}
\begin{centering}
\includegraphics{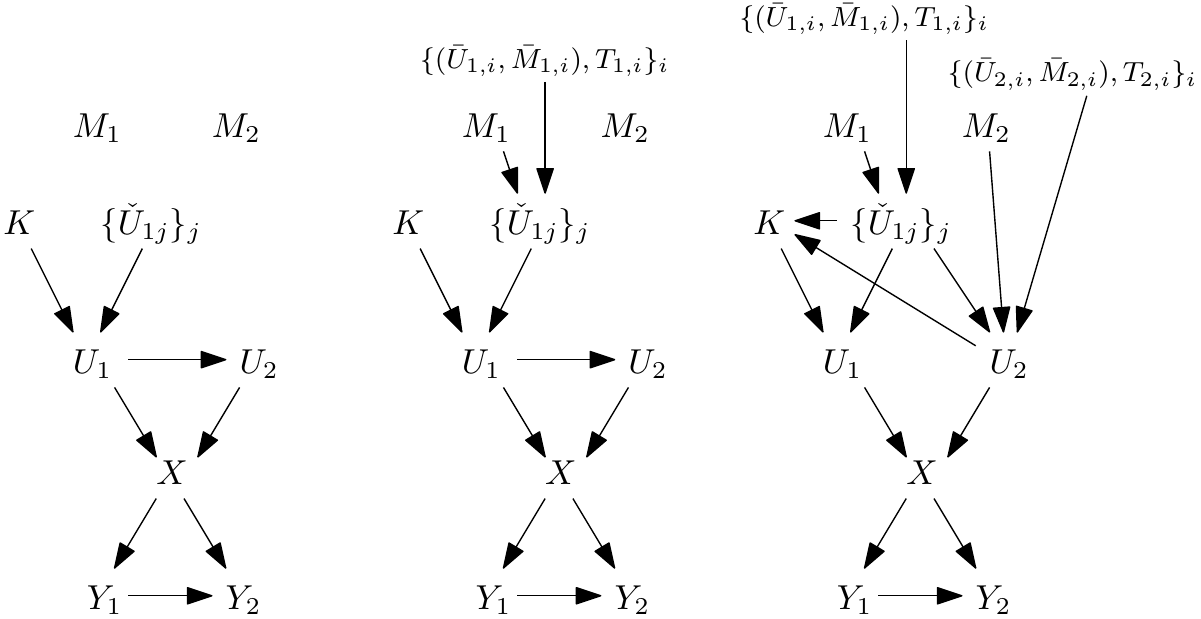}
\par\end{centering}
\caption{\label{fig:bc_bayes}Left: The Bayesian network described in \eqref{eq:bc_dist}.
Middle: The Bayesian network deduced from \eqref{eq:bc_dist} and
$\check{U}_{1j}=(\tilde{U}_{1})_{P_{U_{1}}\times\delta_{M_{1}}}(j)$.
Right: The Bayesian network describing the encoding scheme. Note that
all three are valid Bayesian networks, and the desired conditional
independence relations can be deduced using d-separation.}

\end{figure}

In the conventional mutual covering approach in \cite{yassaee2013oneshot,liu2015oneshotmutual},
sub-codebooks for both $U_{1}$ and $U_{2}$ are generated, whereas
in our approach we generate a sub-codebook only for $U_{1}$, and
the codebook of $U_{2}$ adapts to the sub-codebook automatically,
eliminating the need for a sub-codebook for $U_{2}$.
\begin{thm}
\label{thm:bc_ncm}Fix any $P_{U_{1},U_{2}}$ and function $x:\mathcal{U}_{1}\times\mathcal{U}_{2}\to\mathcal{X}$.
For any $\mathsf{J}\in\mathbb{N}$, there exists a code for the broadcast
channel $P_{Y_{1},Y_{2}|X}$ for independent private messages $M_{j}\sim\mathrm{Unif}[1:\mathsf{L}_{j}]$,
$j=1,2$, with the error probability bounded by
\begin{align*}
P_{e} & \le\mathbf{E}\biggl[\min\biggl\{\mathsf{L}_{1}\mathsf{J}2^{-\iota_{U_{1};Y_{1}}(U_{1};Y_{1})}+\mathsf{L}_{2}(1-\mathsf{J}^{-1})2^{-\iota_{U_{2};Y_{2}}(U_{2};Y_{2})}+\mathsf{L}_{2}\mathsf{J}^{-1}2^{\iota_{U_{1};U_{2}}(U_{1};U_{2})-\iota_{U_{2};Y_{2}}(U_{2};Y_{2})},\,1\biggr\}\biggr]
\end{align*}
if all the information density terms are defined, where $(U_{1},U_{2},X,Y_{1},Y_{2})\sim P_{U_{1}U_{2}}\delta_{x(U_{1},U_{2})}P_{Y_{1},Y_{2}|X}$.
\end{thm}
\begin{IEEEproof}
Let $\{(\bar{U}_{1,i},\bar{M}_{1,i}),T_{1,i}\}_{i\in\mathbb{N}}$,
$\{(\bar{U}_{2,i},\bar{M}_{2,i}),T_{2,i}\}_{i\in\mathbb{N}}$ be two
independent Poisson processes with intensity measures $P_{U_{1}}\times P_{M_{1}}\times\lambda_{\mathbb{R}_{\ge0}}$
and $P_{U_{2}}\times P_{M_{2}}\times\lambda_{\mathbb{R}_{\ge0}}$
respectively, independent of $M_{1},M_{2}$.

The encoder would generate $X$ such that
\begin{equation}
(M_{1},M_{2},K,\{\check{U}_{1j}\}_{j\in[1:\mathsf{J}]},U_{1},U_{2},X)\sim P_{M_{1}}\times P_{M_{2}}\times P_{K}P_{U_{1}}^{\otimes\mathsf{J}}\delta_{\check{U}_{1K}}P_{U_{2}|U_{1}}\delta_{x(U_{1},U_{2})},\label{eq:bc_dist}
\end{equation}
where $P_{K}=\mathrm{Unif}[1:\mathsf{J}]$, and $\{\check{U}_{1j}\}_{j\in[1:\mathsf{J}]}\in\mathcal{U}_{1}^{\mathsf{J}}$
is an intermediate list (which can be regarded as a sub-codebook).
The term $P_{U_{1}}^{\otimes\mathsf{J}}\delta_{\check{U}_{1K}}$ in
\eqref{eq:bc_dist} means that $\{\check{U}_{1j}\}_{j}$ are i.i.d.
$P_{U_{1}}$, and $U_{1}=\check{U}_{1K}$. To accomplish this, the
encoder computes $\check{U}_{1j}=(\tilde{U}_{1})_{P_{U_{1}}\times\delta_{M_{1}}}(j)$
for $j=1,\ldots,\mathsf{J}$ (which Poisson process we are referring
to can be deduced from whether we are discussing $U_{1}$ or $U_{2}$),
$U_{2}=(\tilde{U}_{2})_{\mathsf{J}^{-1}\sum_{j=1}^{\mathsf{J}}P_{U_{2}|U_{1}}(\cdot|\check{U}_{1j})\times\delta_{M_{2}}}$,
and $(K,U_{1})|(\{\check{U}_{1j}\}_{j},U_{2})\sim P_{K,U_{1}|\{\check{U}_{1j}\}_{j},U_{2}}$
(where $P_{K,U_{1}|\{\check{U}_{1j}\}_{j},U_{2}}$ is derived from
\eqref{eq:bc_dist}), and outputs $X=x(U_{1},U_{2})$. It can be verified
that \eqref{eq:bc_dist} is satisfied.

The decoding functions are $\hat{M}_{1}=(\tilde{M}_{1})_{P_{U_{1}|Y_{1}}(\cdot|Y_{1})\times P_{M_{1}}}$,
$\hat{M}_{2}=(\tilde{M}_{2})_{P_{U_{2}|Y_{2}}(\cdot|Y_{2})\times P_{M_{2}}}$.
We have the following almost surely:
\begin{align*}
 & \mathbf{P}\biggl\{(\tilde{U}_{1},\tilde{M}_{1})_{P_{U_{1}|Y_{1}}(\cdot|Y_{1})\times P_{M_{1}}}\neq(U_{1},M_{1})\,\biggl|\,U_{1},U_{2},Y_{1},Y_{2},M_{1},K\biggr\}\\
 & \stackrel{(a)}{=}\mathbf{P}\biggl\{(\tilde{U}_{1},\tilde{M}_{1})_{P_{U_{1}|Y_{1}}(\cdot|Y_{1})\times P_{M_{1}}}\neq(U_{1},M_{1})\,\biggl|\,U_{1},Y_{1},M_{1},K\biggr\}\\
 & \stackrel{(b)}{\le}K\frac{dP_{U_{1}}\times\delta_{M_{1}}}{dP_{U_{1}|Y_{1}}(\cdot|Y_{1})\times P_{M_{1}}}(U_{1},M_{1})\\
 & \le\mathsf{L}_{1}\mathsf{J}2^{-\iota_{U_{1};Y_{1}}(U_{1};Y_{1})},
\end{align*}
where (a) is by $(U_{2},Y_{2})\leftrightarrow(U_{1},Y_{1},M_{1},K)\leftrightarrow\{(\bar{U}_{1,i},\bar{M}_{1,i}),T_{1,i}\}_{i}$
(see Figure \ref{fig:bc_bayes} middle), and (b) is by the conditional
generalized Poisson matching lemma on $(X,J,U,Y,Q_{U|Y})\leftarrow(M_{1},\,K,\,(U_{1},M_{1}),\,Y_{1},\,P_{U_{1}|Y_{1}}\times P_{M_{1}})$,
since $P_{U_{1},M_{1}|M_{1},K}=P_{U_{1}}\times\delta_{M_{1}}$, $(M_{1},K)\perp\!\!\!\perp\{(\bar{U}_{1,i},\bar{M}_{1,i}),T_{1,i}\}_{i}$,
and $Y_{1}\leftrightarrow(U_{1},M_{1},K)\leftrightarrow\{(\bar{U}_{1,i},\bar{M}_{1,i}),T_{1,i}\}_{i}$,
which can be deduced from \eqref{eq:bc_dist} and $\check{U}_{1j}=(\tilde{U}_{1})_{P_{U_{1}}\times\delta_{M_{1}}}(j)$
(see Figure \ref{fig:bc_bayes} middle).

Also, almost surely,
\begin{align*}
 & \mathbf{P}\left\{ \left.(\tilde{U}_{2},\tilde{M}_{2})_{P_{U_{2}|Y_{2}}(\cdot|Y_{2})\times P_{M_{2}}}\neq(U_{2},M_{2})\,\right|\,U_{1},U_{2},Y_{1},Y_{2},M_{2}\right\} \\
 & \stackrel{(a)}{=}\mathbf{P}\left\{ \left.(\tilde{U}_{2},\tilde{M}_{2})_{P_{U_{2}|Y_{2}}(\cdot|Y_{2})\times P_{M_{2}}}\neq(U_{2},M_{2})\,\right|\,U_{1},U_{2},Y_{2},M_{2}\right\} \\
 & \stackrel{(b)}{\le}\mathbf{E}\left[\left.\frac{d(\mathsf{J}^{-1}\sum_{j=1}^{\mathsf{J}}P_{U_{2}|U_{1}}(\cdot|\check{U}_{1j}))\times\delta_{M_{2}}}{dP_{U_{2}|Y_{2}}(\cdot|Y_{2})\times P_{M_{2}}}(U_{2},M_{2})\,\right|\,U_{1},U_{2},Y_{2},M_{2}\right]\\
 & =\mathbf{E}\left[\left.\mathsf{L}_{2}\mathsf{J}^{-1}\sum_{j=1}^{\mathsf{J}}2^{\iota_{U_{1};U_{2}}(\check{U}_{1j};U_{2})-\iota_{U_{2};Y_{2}}(U_{2};Y_{2})}\,\right|\,U_{1},U_{2},Y_{2},M_{2}\right]\\
 & =\mathbf{E}\left[\left.\mathsf{L}_{2}\mathsf{J}^{-1}2^{-\iota_{U_{2};Y_{2}}(U_{2};Y_{2})}\Bigl(2^{\iota_{U_{1};U_{2}}(U_{1};U_{2})}+\sum_{j\in[1:\mathsf{J}]\backslash K}2^{\iota_{U_{1};U_{2}}(\check{U}_{1j};U_{2})}\Bigr)\,\right|\,U_{1},U_{2},Y_{2},M_{2}\right]\\
 & =\mathbf{E}\left[\left.\mathsf{L}_{2}\mathsf{J}^{-1}2^{-\iota_{U_{2};Y_{2}}(U_{2};Y_{2})}\Bigl(2^{\iota_{U_{1};U_{2}}(U_{1};U_{2})}+\sum_{j=1}^{\mathsf{J}-1}2^{\iota_{U_{1};U_{2}}(\check{U}_{1,j+\mathbf{1}\{j\ge K\}};U_{2})}\Bigr)\,\right|\,U_{1},U_{2},Y_{2},M_{2}\right]\\
 & \stackrel{(c)}{\le}\mathsf{L}_{2}\mathsf{J}^{-1}2^{-\iota_{U_{2};Y_{2}}(U_{2};Y_{2})}(2^{\iota_{U_{1};U_{2}}(U_{1};U_{2})}+\mathsf{J}-1),
\end{align*}
where (a) is by $Y_{1}\leftrightarrow(U_{1},U_{2},Y_{2},M_{2})\leftrightarrow\{(\bar{U}_{2,i},\bar{M}_{2,i}),T_{2,i}\}_{i}$
(see Figure \ref{fig:bc_bayes} right), (b) is by the conditional
Poisson matching lemma on $((\{\check{U}_{1j}\}_{j},M_{2}),\,(U_{2},M_{2}),\,Y_{2},\,P_{U_{2}|Y_{2}}\times P_{M_{2}})$,
and (c) is because $\{\check{U}_{1,j+\mathbf{1}\{j\ge K\}}\}_{j\in[1:\mathsf{J}-1]}$
(the $\check{U}_{1j}$'s not selected as $U_{1}$) are independent
of $(U_{1},U_{2},Y_{2},M_{2})$, $\mathbf{E}[2^{\iota_{U_{1};U_{2}}(\check{U}_{1,j+\mathbf{1}\{j\ge K\}};U_{2})}\,|\,U_{2}]=1$,
and Jensen's inequality. Hence,
\begin{align*}
 & \mathbf{P}\{(M_{1},M_{2})\neq(\hat{M}_{1},\hat{M}_{2})\}\\
 & =\mathbf{E}\left[\mathbf{P}\left\{ \left.(M_{1},M_{2})\neq(\hat{M}_{1},\hat{M}_{2})\,\right|\,U_{1},U_{2},Y_{1},Y_{2}\right\} \right]\\
 & \le\mathbf{E}\left[\min\left\{ \mathbf{P}\left\{ \left.M_{1}\neq\hat{M_{1}}\,\right|\,U_{1},U_{2},Y_{1},Y_{2}\right\} +\mathbf{P}\left\{ \left.M_{2}\neq\hat{M_{2}}\,\right|\,U_{1},U_{2},Y_{1},Y_{2}\right\} ,\,1\right\} \right]\\
 & \le\mathbf{E}\biggl[\min\Bigl\{\mathsf{L}_{1}\mathsf{J}2^{-\iota_{U_{1};Y_{1}}(U_{1};Y_{1})}+\mathsf{L}_{2}\mathsf{J}^{-1}2^{-\iota_{U_{2};Y_{2}}(U_{2};Y_{2})}(2^{\iota_{U_{1};U_{2}}(U_{1};U_{2})}+\mathsf{J}-1),\,1\Bigr\}\biggr].
\end{align*}
 Therefore there exist fixed realizations of the Poisson processes
attaining the desired bound.
\end{IEEEproof}

\section{One-shot Distributed Lossy Source Coding}

The one-shot distributed lossy source coding setting is described
as follows. Let $(X_{1},X_{2})\sim P_{X_{1},X_{2}}$. Upon observing
$X_{j}$, encoder $j$ produces $M_{j}\in[1:\mathsf{L}_{j}]$, $j=1,2$.
The decoder observes $M_{1},M_{2}$ and recovers $\hat{Z}_{1}\in\mathcal{Z}_{1}$,
$\hat{Z}_{2}\in\mathcal{Z}_{2}$ with probability of excess distortion
$P_{e}=\mathbf{P}\{\mathsf{d}_{1}(X_{1},\hat{Z}_{1})>\mathsf{D}_{1}\;\mathrm{or}\;\mathsf{d}_{2}(X_{2},\hat{Z}_{2})>\mathsf{D}_{2}\}$,
where $\mathsf{d}_{j}:\mathcal{X}_{j}\times\mathcal{Z}_{j}\to\mathbb{R}_{\ge0}$
is a distortion measure for $j=1,2$.

We show a one-shot version of the Berger-Tung inner bound \cite{berger1978multiterminal,tung1978multiterminal}.
\begin{thm}
\label{thm:dlsc}Fix any $P_{U_{1}|X_{1}}$, $P_{U_{2}|X_{2}}$ and
functions $z_{j}:\mathcal{U}_{1}\times\mathcal{U}_{2}\to\mathcal{Z}_{j}$,
$j=1,2$. There exists a code for distributed lossy source coding
with sources $P_{X_{1}},P_{X_{2}}$ and message sizes $\mathsf{L}_{1},\mathsf{L}_{2}$,
with probability of excess distortion
\begin{align}
P_{e} & \le\mathbf{E}\biggl[\min\biggl\{\mathbf{1}\{\mathsf{d}_{1}(X_{1},Z_{1})>\mathsf{D}_{1}\;\mathrm{or}\;\mathsf{d}_{2}(X_{2},Z_{2})>\mathsf{D}_{2}\}+\mathsf{L}_{1}^{-1}2^{\iota_{U_{1};X_{1}|U_{2}}(U_{1};X_{1}|U_{2})}\nonumber \\
 & \;\;\;\;+\left(\mathsf{L}_{1}^{-1}\mathsf{L}_{2}^{-1}2^{\iota_{U_{1},U_{2};X_{1},X_{2}}(U_{1},U_{2};X_{1},X_{2})}+\mathsf{L}_{2}^{-1}2^{\iota_{U_{2};X_{2}|U_{1}}(U_{2};X_{2}|U_{1})}\right)\left(\log(\mathsf{L}_{2}2^{-\iota_{U_{2};X_{2}|U_{1}}(U_{2};X_{2}|U_{1})}+1)+1\right)^{2},\,1\biggr\}\biggr]\label{eq:dlsc_pe1}
\end{align}
if all the information density terms are defined, where $(X_{1},X_{2},U_{1},U_{2},Z_{1},Z_{2})\sim P_{X_{1},X_{2}}P_{U_{1}|X_{1}}P_{U_{2}|X_{2}}\delta_{z_{1}(U_{1},U_{2})}\delta_{z_{2}(U_{1},U_{2})}$.
As a result, for $\gamma>0$,
\begin{align}
P_{e} & \le\mathbf{P}\biggl\{\mathsf{d}_{1}(X_{1},Z_{1})>\mathsf{D}_{1}\;\mathrm{or}\;\mathsf{d}_{2}(X_{2},Z_{2})>\mathsf{D}_{2}\;\;\mathrm{or}\;\;\log\mathsf{L}_{1}<\iota_{U_{1};X_{1}|U_{2}}(U_{1};X_{1}|U_{2})+\gamma\nonumber \\
 & \;\;\;\;\mathrm{or}\;\;\log\mathsf{L}_{2}<\iota_{U_{2};X_{2}|U_{1}}(U_{2};X_{2}|U_{1})+\gamma\;\;\mathrm{or}\;\;\log\mathsf{L}_{1}\mathsf{L}_{2}<\iota_{U_{1},U_{2};X_{1},X_{2}}(U_{1},U_{2};X_{1},X_{2})+\gamma\biggr\}\nonumber \\
 & \;\;\;+2^{-\gamma}\left(4\mathbf{E}[(\iota_{U_{1};U_{2}}(U_{1};U_{2}))^{2}]+4\gamma^{2}+29\right).\label{eq:dlsc_pe2}
\end{align}
\end{thm}
The logarithmic term in \eqref{eq:dlsc_pe1} (or the last term in
\eqref{eq:dlsc_pe2}) results in an $O(n^{-1}\log n)$ penalty on
the rate in the finite blocklength regime, and does not affect the
second order result. Ignoring the last term in \eqref{eq:dlsc_pe2},
the error event in \eqref{eq:dlsc_pe2} is a strict subset of that
in \cite[eqn (47)]{yassaee2013oneshot}. This is because the error
event in \cite{yassaee2013oneshot} is a superset of \eqref{eq:dlsc_pe2}
by Fourier-Motzkin elimination on $\mathsf{J}_{1},\mathsf{J}_{2}$
in the error event in \cite{yassaee2013oneshot}, but the reverse
is not true since Fourier-Motzkin elimination only guarantees the
existence of random variables for $\mathsf{J}_{1},\mathsf{J}_{2}$
(that depend on the information density terms) satisfying the bounds,
but $\mathsf{J}_{1},\mathsf{J}_{2}$ must be constants since they
are parameters of the code construction in \cite{yassaee2013oneshot}.

We now prove the result. Unlike previous approaches, our proof does
not require binning. The encoders are the same as those for point-to-point
lossy source coding.
\begin{IEEEproof}
Let $\{(\bar{U}_{1,i},\bar{M}_{1,i}),T_{1,i}\}_{i\in\mathbb{N}}$,
$\{(\bar{U}_{2,i},\bar{M}_{2,i}),T_{2,i}\}_{i\in\mathbb{N}}$ be two
independent Poisson processes with intensity measures $P_{U_{1}}\times P_{M_{1}}\times\lambda_{\mathbb{R}_{\ge0}}$
and $P_{U_{2}}\times P_{M_{2}}\times\lambda_{\mathbb{R}_{\ge0}}$
respectively, independent of $X_{1},X_{2}$. The encoding functions
are $M_{j}=(\tilde{M}_{j})_{P_{U_{j}|X_{j}}(\cdot|X_{j})\times P_{M_{j}}}$,
$j=1,2$ (which Poisson process we are referring to can be deduced
from whether we are discussing $M_{1}$ or $M_{2}$). Also define
$U_{j}=(\tilde{U}_{j})_{P_{U_{j}|X_{j}}(\cdot|X_{j})\times P_{M_{j}}}$,
$Z_{j}=z_{j}(U_{1},U_{2})$, $j=1,2$. For the decoding function,
let $\check{U}_{1k}=(\tilde{U}_{1})_{P_{U_{1}}\times\delta_{M_{1}}}(k)$
for $k\in\mathbb{N}$, $\hat{U}_{2}=(\tilde{U}_{2})_{\sum_{k=1}^{\infty}\phi(k)P_{U_{2}|U_{1}}(\cdot|\check{U}_{1k})\times\delta_{M_{2}}}$
where $\phi(k)\propto k^{-1}(\log(k+2))^{-2}$ with $\sum_{k=1}^{\infty}\phi(k)=1$,
and $\hat{U}_{1}=(\tilde{U}_{1})_{P_{U_{1}|U_{2}}(\cdot|\hat{U}_{2})\times\delta_{M_{1}}}$,
$\hat{Z}_{j}=z_{j}(\hat{U}_{1},\hat{U}_{2})$, $j=1,2$. Note that
$(M_{1},M_{2},X_{1},X_{2},U_{1},U_{2},Z_{1},Z_{2})\sim P_{M_{1}}\times P_{M_{2}}\times P_{X_{1},X_{2}}P_{U_{1}|X_{1}}P_{U_{2}|X_{2}}\delta_{z_{1}(U_{1},U_{2})}\delta_{z_{2}(U_{1},U_{2})}$.

Let $K=\Upsilon_{P_{U_{1}|X_{1}}(\cdot|X_{1})\times P_{M_{1}}\Vert P_{U_{1}}\times\delta_{M_{1}}}(1)$
(using the Poisson process $\{(\bar{U}_{1,i},\bar{M}_{1,i}),T_{1,i}\}_{i\in\mathbb{N}}$).
By the conditional generalized Poisson matching lemma on $(X_{1},\,1,\,(U_{1},M_{1}),\,M_{1},\,P_{U_{1}}\times\delta_{M_{1}})$
(note that $P_{U_{1},M_{1}|X_{1}}=P_{U_{1}|X_{1}}\times P_{M_{1}}$),
almost surely,
\begin{align}
\mathbf{E}\left[\left.K\,\right|\,X_{1},U_{1},M_{1}\right] & \le\frac{dP_{U_{1}|X_{1}}(\cdot|X_{1})\times P_{M_{1}}}{dP_{U_{1}}\times\delta_{M_{1}}}(U_{1},M_{1})+1\nonumber \\
 & =\mathsf{L}_{1}^{-1}2^{\iota_{U_{1};X_{1}}(U_{1};X_{1})}+1.\label{eq:dlsc_ek}
\end{align}
Since $\{\check{U}_{1k}\}_{k}$ is a function of $\{(\bar{U}_{1,i},\bar{M}_{1,i}),T_{1,i}\}_{i}$
and $M_{1}$, we have $\{\check{U}_{1k}\}_{k}\leftrightarrow(X_{1},X_{2},U_{1},U_{2},M_{2})\leftrightarrow\{(\bar{U}_{2,i},\bar{M}_{2,i}),T_{2,i}\}_{i}$.
By the conditional Poisson matching lemma on $(X_{2},\,(U_{2},M_{2}),\,(\{\check{U}_{1k}\}_{k},M_{2}),\,\sum_{k=1}^{\infty}\phi(k)P_{U_{2}|U_{1}}(\cdot|\check{U}_{1k})\times\delta_{M_{2}})$
(note that $P_{U_{2},M_{2}|X_{2}}=P_{U_{2}|X_{2}}\times P_{M_{2}}$),
almost surely,
\begin{align*}
 & \mathbf{P}\biggl\{\left.(\tilde{U}_{2},\tilde{M}_{2})_{\sum_{k=1}^{\infty}\phi(k)P_{U_{2}|U_{1}}(\cdot|\check{U}_{1k})\times\delta_{M_{2}}}\neq(U_{2},M_{2})\,\right|\,X_{1},X_{2},U_{1},U_{2},M_{2}\biggr\}\\
 & \le\mathbf{E}\left[\left.\min\biggl\{\frac{dP_{U_{2}|X_{2}}(\cdot|X_{2})\times P_{M_{2}}}{d(\sum_{k=1}^{\infty}\phi(k)P_{U_{2}|U_{1}}(\cdot|\check{U}_{1k}))\times\delta_{M_{2}}}(U_{2},M_{2}),\,1\biggr\}\,\right|\,X_{1},X_{2},U_{1},U_{2},M_{2}\right]\\
 & \le\mathbf{E}\left[\left.\min\biggl\{\mathsf{L}_{2}^{-1}\frac{dP_{U_{2}|X_{2}}(\cdot|X_{2})}{\phi(K)dP_{U_{2}|U_{1}}(\cdot|U_{1})}(U_{2}),\,1\biggr\}\,\right|\,X_{1},X_{2},U_{1},U_{2},M_{2}\right]\\
 & =\mathbf{E}\left[\left.\min\{\mathsf{L}_{2}^{-1}(\phi(K))^{-1}2^{\iota_{U_{2};X_{2}|U_{1}}(U_{2};X_{2}|U_{1})},\,1\}\,\right|\,X_{1},X_{2},U_{1},U_{2},M_{2}\right]\\
 & \stackrel{(a)}{\le}\mathbf{E}\left[\left.K\mathsf{L}_{2}^{-1}2^{\iota_{U_{2};X_{2}|U_{1}}(U_{2};X_{2}|U_{1})}\left(\log(\mathsf{L}_{2}2^{-\iota_{U_{2};X_{2}|U_{1}}(U_{2};X_{2}|U_{1})}+1)+1\right)^{2}\,\right|\,X_{1},X_{2},U_{1},U_{2},M_{2}\right]\\
 & \stackrel{(b)}{\le}\left(\mathsf{L}_{1}^{-1}2^{\iota_{U_{1};X_{1}}(U_{1};X_{1})}+1\right)\mathsf{L}_{2}^{-1}2^{\iota_{U_{2};X_{2}|U_{1}}(U_{2};X_{2}|U_{1})}\left(\log(\mathsf{L}_{2}2^{-\iota_{U_{2};X_{2}|U_{1}}(U_{2};X_{2}|U_{1})}+1)+1\right)^{2}\\
 & =\left(\mathsf{L}_{1}^{-1}\mathsf{L}_{2}^{-1}2^{\iota_{U_{1},U_{2};X_{1},X_{2}}(U_{1},U_{2};X_{1},X_{2})}+\mathsf{L}_{2}^{-1}2^{\iota_{U_{2};X_{2}|U_{1}}(U_{2};X_{2}|U_{1})}\right)\left(\log(\mathsf{L}_{2}2^{-\iota_{U_{2};X_{2}|U_{1}}(U_{2};X_{2}|U_{1})}+1)+1\right)^{2},
\end{align*}
where (a) is by Proposition \ref{prop:phi_ineq}, and (b) is by $K\leftrightarrow(U_{1},X_{1})\leftrightarrow(X_{2},U_{2},M_{2})$,
\eqref{eq:dlsc_ek} and Jensen's inequality. By the conditional Poisson
matching lemma on $(X_{1},\,(U_{1},M_{1}),\,(U_{2},M_{1}),\,P_{U_{1}|U_{2}}\times\delta_{M_{1}})$
(note that $P_{U_{1},M_{1}|X_{1}}=P_{U_{1}|X_{1}}\times P_{M_{1}}$),
and $X_{2}\leftrightarrow(X_{1},U_{1},U_{2},M_{1})\leftrightarrow\{(\bar{U}_{1,i},\bar{M}_{1,i}),T_{1,i}\}_{i}$,
almost surely,
\begin{align*}
 & \mathbf{P}\biggl\{(\tilde{U}_{1},\tilde{M}_{1})_{P_{U_{1}|U_{2}}(\cdot|U_{2})\times\delta_{M_{1}}}\neq(U_{1},M_{1})\,\biggl|\,X_{1},X_{2},U_{1},U_{2},M_{1}\biggr\}\\
 & \le\frac{dP_{U_{1}|X_{1}}(\cdot|X_{1})\times P_{M_{1}}}{dP_{U_{1}|U_{2}}(\cdot|U_{2})\times\delta_{M_{1}}}(U_{1},M_{1})\\
 & =\mathsf{L}_{1}^{-1}2^{\iota_{U_{1};X_{1}|U_{2}}(U_{1};X_{1}|U_{2})}.
\end{align*}
We have
\begin{align*}
 & \mathbf{P}\{\mathsf{d}_{1}(X_{1},\hat{Z}_{1})>\mathsf{D}_{1}\;\mathrm{or}\;\mathsf{d}_{2}(X_{2},\hat{Z}_{2})>\mathsf{D}_{2}\}\\
 & \le\mathbf{E}\biggl[\mathbf{P}\biggl\{\mathsf{d}_{1}(X_{1},Z_{1})>\mathsf{D}_{1}\;\mathrm{or}\;\mathsf{d}_{2}(X_{2},Z_{2})>\mathsf{D}_{2}\;\mathrm{or}\;\hat{U}_{2}\neq U_{2}\\
 & \;\;\;\;\;\mathrm{or}\;(\hat{U}_{2}=U_{2}\;\mathrm{and}\;\hat{U}_{1}\neq U_{1})\,\biggl|\,X_{1},X_{2},U_{1},U_{2}\biggr\}\biggr]\\
 & \le\mathbf{E}\biggl[\min\biggl\{\mathbf{1}\{\mathsf{d}_{1}(X_{1},Z_{1})>\mathsf{D}_{1}\;\mathrm{or}\;\mathsf{d}_{2}(X_{2},Z_{2})>\mathsf{D}_{2}\}+\mathsf{L}_{1}^{-1}2^{\iota_{U_{1};X_{1}|U_{2}}(U_{1};X_{1}|U_{2})}\\
 & \;\;\;\;+\left(\mathsf{L}_{1}^{-1}\mathsf{L}_{2}^{-1}2^{\iota_{U_{1},U_{2};X_{1},X_{2}}(U_{1},U_{2};X_{1},X_{2})}+\mathsf{L}_{2}^{-1}2^{\iota_{U_{2};X_{2}|U_{1}}(U_{2};X_{2}|U_{1})}\right)\left(\log(\mathsf{L}_{2}2^{-\iota_{U_{2};X_{2}|U_{1}}(U_{2};X_{2}|U_{1})}+1)+1\right)^{2},\,1\biggr\}\biggr]
\end{align*}
Therefore there exist fixed values of the Poisson processes attaining
the desired bound.

For \eqref{eq:dlsc_pe2}, if the event in \eqref{eq:dlsc_pe2} does
not occur, by Proposition \ref{prop:phi_ineq} with $\alpha=\gamma-1$,
$\tilde{\alpha}=\gamma$, $\beta=\iota_{U_{1};U_{2}}(U_{1};U_{2})-\gamma$,
\begin{align*}
 & \mathsf{L}_{1}^{-1}2^{\iota_{U_{1};X_{1}|U_{2}}(U_{1};X_{1}|U_{2})}\\
 & \;\;\;+\left(\mathsf{L}_{1}^{-1}\mathsf{L}_{2}^{-1}2^{\iota_{U_{1},U_{2};X_{1},X_{2}}(U_{1},U_{2};X_{1},X_{2})}+\mathsf{L}_{2}^{-1}2^{\iota_{U_{2};X_{2}|U_{1}}(U_{2};X_{2}|U_{1})}\right)\left(\log(\mathsf{L}_{2}2^{-\iota_{U_{2};X_{2}|U_{1}}(U_{2};X_{2}|U_{1})}+1)+1\right)^{2}\\
 & \le2^{-\gamma}+2^{1-\gamma}\left(2(\iota_{U_{1};U_{2}}(U_{1};U_{2}))^{2}+2\gamma^{2}+14\right)\\
 & =2^{-\gamma}\left(4(\iota_{U_{1};U_{2}}(U_{1};U_{2}))^{2}+4\gamma^{2}+29\right).
\end{align*}
\end{IEEEproof}
\begin{rem}
The reason for the logarithmic term is that we want to translate a
bound on $\mathbf{E}[K]$ (given by the generalized Poisson matching
lemma) into a bound on $\mathbf{E}[(\phi(K))^{-1}]$ for some distribution
$\phi$ over $\mathbb{N}$. Ideally, we wish $(\phi(k))^{-1}\propto k$,
but this is impossible since the harmonic series diverges. Therefore
we use a slow converging series $\phi(k)\propto k^{-1}(\log(k+2))^{-2}$
instead, resulting in a logarithmic penalty.

If we use $\mathsf{J}^{-1}\mathbf{1}\{k\le\mathsf{J}\}$ instead of
$\phi(k)$ in the proof, we can obtain the following bound for any
$\mathsf{J}\in\mathbb{N}$:
\begin{align*}
P_{e} & \le\mathbf{E}\biggl[\min\biggl\{\mathbf{1}\{\mathsf{d}_{1}(X_{1},Z_{1})>\mathsf{D}_{1}\;\mathrm{or}\;\mathsf{d}_{2}(X_{2},Z_{2})>\mathsf{D}_{2}\}\\
 & \;\;\;+\mathsf{L}_{1}^{-1}\mathsf{J}^{-1}2^{\iota_{U_{1};X_{1}}(U_{1};X_{1})}+\mathsf{L}_{2}^{-1}\mathsf{J}2^{\iota_{U_{2};X_{2}|U_{1}}(U_{2};X_{2}|U_{1})}+\mathsf{L}_{1}^{-1}2^{\iota_{U_{1};X_{1}|U_{2}}(U_{1};X_{1}|U_{2})},\,1\biggr\}\biggr].
\end{align*}
Compared to Theorem \ref{thm:dlsc}, this does not contain the logarithmic
term, but requires optimizing over $\mathsf{J}$, and may give a worse
second order result.

Another choice is to use $g(k)\propto k^{-1}\mathbf{1}\{k\le\mathsf{J}\}$
instead of $\phi(k)$. We can obtain the following bound for any $\mathsf{J}\in\mathbb{N}$:
\begin{align*}
P_{e} & \le\mathbf{E}\biggl[\min\biggl\{\mathbf{1}\{\mathsf{d}_{1}(X_{1},Z_{1})>\mathsf{D}_{1}\;\mathrm{or}\;\mathsf{d}_{2}(X_{2},Z_{2})>\mathsf{D}_{2}\}+\mathsf{L}_{1}^{-1}\mathsf{J}^{-1}2^{\iota_{U_{1};X_{1}}(U_{1};X_{1})}\\
 & \;\;+\mathsf{L}_{1}^{-1}\mathsf{L}_{2}^{-1}(\ln\mathsf{J}+1)2^{\iota_{U_{1},U_{2};X_{1},X_{2}}(U_{1},U_{2};X_{1},X_{2})}+\mathsf{L}_{2}^{-1}(\ln\mathsf{J}+1)2^{\iota_{U_{2};X_{2}|U_{1}}(U_{2};X_{2}|U_{1})}+\mathsf{L}_{1}^{-1}2^{\iota_{U_{1};X_{1}|U_{2}}(U_{1};X_{1}|U_{2})},\,1\biggr\}\biggr].
\end{align*}
which gives the same second order result as Theorem \ref{thm:dlsc}.
Nevertheless, we prefer using $\phi(k)$ which eliminates the need
for a parameter $\mathsf{J}$ at the decoder.
\end{rem}

\section{One-shot Coding for Multiple Access Channels}

The one-shot coding setting for the multiple access channel is described
as follows. Upon observing $M_{j}\sim\mathrm{Unif}[1:\mathsf{L}_{j}]$
($M_{1},M_{2}$ independent), encoder $j$ produces $X_{j}$, $j=1,2$.
The decoder observes the output $Y$ of the channel $P_{Y|X_{1},X_{2}}$
and recovers $(\hat{M}_{1},\hat{M}_{2})$. The error probability is
$P_{e}=\mathbf{P}\{(M_{1},M_{2})\neq(\hat{M}_{1},\hat{M}_{2})\}$.

We present a one-shot achievability result for the capacity region
in \cite{ahlswede1971multi,liao1972multiple,ahlswede1974capacity}.
While this result is slightly weaker than that in \cite{verdu2012nonasymp},
we include it to illustrate the use of the generalized Poisson matching
lemma in simultaneous decoding. Note that the logarithmic term results
in an $O(n^{-1}\log n)$ penalty on the rate in the finite blocklength
regime, and does not affect the second order result.
\begin{thm}
\label{thm:mac}Fix any $P_{X_{1}},P_{X_{2}}$. There exists a code
for the multiple access channel $P_{Y|X_{1},X_{2}}$ for messages
$M_{j}\sim\mathrm{Unif}[1:\mathsf{L}_{j}]$, $j=1,2$, with the error
probability bounded by
\begin{align*}
P_{e} & \le\mathbf{E}\biggl[\min\biggl\{\left(\mathsf{L}_{1}\mathsf{L}_{2}2^{-\iota_{X_{1},X_{2};Y}(X_{1},X_{2};Y)}+\mathsf{L}_{2}2^{-\iota_{X_{2};X_{1},Y}(X_{2};X_{1},Y)}\right)\left(\log(\mathsf{L}_{2}^{-1}2^{\iota_{X_{2};X_{1},Y}(X_{2};X_{1},Y)}+1)+1\right)^{2}\\
 & \;\;\;\;\;\;\;\;\;\;+\mathsf{L}_{1}2^{-\iota_{X_{1};X_{2},Y}(X_{1};X_{2},Y)},\,1\biggr\}\biggr]
\end{align*}
if $P_{X_{1}X_{2}Y}\ll P_{X_{1}}\times P_{X_{2}}\times P_{Y}$, where
$(X_{1},X_{2},Y)\sim P_{X_{1}}P_{X_{2}}P_{Y|X_{1},X_{2}}$. As a result,
for $\gamma>0$,
\begin{align}
P_{e} & \le\mathbf{P}\biggl\{\log\mathsf{L}_{1}>\iota_{X_{1};X_{2},Y}(X_{1};X_{2},Y)-\gamma\;\;\mathrm{or}\;\;\log\mathsf{L}_{2}>\iota_{X_{2};X_{1},Y}(X_{2};X_{1},Y)-\gamma\nonumber \\
 & \;\;\;\;\mathrm{or}\;\;\log\mathsf{L}_{1}\mathsf{L}_{2}>\iota_{X_{1},X_{2};Y}(X_{1},X_{2};Y)-\gamma\biggr\}+2^{-\gamma}\left(4\mathbf{E}[(\iota_{X_{1};X_{2}|Y}(X_{1};X_{2}|Y))^{2}]+4\gamma^{2}+29\right).\label{eq:mac_pe2}
\end{align}
\end{thm}
\begin{IEEEproof}
Let $\{(\bar{X}_{1,i},\bar{M}_{1,i}),T_{1,i}\}_{i\in\mathbb{N}}$,
$\{(\bar{X}_{2,i},\bar{M}_{2,i}),T_{2,i}\}_{i\in\mathbb{N}}$ be two
independent Poisson processes with intensity measures $P_{X_{1}}\times P_{M_{1}}\times\lambda_{\mathbb{R}_{\ge0}}$
and $P_{X_{2}}\times P_{M_{2}}\times\lambda_{\mathbb{R}_{\ge0}}$
respectively, independent of $M_{1},M_{2}$. The encoding functions
are $X_{1}=(\tilde{X}_{1})_{P_{X_{1}}\times\delta_{M_{1}}}$, $X_{2}=(\tilde{X}_{2})_{P_{X_{2}}\times\delta_{M_{2}}}$
(which Poisson process we are referring to can be deduced from whether
we are discussing $X_{1}$ or $X_{2}$). For the decoding function,
let $\check{X}_{1k}=(\tilde{X}_{1})_{P_{X_{1}|Y}(\cdot|Y)\times P_{M_{1}}}(k)$
for $k\in\mathbb{N}$, $(\hat{X}_{2},\hat{M}_{2})=(\tilde{X}_{2},\tilde{M}_{2})_{\sum_{k=1}^{\infty}\phi(k)P_{X_{2}|X_{1},Y}(\cdot|\check{X}_{1k},Y)\times P_{M_{2}}}$
where $\phi(k)\propto k^{-1}(\log(k+2))^{-2}$ with $\sum_{k=1}^{\infty}\phi(k)=1$,
and $\hat{M}_{1}=(\tilde{M}_{1})_{P_{X_{1}|X_{2},Y}(\cdot|\hat{X}_{2},Y)\times P_{M_{1}}}$.

Let $K=\Upsilon_{P_{X_{1}}\times\delta_{M_{1}}\Vert P_{X_{1}|Y}(\cdot|Y)\times P_{M_{1}}}(1)$
(using the Poisson process $\{(\bar{X}_{1,i},\bar{M}_{1,i}),T_{1,i}\}_{i\in\mathbb{N}}$).
By the conditional generalized Poisson matching lemma on $(M_{1},\,1,\,(X_{1},M_{1}),\,Y,\,P_{X_{1}|Y}\times P_{M_{1}})$
(note that $P_{X_{1},M_{1}|M_{1}}=P_{X_{1}}\times\delta_{M_{1}}$),
almost surely,
\begin{align}
\mathbf{E}\left[\left.K\,\right|\,X_{1},Y,M_{1}\right] & \le\frac{dP_{X_{1}}\times\delta_{M_{1}}}{dP_{X_{1}|Y}(\cdot|Y)\times P_{M_{1}}}(X_{1},M_{1})+1\nonumber \\
 & =\mathsf{L}_{1}2^{-\iota_{X_{1};Y}(X_{1};Y)}+1.\label{eq:mac_ek}
\end{align}
Since $\{\check{X}_{1k}\}_{k}$ is a function of $\{(\bar{X}_{1,i},\bar{M}_{1,i}),T_{1,i}\}_{i}$
and $Y$, we have $\{\check{X}_{1k}\}_{k}\leftrightarrow(X_{1},X_{2},Y,M_{2})\leftrightarrow\{(\bar{X}_{2,i},\bar{M}_{2,i}),T_{2,i}\}_{i}$.
By the conditional Poisson matching lemma on $(M_{2},\,(X_{2},M_{2}),\,(\{\check{X}_{1k}\}_{k},Y),\,\sum_{k=1}^{\infty}\phi(k)P_{X_{2}|X_{1},Y}(\cdot|\check{X}_{1k},Y)\times P_{M_{2}})$
(note that $P_{X_{2},M_{2}|M_{2}}=P_{X_{2}}\times\delta_{M_{2}}$),
almost surely,
\begin{align*}
 & \mathbf{P}\biggl\{\left.(\tilde{X}_{2},\tilde{M}_{2})_{\sum_{k=1}^{\infty}\phi(k)P_{X_{2}|X_{1},Y}(\cdot|\check{X}_{1k},Y)\times P_{M_{2}}}\neq(X_{2},M_{2})\,\right|\,X_{1},X_{2},Y,M_{2}\biggr\}\\
 & \le\mathbf{E}\left[\left.\min\biggl\{\frac{dP_{X_{2}}\times\delta_{M_{2}}}{d(\sum_{k=1}^{\infty}\phi(k)P_{X_{2}|X_{1},Y}(\cdot|\check{X}_{1k},Y))\times P_{M_{2}}}(X_{2},M_{2}),\,1\biggr\}\,\right|\,X_{1},X_{2},Y,M_{2}\right]\\
 & \le\mathbf{E}\left[\left.\min\biggl\{\mathsf{L}_{2}\frac{dP_{X_{2}}}{\phi(K)dP_{X_{2}|X_{1},Y}(\cdot|X_{1},Y)}(X_{2}),\,1\biggr\}\,\right|\,X_{1},X_{2},Y,M_{2}\right]\\
 & =\mathbf{E}\left[\left.\min\{\mathsf{L}_{2}(\phi(K))^{-1}2^{-\iota_{X_{2};X_{1},Y}(X_{2};X_{1},Y)},\,1\}\,\right|\,X_{1},X_{2},Y,M_{2}\right]\\
 & \stackrel{(a)}{\le}\mathbf{E}\left[\left.K\mathsf{L}_{2}2^{-\iota_{X_{2};X_{1},Y}(X_{2};X_{1},Y)}\left(\log(\mathsf{L}_{2}^{-1}2^{\iota_{X_{2};X_{1},Y}(X_{2};X_{1},Y)}+1)+1\right)^{2}\,\right|\,X_{1},X_{2},Y,M_{2}\right]\\
 & \stackrel{(b)}{\le}\left(\mathsf{L}_{1}2^{-\iota_{X_{1};Y}(X_{1};Y)}+1\right)\mathsf{L}_{2}2^{-\iota_{X_{2};X_{1},Y}(X_{2};X_{1},Y)}\left(\log(\mathsf{L}_{2}^{-1}2^{\iota_{X_{2};X_{1},Y}(X_{2};X_{1},Y)}+1)+1\right)^{2}\\
 & =\left(\mathsf{L}_{1}\mathsf{L}_{2}2^{-\iota_{X_{1},X_{2};Y}(X_{1},X_{2};Y)}+\mathsf{L}_{2}2^{-\iota_{X_{2};X_{1},Y}(X_{2};X_{1},Y)}\right)\left(\log(\mathsf{L}_{2}^{-1}2^{\iota_{X_{2};X_{1},Y}(X_{2};X_{1},Y)}+1)+1\right)^{2},
\end{align*}
where (a) is by Proposition \ref{prop:phi_ineq}, and (b) is by $K\leftrightarrow(X_{1},Y)\leftrightarrow X_{2}$,
\eqref{eq:mac_ek} and Jensen's inequality. By the conditional Poisson
matching lemma on $(M_{1},\,(X_{1},M_{1}),\,(X_{2},Y),\,P_{X_{1}|X_{2},Y}\times P_{M_{1}})$
(note that $P_{X_{1},M_{1}|M_{1}}=P_{X_{1}}\times\delta_{M_{1}}$),
almost surely,
\begin{align*}
 & \mathbf{P}\biggl\{(\tilde{X}_{1},\tilde{M}_{1})_{P_{X_{1}|X_{2},Y}(\cdot|X_{2},Y)\times P_{M_{1}}}\neq(X_{1},M_{1})\,\biggl|\,X_{1},X_{2},Y,M_{1}\biggr\}\\
 & \le\frac{dP_{X_{1}}\times\delta_{M_{1}}}{dP_{X_{1}|X_{2},Y}(\cdot|X_{2},Y)\times P_{M_{1}}}(X_{1},M_{1})\\
 & =\mathsf{L}_{1}2^{-\iota_{X_{1};X_{2},Y}(X_{1};X_{2},Y)}.
\end{align*}
Therefore there exist fixed values of the Poisson processes attaining
the desired bound.

For \eqref{eq:mac_pe2}, if the event in \eqref{eq:mac_pe2} does
not occur, by Proposition \ref{prop:phi_ineq} with $\alpha=\gamma-1$,
$\tilde{\alpha}=\gamma$, $\beta=\iota_{X_{1};X_{2}|Y}(X_{1};X_{2}|Y)-\gamma$,
\begin{align*}
 & \left(\mathsf{L}_{1}\mathsf{L}_{2}2^{-\iota_{X_{1},X_{2};Y}(X_{1},X_{2};Y)}+\mathsf{L}_{2}2^{-\iota_{X_{2};X_{1},Y}(X_{2};X_{1},Y)}\right)\left(\log(\mathsf{L}_{2}^{-1}2^{\iota_{X_{2};X_{1},Y}(X_{2};X_{1},Y)}+1)+1\right)^{2}+\mathsf{L}_{1}2^{-\iota_{X_{1};X_{2},Y}(X_{1};X_{2},Y)}\\
 & \le2^{1-\gamma}\left(2(\iota_{X_{1};X_{2}|Y}(X_{1};X_{2}|Y))^{2}+2\gamma^{2}+14\right)+2^{-\gamma}\\
 & =2^{-\gamma}\left(4(\iota_{X_{1};X_{2}|Y}(X_{1};X_{2}|Y))^{2}+4\gamma^{2}+29\right).
\end{align*}
\end{IEEEproof}
\begin{rem}
If we use $\mathsf{J}^{-1}\mathbf{1}\{k\le\mathsf{J}\}$ instead of
$\phi(k)$ in the proof, we can obtain the following bound for any
$\mathsf{J}\in\mathbb{N}$:
\begin{align*}
P_{e} & \le\mathbf{E}\biggl[\min\biggl\{\mathsf{L}_{1}\mathsf{J}^{-1}2^{-\iota_{X_{1};Y}(X_{1};Y)}+\mathsf{L}_{2}\mathsf{J}2^{-\iota_{X_{2};X_{1},Y}(X_{2};X_{1},Y)}+\mathsf{L}_{1}2^{-\iota_{X_{1};X_{2},Y}(X_{1};X_{2},Y)},\,1\biggr\}\biggr].
\end{align*}
Compared to Theorem \ref{thm:mac}, this does not contain the logarithmic
term, but requires optimizing over $\mathsf{J}$, and may give a worse
second order result.

Another choice is to use $g(k)\propto k^{-1}\mathbf{1}\{k\le\mathsf{J}\}$
instead of $\phi(k)$. We can obtain the following bound for any $\mathsf{J}\in\mathbb{N}$:
\begin{align*}
P_{e} & \le\mathbf{E}\biggl[\min\biggl\{\mathsf{L}_{1}\mathsf{L}_{2}(\ln\mathsf{J}+1)2^{-\iota_{X_{1},X_{2};Y}(X_{1},X_{2};Y)}+\mathsf{L}_{2}(\ln\mathsf{J}+1)2^{-\iota_{X_{2};X_{1},Y}(X_{2};X_{1},Y)}\\
 & \;\;\;\;\;\;\;\;\;\;+\mathsf{L}_{1}2^{-\iota_{X_{1};X_{2},Y}(X_{1};X_{2},Y)}+\mathsf{L}_{1}\mathsf{J}^{-1}2^{-\iota_{X_{1};Y}(X_{1};Y)},\,1\biggr\}\biggr],
\end{align*}
which gives the same second order result as Theorem \ref{thm:mac}.
Nevertheless, we prefer using $\phi(k)$ which eliminates the need
for a parameter $\mathsf{J}$ at the decoder.
\end{rem}

\section{One-shot Channel Resolvability and Soft Covering}

The one-shot channel resolvability setting \cite{han1993approximation}
is described as follows. Fix a channel $P_{Y|X}$ and input distribution
$P_{X}$. Upon observing an integer $M\sim\mathrm{Unif}[1:\mathsf{L}]$,
the encoder applies a deterministic mapping $g:[1:\mathsf{L}]\to\mathcal{X}$
on $M$ to produce $\hat{X}=g(M)$, which is sent through the channel
$P_{Y|X}$ and gives the output $\hat{Y}$. The goal is to minimize
the total variation distance between $P_{\hat{Y}}$ and $P_{Y}$ ($Y$-marginal
of $P_{X}P_{Y|X}$), i.e., $\epsilon:=\Vert\mathsf{L}^{-1}\sum_{m=1}^{\mathsf{L}}P_{Y|X}(\cdot|g(m))-P_{Y}(\cdot)\Vert_{\mathrm{TV}}$.

We show a one-shot channel resolvability result using the the Poisson
matching lemma. This result can also be regarded as a one-shot soft
covering lemma \cite{cuff2013synthesis}.
\begin{prop}
\label{prop:resolve}Given channel $P_{Y|X}$ and input distribution
$P_{X}$ with $P_{XY}\ll P_{X}\times P_{Y}$. Let $\{\check{X}_{m}\}_{m\in[1:\mathsf{L}]}\stackrel{iid}{\sim}P_{X}$,
then for any $\mathsf{J}\in\mathbb{N}$,
\begin{align}
 & \mathbf{E}\left[\Bigl\Vert\mathsf{L}^{-1}\sum_{m=1}^{\mathsf{L}}P_{Y|X}(\cdot|\check{X}_{m})-P_{Y}(\cdot)\Bigr\Vert_{\mathrm{TV}}\right]\nonumber \\
 & \le\mathbf{E}\left[(1+2^{-\iota_{X;Y}(X;Y)})^{-\mathsf{J}}\right]+\frac{1}{2}\sqrt{\mathsf{J}\mathsf{L}^{-1}}.\label{eq:resolve_pe1}
\end{align}
As a result, for any $0<\gamma\le\log\mathsf{L}$,
\begin{align}
 & \mathbf{E}\left[\Bigl\Vert\mathsf{L}^{-1}\sum_{m=1}^{\mathsf{L}}P_{Y|X}(\cdot|\check{X}_{m})-P_{Y}(\cdot)\Bigr\Vert_{\mathrm{TV}}\right]\nonumber \\
 & \le\mathbf{P}\left\{ \iota_{X;Y}(X;Y)>\log\mathsf{L}-\gamma\right\} +2^{-\gamma/2}\left(1+\frac{1}{2}\sqrt{\gamma}\right)+\frac{1}{2}\sqrt{\mathsf{L}^{-1}}.\label{eq:resolve_pe2}
\end{align}
Hence there exists a code for channel resolvability satisfying the
above bounds. 
\end{prop}
\begin{IEEEproof}
Let $\mathfrak{P}=\{\bar{Y}_{i},T_{i}\}_{i\in\mathbb{N}}$ be the
points of a Poisson process with intensity measure $P_{Y}\times\lambda_{\mathbb{R}_{\ge0}}$.
Let $M\sim\mathrm{Unif}[1:\mathsf{L}]$, $\{\check{X}_{m}\}_{m\in[1:\mathsf{L}]}\stackrel{iid}{\sim}P_{X}$
($M\perp\!\!\!\perp\{\check{X}_{j}\}_{j}\perp\!\!\!\perp\mathfrak{P}$),
and $X=\check{X}_{M}$. Let $Y=\tilde{Y}_{P_{Y|X}(\cdot|X)}$, and
$\hat{Y}_{j}=\tilde{Y}_{P_{Y}}(j)$ for $j\in\mathbb{N}$. We have
\begin{align*}
 & \mathbf{E}\left[\Vert P_{Y|\{\check{X}_{m}\}_{m}}(\cdot|\{\check{X}_{m}\}_{m})-P_{Y}(\cdot)\Vert_{\mathrm{TV}}\right]\\
 & \stackrel{(a)}{\le}\mathbf{E}\left[\Vert P_{Y|\{\check{X}_{m}\}_{m},\mathfrak{P}}(\cdot|\{\check{X}_{m}\}_{m},\mathfrak{P})-P_{Y|\mathfrak{P}}(\cdot|\mathfrak{P})\Vert_{\mathrm{TV}}\right]\\
 & \stackrel{(b)}{=}\frac{1}{2}\mathbf{E}\left[\sum_{y\in\{\hat{Y}_{j}\}_{j\in\mathbb{N}}}\left|P_{Y|\mathfrak{P}}(y|\mathfrak{P})-P_{Y|\{\check{X}_{m}\}_{m},\mathfrak{P}}(y|\{\check{X}_{m}\}_{m},\mathfrak{P})\right|\right]\\
 & \le\frac{1}{2}\mathbf{E}\biggl[\sum_{y\in\{\hat{Y}_{j}\}_{j\in[1:\mathsf{J}]}}\left|P_{Y|\mathfrak{P}}(y|\mathfrak{P})-P_{Y|\{\check{X}_{m}\}_{m},\mathfrak{P}}(y|\{\check{X}_{m}\}_{m},\mathfrak{P})\right|\\
 & \;\;\;\;\;\;\;\;+\sum_{y\in\{\hat{Y}_{j}\}_{j\in\mathbb{N}}\backslash\{\hat{Y}_{j}\}_{j\in[1:\mathsf{J}]}}\left(P_{Y|\mathfrak{P}}(y|\mathfrak{P})+P_{Y|\{\check{X}_{m}\}_{m},\mathfrak{P}}(y|\{\check{X}_{m}\}_{m},\mathfrak{P})\right)\biggr]\\
 & =\frac{1}{2}\mathbf{E}\left[\sum_{y\in\{\hat{Y}_{j}\}_{j\in[1:\mathsf{J}]}}\left|P_{Y|\mathfrak{P}}(y|\mathfrak{P})-\mathsf{L}^{-1}\sum_{m=1}^{\mathsf{L}}P_{Y|X,\mathfrak{P}}(y|\check{X}_{m},\mathfrak{P})\right|\right]\\
 & \;\;\;\;\;\;\;\;+\frac{1}{2}\mathbf{E}\left[P_{Y|\mathfrak{P}}(\mathcal{Y}\backslash\{\hat{Y}_{j}\}_{j\in[1:\mathsf{J}]}|\mathfrak{P})+P_{Y|\{\check{X}_{m}\}_{m},\mathfrak{P}}(\mathcal{Y}\backslash\{\hat{Y}_{j}\}_{j\in[1:\mathsf{J}]}\,|\,\{\check{X}_{m}\}_{m},\mathfrak{P})\right]\\
 & =\frac{1}{2}\mathbf{E}\left[\sum_{y\in\{\hat{Y}_{j}\}_{j\in[1:\mathsf{J}]}}\left|P_{Y|\mathfrak{P}}(y|\mathfrak{P})-\mathsf{L}^{-1}\sum_{m=1}^{\mathsf{L}}P_{Y|X,\mathfrak{P}}(y|\check{X}_{m},\mathfrak{P})\right|\right]+\mathbf{P}\left\{ Y\notin\{\hat{Y}_{j}\}_{j\in[1:\mathsf{J}]}\right\} ,
\end{align*}
where (a) is by the convexity of the total variation distance, and
(b) is because $Y\in\{\hat{Y}_{j}\}_{j\in\mathbb{N}}$ almost surely
(note that the summation $\sum_{y\in\{\hat{Y}_{j}\}_{j\in\mathbb{N}}}$
ignores multiplicity of elements in $\{\hat{Y}_{j}\}_{j\in\mathbb{N}}$).
For the first term, note that since $Y$ is a function of $(X,\mathfrak{P})$,
we have $P_{Y|X,\mathfrak{P}}(y|\check{X}_{m},\mathfrak{P})\in\{0,1\}$,
and hence 
\[
\left(\sum_{m=1}^{\mathsf{L}}P_{Y|X,\mathfrak{P}}(y|\check{X}_{m},\mathfrak{P})\right)\,\biggl|\,\mathfrak{P}\sim\mathrm{Bin}(\mathsf{L},\,P_{Y|\mathfrak{P}}(y|\mathfrak{P})).
\]
We have
\begin{align*}
 & \frac{1}{2}\mathbf{E}\left[\sum_{y\in\{\hat{Y}_{j}\}_{j\in[1:\mathsf{J}]}}\left|P_{Y|\mathfrak{P}}(y|\mathfrak{P})-\mathsf{L}^{-1}\sum_{m=1}^{\mathsf{L}}P_{Y|X,\mathfrak{P}}(y|\check{X}_{m},\mathfrak{P})\right|\right]\\
 & =\frac{1}{2}\mathbf{E}\left[\sum_{y\in\{\hat{Y}_{j}\}_{j\in[1:\mathsf{J}]}}\mathbf{E}\left[\left.\left|P_{Y|\mathfrak{P}}(y|\mathfrak{P})-\mathsf{L}^{-1}\sum_{m=1}^{\mathsf{L}}P_{Y|X,\mathfrak{P}}(y|\check{X}_{m},\mathfrak{P})\right|\,\right|\,\mathfrak{P}\right]\right]\\
 & \le\frac{1}{2}\mathbf{E}\left[\sum_{y\in\{\hat{Y}_{j}\}_{j\in[1:\mathsf{J}]}}\sqrt{\mathrm{Var}\left[\left.\mathsf{L}^{-1}\sum_{m=1}^{\mathsf{L}}P_{Y|X,\mathfrak{P}}(y|\check{X}_{m},\mathfrak{P})\,\right|\,\mathfrak{P}\right]}\right]\\
 & \le\frac{1}{2}\mathbf{E}\left[\sum_{y\in\{\hat{Y}_{j}\}_{j\in[1:\mathsf{J}]}}\sqrt{\mathsf{L}^{-1}P_{Y|\mathfrak{P}}(y|\mathfrak{P})}\right]\\
 & \le\frac{1}{2}\mathbf{E}\left[\sqrt{\mathsf{J}\sum_{y\in\{\hat{Y}_{j}\}_{j\in[1:\mathsf{J}]}}\mathsf{L}^{-1}P_{Y|\mathfrak{P}}(y|\mathfrak{P})}\right]\\
 & \le\frac{1}{2}\sqrt{\mathsf{J}\mathsf{L}^{-1}}.
\end{align*}
For the second term, by the conditional generalized Poisson matching
lemma on $(X,\,1,\,Y,\,\emptyset,\,P_{Y})$,
\begin{align*}
 & \mathbf{P}\{Y\notin\{\hat{Y}_{j}\}_{j\in[1:\mathsf{J}]}\}\\
 & \le\mathbf{E}\left[\left(1-\left(1+\frac{dP_{Y|X}(\cdot|X)}{dP_{Y}}(Y)\right)^{-1}\right)^{\mathsf{J}}\right]\\
 & =\mathbf{E}\left[(1-(1+2^{\iota_{X;Y}(X;Y)})^{-1})^{\mathsf{J}}\right].
\end{align*}
Hence,
\begin{align*}
 & \mathbf{E}\left[\Vert P_{Y|\{\check{X}_{j}\}_{j}}(\cdot|\{\check{X}_{m}\}_{m})-P_{Y}(\cdot)\Vert_{\mathrm{TV}}\right]\\
 & \le\mathbf{E}\left[(1-(1+2^{\iota_{X;Y}(X;Y)})^{-1})^{\mathsf{J}}\right]+\frac{1}{2}\sqrt{\mathsf{J}\mathsf{L}^{-1}}\\
 & =\mathbf{E}\left[(1+2^{-\iota_{X;Y}(X;Y)})^{-\mathsf{J}}\right]+\frac{1}{2}\sqrt{\mathsf{J}\mathsf{L}^{-1}}.
\end{align*}
For \eqref{eq:resolve_pe2}, substitute $\mathsf{J}=\lceil\gamma2^{-\gamma}\mathsf{L}\rceil$,
\begin{align*}
 & \mathbf{E}\left[(1-(1+2^{\iota_{X;Y}(X;Y)})^{-1})^{\mathsf{J}}\right]+\frac{1}{2}\sqrt{\mathsf{J}\mathsf{L}^{-1}}\\
 & \stackrel{(a)}{\le}\mathbf{E}\left[(1-(1+(2\mathsf{L}2^{-\gamma})^{-1}2^{\iota_{X;Y}(X;Y)})^{-1})^{\mathsf{J}(2\mathsf{L}2^{-\gamma})^{-1}}\right]+\frac{1}{2}\sqrt{\mathsf{J}\mathsf{L}^{-1}}\\
 & \le\mathbf{P}\left\{ \iota_{X;Y}(X;Y)>\log\mathsf{L}-\gamma\right\} +2^{-\mathsf{J}(2\mathsf{L}2^{-\gamma})^{-1}}+\frac{1}{2}\sqrt{(\gamma2^{-\gamma}\mathsf{L}+1)\mathsf{L}^{-1}}\\
 & \le\mathbf{P}\left\{ \iota_{X;Y}(X;Y)>\log\mathsf{L}-\gamma\right\} +2^{-\gamma/2}+\frac{1}{2}\sqrt{\gamma2^{-\gamma}}+\frac{1}{2}\sqrt{\mathsf{L}^{-1}}\\
 & =\mathbf{P}\left\{ \iota_{X;Y}(X;Y)>\log\mathsf{L}-\gamma\right\} +2^{-\gamma/2}\left(1+\frac{1}{2}\sqrt{\gamma}\right)+\frac{1}{2}\sqrt{\mathsf{L}^{-1}},
\end{align*}
where (a) is because $\gamma\le\log\mathsf{L}$, $2\mathsf{L}2^{-\gamma}>1$
and $(1-(1+\alpha)^{-1})^{\beta}\le1-(1+\beta^{-1}\alpha)^{-1}$ for
$\alpha\ge0$, $\beta\ge1$.

\end{IEEEproof}
\medskip{}

Compare this to Theorem 2 in \cite{hayashi2006resolvability} (weakened
by substituting $\delta'_{p,W,C}\le C$): for any $\alpha>0$,
\[
\epsilon\le\mathbf{P}\left\{ \iota_{X;Y}(X;Y)>\log\alpha\right\} +\frac{1}{2}\sqrt{\alpha\mathsf{L}^{-1}}.
\]
If we assume $1\le\alpha\le\mathsf{L}$ and substitute $\gamma=\log(\mathsf{L}/\alpha)$
in \eqref{eq:resolve_pe2}, we obtain the following slightly weaker
bound (within a logarithmic gap from that in \cite{hayashi2006resolvability}):
\begin{align*}
\epsilon & \le\mathbf{P}\left\{ \iota_{X;Y}(X;Y)>\log\alpha\right\} +\sqrt{\alpha\mathsf{L}^{-1}}\left(1+\frac{1}{2}\sqrt{\log(\mathsf{L}/\alpha)}\right)+\frac{1}{2}\sqrt{\mathsf{L}^{-1}}.
\end{align*}
Nevertheless, the bound in \cite{hayashi2006resolvability} does not
imply \eqref{eq:resolve_pe1}, so neither bound is stronger than the
other.

The channel resolvability or soft covering bound in Proposition \ref{prop:resolve}
can be applied to prove various secrecy and coordination results,
e.g. one-shot coding for wiretap channels \cite{wyner1975wire}, one-shot
channel synthesis \cite{cuff2013synthesis}, and one-shot distributed
source simulation \cite{wyner1975common}. Hence these results can
also be proved using the Poisson matching lemma alone. In the next
section, we will prove a one-shot result for wiretap channels.

\section{One-shot Coding for Wiretap Channels}

The one-shot version of the wiretap channel setting \cite{wyner1975wire}
 is described as follows. Upon observing $M\sim\mathrm{Unif}[1:\mathsf{L}]$,
the encoder produces $X$, which is sent through the broadcast channel
$P_{Y,Z|X}$. The legitimate decoder observes $Y$ and recovers $\hat{M}$
with error probability $P_{e}=\mathbf{P}\{M\neq\hat{M}\}$. The eavesdropper
observes $Z$. Secrecy is measured by the total variation distance
$\epsilon:=\Vert P_{M,Z}-P_{M}\times P_{Z}\Vert_{\mathrm{TV}}$.

The following bound is a direct result of the generalized Poisson
matching lemma and Proposition \ref{prop:resolve}. It is included
for demonstration purposes. See \cite{hayashi2006resolvability,yassaee2015one,liu2017resolvability}
for other one-shot bounds (that are not strictly stronger or weaker
than ours).
\begin{prop}
\label{prop:wire}Fix any $P_{U,X}$. For any $\nu\ge0$, $\mathsf{K},\mathsf{J}\in\mathbb{N}$,
there exists a code for the wiretap channel $P_{Y,Z|X}$, with message
$M\sim\mathrm{Unif}[1:\mathsf{L}]$, with average error probability
$P_{e}$ and secrecy measure $\epsilon$ satisfying
\begin{align*}
P_{e}+\nu\epsilon & \le\mathbf{E}\left[\min\{\mathsf{L}\mathsf{K}2^{-\iota_{U;Y}(U;Y)},\,1\}\right]\\
 & \;\;\;+\nu\left(2\mathbf{E}\left[(1+2^{-\iota_{U;Z}(U;Z)})^{-\mathsf{J}}\right]+\sqrt{\mathsf{J}\mathsf{K}^{-1}}\right)
\end{align*}
if $P_{UY}\ll P_{U}\times P_{Y}$ and $P_{UZ}\ll P_{U}\times P_{Z}$.
\end{prop}
\begin{IEEEproof}
Let $\mathfrak{P}=\{(\bar{U}_{i},\bar{M}_{i}),T_{i}\}_{i\in\mathbb{N}}$
be the points of a Poisson process with intensity measure $P_{U}\times P_{M}\times\lambda_{\mathbb{R}_{\ge0}}$
independent of $M$. Let $K\sim\mathrm{Unif}[1:\mathsf{K}]$ independent
of $(M,\mathfrak{P})$. The encoder computes $U=\tilde{U}_{P_{U}\times\delta_{M}}(K)$
and generates $X|U\sim P_{X|U}$. The decoder recovers $\hat{M}=\tilde{M}_{P_{U|Y}(\cdot|Y)\times P_{M}}$.
We have $(M,K,U,X,Y,Z)\sim P_{M}\times P_{K}\times P_{U,X}P_{Y,Z|X}$.
By the conditional generalized Poisson matching lemma on $(M,\,K,\,(U,M),\,Y,\,P_{U|Y}\times P_{M})$
(note that $P_{U,M|M,K}=P_{U}\times\delta_{M}$),
\begin{align*}
 & \mathbf{P}\left\{ M\neq\hat{M}\right\} \\
 & \le\mathbf{E}\left[\mathbf{P}\left\{ (U,M)\neq(\tilde{U},\tilde{M})_{P_{U|Y}(\cdot|Y)\times P_{M}}\,|\,M,K,U,Y\right\} \right]\\
 & \le\mathbf{E}\left[\min\left\{ \mathsf{K}\frac{dP_{U}\times\delta_{M}}{dP_{U|Y}(\cdot|Y)\times P_{M}}(U,M),\,1\right\} \right]\\
 & =\mathbf{E}\left[\min\{\mathsf{L}\mathsf{K}2^{-\iota_{U;Y}(U;Y)},\,1\}\right].
\end{align*}
For the secrecy measure,
\begin{align*}
 & \mathbf{E}\left[\Bigl\Vert P_{M,Z|\mathfrak{P}}(\cdot,\cdot|\mathfrak{P})-P_{M}(\cdot)\times P_{Z|\mathfrak{P}}(\cdot|\mathfrak{P})\Bigr\Vert_{\mathrm{TV}}\right]\\
 & =\mathbf{E}\left[\Bigl\Vert P_{Z|M,\mathfrak{P}}(\cdot|M,\mathfrak{P})-P_{Z|\mathfrak{P}}(\cdot|\mathfrak{P})\Bigr\Vert_{\mathrm{TV}}\right]\\
 & \le\mathbf{E}\left[\Bigl\Vert P_{Z|M,\mathfrak{P}}(\cdot|M,\mathfrak{P})-P_{Z}(\cdot)\Bigr\Vert_{\mathrm{TV}}\right]+\mathbf{E}\left[\Bigl\Vert P_{Z|\mathfrak{P}}(\cdot|\mathfrak{P})-P_{Z}(\cdot)\Bigr\Vert_{\mathrm{TV}}\right]\\
 & \stackrel{(a)}{\le}2\mathbf{E}\left[\Bigl\Vert P_{Z|M,\mathfrak{P}}(\cdot|M,\mathfrak{P})-P_{Z}(\cdot)\Bigr\Vert_{\mathrm{TV}}\right]\\
 & =2\mathbf{E}\left[\Bigl\Vert\mathsf{K}^{-1}\sum_{k=1}^{\mathsf{K}}P_{Z|U}(\cdot|\tilde{U}_{P_{U}\times\delta_{M}}(k))-P_{Z}(\cdot)\Bigr\Vert_{\mathrm{TV}}\right]\\
 & \stackrel{(b)}{\le}2\mathbf{E}\left[(1+2^{-\iota_{U;Z}(U;Z)})^{-\mathsf{J}}\right]+\sqrt{\mathsf{J}\mathsf{K}^{-1}},
\end{align*}
where (a) is by the convexity of total variation distance, and (b)
is by Proposition \ref{prop:resolve} since $\{\tilde{U}_{P_{U}\times\delta_{m}}(k)\}_{k\in[1:\mathsf{K}]}\stackrel{iid}{\sim}P_{U}$
for any $m$. Therefore there exists a fixed set of points for $\mathfrak{P}$
satisfying the desired bound.
\end{IEEEproof}

\section{Strong Functional Representation Lemma and Noncausal Sampling}

The generalized Poisson matching lemma can be applied to give a slight
improvement on the constant in the strong functional representation
lemma in \cite{sfrl_trans}, and hence improves on the variable-length
channel simulation result in \cite{harsha2010communication}, and
the result on minimax remote prediction with a communication constraint
in \cite{li2018minimax}. It also gives an achievability bound on
the moments for the noncausal sampling setting in \cite{liu2018rejection}.
\begin{prop}
\label{prop:concave}Let $\{\bar{U}_{i},T_{i}\}_{i\in\mathbb{N}}$
be the points of a Poisson process with intensity measure $\mu\times\lambda_{\mathbb{R}_{\ge0}}$
over $\mathcal{U}\times\mathbb{R}_{\ge0}$, and $P,Q$ be probability
measures over $\mathcal{U}$ with $P\ll Q\ll\mu$. For any $j\in\mathbb{N}$,
$g:\mathbb{R}_{\ge0}\to\mathbb{R}$ concave nondecreasing, we have
\[
\mathbf{E}\left[g(\Upsilon_{P\Vert Q}(j)-1)\right]\le\mathbf{E}_{U\sim P}\left[g\left(j\frac{dP}{dQ}(U)\right)\right],
\]
i.e., $j(dP/dQ)(U)$ dominates $\Upsilon_{P\Vert Q}(j)-1$ in the
second order. As a result, let 
\[
\mathfrak{C}[xg'(x)](y)=\inf\left\{ \alpha y+\beta:\,xg'(x)\le\alpha x+\beta\;\forall x\ge0\right\} 
\]
be the upper concave envelope of $xg'(x)$, then
\[
\mathbf{E}\left[g(\Upsilon_{P\Vert Q}(j))\right]\le\mathbf{E}_{U\sim P}\left[g\left(j\frac{dP}{dQ}(U)\right)\right]+j^{-1}\mathfrak{C}[xg'(x)](j).
\]
In particular,
\[
\mathbf{E}\left[\log\Upsilon_{P\Vert Q}(j)\right]\le D(P\Vert Q)+\log j+j^{-1}\log e,
\]
and for $\gamma\in(0,1)$,
\begin{align*}
\mathbf{E}\left[(\Upsilon_{P\Vert Q}(j))^{\gamma}\right] & \le j^{\gamma}\mathbf{E}_{U\sim P}\left[\left(\frac{dP}{dQ}(U)\right)^{\gamma}\right]+\gamma j^{\gamma-1}\\
 & =j^{\gamma}2^{\gamma D_{\gamma+1}(P\Vert Q)}+\gamma j^{\gamma-1},
\end{align*}
where $D_{\gamma+1}(P\Vert Q)=\gamma^{-1}\log\mathbf{E}_{U\sim P}\left[\left((dP/dQ)(U)\right)^{\gamma}\right]$
is the R\'enyi divergence.
\end{prop}
\begin{IEEEproof}
For $g:\mathbb{R}_{\ge0}\to\mathbb{R}$ concave nondecreasing, we
have
\begin{align*}
 & \mathbf{E}\left[g(\Upsilon_{P\Vert Q}(j)-1)\right]\\
 & =\int\mathbf{E}\left[\left.g(\Upsilon_{P\Vert Q}(j)-1)\,\right|\,\tilde{U}_{P}(j)=u\right]P(du)\\
 & \stackrel{(a)}{\le}\int g\left(\mathbf{E}\left[\left.\Upsilon_{P\Vert Q}(j)\,\right|\,\tilde{U}_{P}(j)=u\right]-1\right)P(du)\\
 & \stackrel{(b)}{\le}\int g\left(j\frac{dP}{dQ}(u)\right)P(du),
\end{align*}
where (a) is by Jensen's inequality, and (b) is by the generalized
Poisson matching lemma. For any $\alpha,\beta$ such that $xg'(x)\le\alpha x+\beta$
for $x\ge0$,
\begin{align*}
 & \mathbf{E}\left[g(\Upsilon_{P\Vert Q}(j))\right]\\
 & \le\int g\left(j\frac{dP}{dQ}(u)+1\right)P(du)\\
 & \le\int g\left(j\frac{dP}{dQ}(u)\right)P(du)+\int g'\left(j\frac{dP}{dQ}(u)\right)P(du)\\
 & =\int g\left(j\frac{dP}{dQ}(u)\right)P(du)+j^{-1}\int g'\left(j\frac{dP}{dQ}(u)\right)j\frac{dP}{dQ}(u)Q(du)\\
 & \le\int g\left(j\frac{dP}{dQ}(u)\right)P(du)+j^{-1}\int\left(\alpha j\frac{dP}{dQ}(u)+\beta\right)Q(du)\\
 & =\int g\left(j\frac{dP}{dQ}(u)\right)P(du)+j^{-1}(\alpha j+\beta).
\end{align*}
For $g(x)=\log x$, $xg'(x)=\log e$, and hence
\[
\mathbf{E}\left[\log\Upsilon_{P\Vert Q}(j)\right]\le D(P\Vert Q)+\log j+j^{-1}\log e.
\]
For $g(x)=x^{\gamma}$, $\gamma\in(0,1)$, $xg'(x)=\gamma x^{\gamma}$
is concave, and hence
\begin{align*}
\mathbf{E}\left[(\Upsilon_{P\Vert Q}(j))^{\gamma}\right] & \le\mathbf{E}_{U\sim P}\left[\left(j\frac{dP}{dQ}(U)\right)^{\gamma}\right]+j^{-1}\gamma j^{\gamma}\\
 & =j^{\gamma}\mathbf{E}_{U\sim P}\left[\left(\frac{dP}{dQ}(U)\right)^{\gamma}\right]+\gamma j^{\gamma-1}.
\end{align*}
\end{IEEEproof}
\medskip{}
Consider the setting in the strong functional representation lemma
\cite{sfrl_trans}: given $(X,Y)$, we want to find a random variable
$Z$ independent of $X$ such that $Y$ is a function of $(X,Z)$,
and $H(Y|Z)$ is minimized. Take $Z=\{\bar{Y}_{i},T_{i}\}_{i\in\mathbb{N}}$.
Applying Proposition \ref{prop:concave} on $P=P_{Y|X}(\cdot|X)$,
$Q=P_{Y}$, we obtain
\begin{align*}
\mathbf{E}\left[\log\Upsilon_{P_{Y|X}(\cdot|X)\Vert P_{Y}}(1)\right] & \le\mathbf{E}\left[D(P_{Y|X}(\cdot|X)\Vert P_{Y})\right]\\
 & =I(X;Y).
\end{align*}
Using Proposition 4 in \cite{sfrl_trans},
\begin{align*}
 & H(Y|Z)\\
 & \le H(\Upsilon_{P_{Y|X}(\cdot|X)\Vert P_{Y}}(1))\\
 & \le\mathbf{E}\left[\log\Upsilon_{P_{Y|X}(\cdot|X)\Vert P_{Y}}(1)\right]+\log\left(\mathbf{E}\left[\log\Upsilon_{P_{Y|X}(\cdot|X)\Vert P_{Y}}(1)\right]+1\right)+1\\
 & \le I(X;Y)+\log e+\log\left(I(X;Y)+\log e+1\right)+1\\
 & \le I(X;Y)+\log\left(I(X;Y)+1\right)+\log e+1+\log\left(\log e+1\right)\\
 & \le I(X;Y)+\log\left(I(X;Y)+1\right)+3.732.
\end{align*}
The constant $3.732$ is smaller than that in \cite{sfrl_trans}:
\[
e^{-1}\log e+2+\log\left(e^{-1}\log e+2\right)\approx3.870.
\]

\section{Conclusions and Discussion}

In this paper, we introduced a simple yet versatile approach to achievability
proofs via the Poisson matching lemma. By reducing the uses of sub-codebooks
and binning, we improved upon existing one-shot bounds on channels
with state information at the encoder, lossy source coding with side
information at the decoder, broadcast channels, and distributed lossy
source coding. The Poisson matching lemma can replace the packing
lemma, covering lemma and soft covering lemma to be the only tool
needed to prove a wide range of results in network information theory.

In the proofs, random variables (e.g. the channel input and message
in channel coding settings, the source and description in source coding
settings, the channel output in channel resolvability) are regarded
as points in a Poisson process. The Poisson functional representation
is applied to map the Poisson process to give the correct conditional
distribution. Viewing every random variable in the operational setting
as a Poisson process gives a simple, unified and systematic approach
to code constructions.

A possible extension is to generalize the Poisson functional representation
to the multivariate case. In the proof of Marton's inner bound for
broadcast channels, we had two independent Poisson processes for $U_{1}$
and $U_{2}$ respectively. We first used the process for $U_{1}$
to obtain a list of values for $U_{1}$, then used the list to index
into the process for $U_{2}$. A more symmetric approach where we
select $(U_{1},U_{2})$ together (similar to the conventional mutual
covering approach) using a multivariate version of the Poisson functional
representation may be possible. Similarly, for distributed lossy source
coding and the multiple access channel, it may be possible to decode
both sources/messages simultaneously. While it can be argued that
the gain we obtained in broadcast channels and distributed lossy source
coding over conventional approaches comes from the asymmetry of our
construction (our bounds are asymmetric unlike previous bounds), a
symmetric treatment that does not result in a looser bound may be
developed in the future.

\section{Acknowledgements}

The authors acknowledge support from the NSF grants CNS-1527846, CCF-1618145,
the NSF Science \& Technology Center grant CCF-0939370 (Science of
Information), and the William and Flora Hewlett Foundation supported
Center for Long Term Cybersecurity at Berkeley.

\appendix

\subsection{Proof of Lemmas \ref{lem:phidiv} and \ref{lem:phidiv_gen}\label{subsec:pf_phidiv}}

We first prove Lemma \ref{lem:phidiv_gen}. For notational simplicity,
we use $\{X_{i}\}_{i\in\mathbb{N}}\sim\mathfrak{P}(\mu)$ to denote
that $\{X_{i}\}_{i\in\mathbb{N}}$ is the set of points of a Poisson
process with intensity measure $\mu$ (the ordering of the points
is ignored). Let $f(u)=(dP/d\mu)(u)$, $g(u)=(dQ/d\mu)(u)$. Let $\{\bar{U}_{i},T_{i}\}_{i\in\mathbb{N}}\sim\mathfrak{P}(\mu\times\lambda_{\mathbb{R}_{\ge0}})$.
Let $\{\check{U}_{k},\check{T}_{k}\}_{k\in\mathbb{N}}$ be the points
$(\bar{U}_{i},T_{i})$ where $f(\bar{U}_{i})=0$. By the mapping theorem
\cite{kingman1992poisson,last2017lectures} on the mapping 
\[
\psi(u,t)=\begin{cases}
(1,\,u,\,t/f(u)) & \mathrm{if}\;f(u)>0\\
(0,\,u,\,t) & \mathrm{if}\;f(u)=0,
\end{cases}
\]
we have $\{\psi(\bar{U}_{i},T_{i})\}_{i\in\mathbb{N}}\sim\mathfrak{P}(\delta_{1}\times P\times\lambda_{\mathbb{R}_{\ge0}}+\delta_{0}\times\mu_{\{f(u)=0\}}\times\lambda_{\mathbb{R}_{\ge0}})$
(where $\mu_{\{f(u)=0\}}$ denotes $\mu$ restricted to the set $\{u:\,f(u)=0\}$),
and hence $\{\tilde{U}_{P}(k),\tilde{T}_{P}(k)\}_{k\in\mathbb{N}}\sim\mathfrak{P}(P\times\lambda_{\mathbb{R}_{\ge0}})$
(the points in $\{\psi(\bar{U}_{i},T_{i})\}_{i\in\mathbb{N}}$ with
$f(\bar{U}_{i})>0$) is independent of $\{\check{U}_{k},\check{T}_{k}\}_{k\in\mathbb{N}}\sim\mathfrak{P}(\mu_{\{f(u)=0\}}\times\lambda_{\mathbb{R}_{\ge0}})$
(the points in $\{\psi(\bar{U}_{i},T_{i})\}_{k\in\mathbb{N}}$ with
$f(\bar{U}_{i})=0$).

Condition on $\tilde{U}_{P}(j)=u$ and $\tilde{T}_{P}(j)=t$ unless
otherwise stated. Assume $f(u)>0$ (which happens almost surely since
$\tilde{U}_{P}(j)\sim P$) and $g(u)>0$ (otherwise the inequalities
in the lemmas trivially hold). Recall that $\tilde{T}_{P}(1)\le\tilde{T}_{P}(2)\le\cdots$
by definition. It is straightforward to check that $\{\tilde{U}_{P}(k),\tilde{T}_{P}(k)\}_{k>j}\sim\mathfrak{P}(P\times\lambda_{[t,\infty)})$
independent of $\{\tilde{U}_{P}(k)\}_{k<j}\stackrel{iid}{\sim}P$
independent of $\{\tilde{T}_{P}(k)\}_{k<j}\sim\mathrm{Unif}(t\Delta_{*}^{j-1})$,
the uniform distribution over the ordered simplex $t\Delta_{*}^{j-1}=\{s^{j-1}:\,0\le s_{1}\le\cdots\le s_{j-1}\le t\}$
(i.e., $\{\tilde{U}_{P}(k),\tilde{T}_{P}(k)\}_{k<j}$ has the same
distribution as $j-1$ i.i.d. points following $P\times\mathrm{Unif}[0,t]$
sorted in ascending order of the second coordinate). We have
\begin{align*}
 & \Upsilon_{P\Vert Q}(j)-1\\
 & =\left|\left\{ k:\,T_{k}/g(\bar{U}_{k})<tf(u)/g(u)\right\} \right|\\
 & =\left|\left\{ k:\,f(\bar{U}_{k})=0\;\mathrm{and}\;T_{k}/g(\bar{U}_{k})<tf(u)/g(u)\right\} \right|\\
 & \;\;\;\;+\left|\left\{ k:\,\,f(\bar{U}_{k})>0\;\mathrm{and}\;T_{k}/g(\bar{U}_{k})<tf(u)/g(u)\right\} \right|\\
 & =\left|\left\{ k:\,\check{T}_{k}/g(\check{U}_{k})<tf(u)/g(u)\right\} \right|\\
 & \;\;\;\;+\left|\left\{ k:\,\tilde{T}_{P}(k)f(\tilde{U}_{P}(k))/g(\tilde{U}_{P}(k))<tf(u)/g(u)\right\} \right|\\
 & =A_{0}+A_{1}+B,
\end{align*}
where
\begin{align*}
A_{0} & :=\left|\left\{ k:\,\check{T}_{k}/g(\check{U}_{k})<tf(u)/g(u)\right\} \right|,\\
A_{1} & :=\left|\left\{ k>j:\,\tilde{T}_{P}(k)f(\tilde{U}_{P}(k))/g(\tilde{U}_{P}(k))<tf(u)/g(u)\right\} \right|,\\
B & :=\left|\left\{ k<j:\,\tilde{T}_{P}(k)f(\tilde{U}_{P}(k))/g(\tilde{U}_{P}(k))<tf(u)/g(u)\right\} \right|.
\end{align*}
Due to the aforementioned independence between $\{\check{U}_{k},\check{T}_{k}\}_{k\in\mathbb{N}}$,
$\{\tilde{U}_{P}(k),\tilde{T}_{P}(k)\}_{k>j}$ and $\{\tilde{U}_{P}(k),\tilde{T}_{P}(k)\}_{k<j}$,
we have $A_{0}{\perp\!\!\!\perp}A_{1}{\perp\!\!\!\perp}B$. For $A_{0}$,
since $\{\tilde{U}_{P}(k),\tilde{T}_{P}(k)\}_{k}{\perp\!\!\!\perp}\{\check{U}_{k},\check{T}_{k}\}_{k}$,
conditioning on $(\tilde{U}_{P}(j),\tilde{T}_{P}(j))=(u,t)$ does
not affect the distribution of $\{\check{U}_{k},\check{T}_{k}\}_{k}$,
and hence $A_{0}$ follows the Poisson distribution with rate
\begin{align*}
 & (\mu_{\{f(u)=0\}}\times\lambda)\left(\left\{ (v,s):\,s/g(v)<tf(u)/g(u)\right\} \right)\\
 & =\int\mathbf{1}\{f(v)=0\}\frac{tg(v)f(u)}{g(u)}\mu(dv).
\end{align*}
For $A_{1}$, since $\{\tilde{U}_{P}(k),\tilde{T}_{P}(k)\}_{k>j}\sim\mathfrak{P}(P\times\lambda_{[t,\infty)})$,
$A_{1}$ follows the Poisson distribution with rate
\begin{align*}
 & (P\times\lambda_{[t,\infty)})\left(\left\{ (v,s):\,sf(v)/g(v)<tf(u)/g(u)\right\} \right)\\
 & =\int\max\left\{ \frac{tg(v)f(u)}{f(v)g(u)}-t,\,0\right\} f(v)\mu(dv)\\
 & =t\int\mathbf{1}\{f(v)>0\}\max\left\{ \frac{g(v)f(u)}{g(u)}-f(v),\,0\right\} \mu(dv).
\end{align*}
Hence $A:=A_{0}+A_{1}$ follows the Poisson distribution with rate
\begin{align*}
 & t\int\left(\mathbf{1}\{f(v)=0\}\frac{g(v)f(u)}{g(u)}+\mathbf{1}\{f(v)>0\}\max\left\{ \frac{g(v)f(u)}{g(u)}-f(v),\,0\right\} \right)\mu(dv)\\
 & =t\int\max\left\{ \frac{g(v)f(u)}{g(u)}-f(v),\,0\right\} \mu(dv)\\
 & =tf(u)\int\max\left\{ \frac{g(v)}{g(u)}-\frac{f(v)}{f(u)},\,0\right\} \mu(dv)\\
 & =:\,t\alpha(u).
\end{align*}
For $B$, since $\{\tilde{U}_{P}(k),\tilde{T}_{P}(k)\}_{k<j}$ has
the same distribution as $j-1$ i.i.d. points following $P\times\mathrm{Unif}[0,t]$
sorted in ascending order of the second coordinate, $B$ follows the
binomial distribution with number of trials $j-1$ and success probability
\begin{align*}
 & (P\times\mathrm{Unif}[0,t])\left(\left\{ (v,s):\,sf(v)/g(v)<tf(u)/g(u)\right\} \right)\\
 & =t^{-1}\int\min\left\{ \frac{tg(v)f(u)}{f(v)g(u)},\,t\right\} f(v)\mu(dv)\\
 & =f(u)\int\min\left\{ \frac{g(v)}{g(u)},\,\frac{f(v)}{f(u)}\right\} \mu(dv)\\
 & =:\,\beta(u).
\end{align*}

Conditioned on $\tilde{U}_{P}(j)=u$ (without conditioning on $\tilde{T}_{P}(j)$),
we have $\tilde{T}_{P}(j)\sim\mathrm{Erlang}(j,1)$, and $(A,B)|\{\tilde{T}_{P}(j)=t\}\sim\mathrm{Poi}(t\alpha(u))\times\mathrm{Bin}(j-1,\beta(u))$.
Hence, conditioned on $\tilde{U}_{P}(j)=u$, the distribution of $\Upsilon_{P\Vert Q}(j)-1=A+B$
is
\begin{equation}
\mathrm{NegBin}\left(j,\,1-\frac{1}{1+\alpha(u)}\right)+\mathrm{Bin}(j-1,\beta(u)),\label{eq:dist_exact}
\end{equation}
i.e., the sum of a negative binomial random variable and an independent
binomial random variable. The mean is
\begin{align*}
 & \mathbf{E}\left[\left.\Upsilon_{P\Vert Q}(j)\,\right|\,\tilde{U}_{P}(j)=u\right]-1\\
 & =j\alpha(u)+(j-1)\beta(u)\\
 & =jf(u)\int\max\left\{ \frac{g(v)}{g(u)}-\frac{f(v)}{f(u)},\,0\right\} \mu(dv)+(j-1)f(u)\int\min\left\{ \frac{g(v)}{g(u)},\,\frac{f(v)}{f(u)}\right\} \mu(dv)\\
 & =jf(u)\int\frac{g(v)}{g(u)}\mu(dv)-f(u)\int\min\left\{ \frac{g(v)}{g(u)},\,\frac{f(v)}{f(u)}\right\} \mu(dv)\\
 & \le j\frac{f(u)}{g(u)}\\
 & =j\frac{dP}{dQ}(u).
\end{align*}
Also,
\begin{align*}
 & \mathbf{P}\left\{ \left.\Upsilon_{P\Vert Q}(1)>1\,\right|\,\tilde{U}_{P}(j)=u\right\} \\
 & =1-\mathbf{P}\left\{ \left.A=0\;\mathrm{and}\;B=0\,\right|\,\tilde{U}_{P}(j)=u\right\} \\
 & =1-\frac{(1-\beta(u))^{j-1}}{(1+\alpha(u))^{j}}\\
 & \le1-\left(\frac{1-\beta(u)}{1+\alpha(u)}\right)^{j}\\
 & \le1-\left(1-\min\{\alpha(u)+\beta(u),1\}\right)^{j}\\
 & =1-\left(1-\min\left\{ f(u)\int\frac{g(v)}{g(u)}\mu(dv),\,1\right\} \right)^{j}\\
 & =1-\left(1-\min\left\{ \frac{f(u)}{g(u)},\,1\right\} \right)^{j}\\
 & =1-\left(1-\min\left\{ \frac{dP}{dQ}(u),\,1\right\} \right)^{j}.
\end{align*}
For $j=1$,
\begin{align}
 & \mathbf{P}\left\{ \left.\Upsilon_{P\Vert Q}(1)>k\,\right|\,\tilde{U}_{P}(1)=u\right\} \nonumber \\
 & =\left(1-(1+\alpha(u))^{-1}\right)^{k}\nonumber \\
 & =\left(1-\left(1+f(u)\int\max\left\{ \frac{g(v)}{g(u)}-\frac{f(v)}{f(u)},\,0\right\} \mu(dv)\right)^{-1}\right)^{k}\label{eq:prob_exact}\\
 & \le\left(1-\left(1+f(u)\int\frac{g(v)}{g(u)}\mu(dv)\right)^{-1}\right)^{k}\nonumber \\
 & =\left(1-(1+f(u)/g(u))^{-1}\right)^{k}\nonumber \\
 & \le\Bigl(1-\Bigl(1+\frac{dP}{dQ}(u)\Bigr)^{-1}\Bigr)^{k}\nonumber \\
 & =\exp\Bigl(-k\ln\Bigl(\Bigl(\frac{dP}{dQ}(u)\Bigr)^{-1}+1\Bigr)\Bigr)\nonumber \\
 & \le\exp\Bigl(-\ln\Bigl(k\Bigl(\frac{dP}{dQ}(u)\Bigr)^{-1}+1\Bigr)\Bigr)\nonumber \\
 & =1-\Bigl(1+k^{-1}\frac{dP}{dQ}(u)\Bigr)^{-1}.\nonumber 
\end{align}

\subsection{Proof of the Conditional Poisson Matching Lemma\label{subsec:phidiv_cond}}

The conditional Poisson matching lemma is intuitively obvious. The
Poisson matching lemma can be equivalently stated as: for any probability
measures $\nu,\xi\ll\mu$, the following holds for $\nu$-almost all
$u$:
\[
\mathbf{P}_{\{\bar{U}_{i},T_{i}\}_{i}\sim P_{\{\bar{U}_{i},T_{i}\}_{i}\,|\,\tilde{U}_{\nu}=u}}\{\tilde{U}_{\xi}(\{\bar{U}_{i},T_{i}\}_{i})\neq u\}\le1-\left(1+\frac{d\nu}{d\xi}(u)\right)^{-1},
\]
where $P_{\{\bar{U}_{i},T_{i}\}_{i}\,|\,\tilde{U}_{\nu}=u}$ is the
conditional distribution of the Poisson process given $\tilde{U}_{\nu}=u$.
Intuitively, we can consider the Poisson matching lemma to be a statement
with 3 parameters $\nu,\xi,u$ (ignore the almost-all condition on
$u$ for the moment). Since the statement holds for (almost) any $(\nu,\xi,u)$,
it also holds for any random choice of $(\nu,\xi,u)$. In particular,
it holds for $(\nu,\xi,u)=(P_{U|X}(\cdot|X),\,Q_{U|Y}(\cdot|Y),\,U)$,
where $(X,U,Y)\sim P_{X,U,Y}$, which gives the conditional Poisson
matching lemma. Note that the probability in the conditional Poisson
matching lemma is conditional on $(X,U,Y)$, where $(X,U,Y)\leftrightarrow(\nu,\xi,u)\leftrightarrow\{\bar{U}_{i},T_{i}\}_{i}$,
and hence conditioning on $(X,U,Y)$ has the same effect on $\{\bar{U}_{i},T_{i}\}_{i}$
as conditioning on the parameters $(\nu,\xi,u)$.

\smallskip{}

We now prove the conditional Poisson matching lemma rigorously. Let
$(\Omega,\mathcal{F},P_{\{\bar{U}_{i},T_{i}\}_{i}})$ be the probability
space for $\{\bar{U}_{i},T_{i}\}_{i}$, the points of a Poisson process
with intensity measure $\mu\times\lambda_{\mathbb{R}_{\ge0}}$ on
$\mathcal{U}\times\mathbb{R}_{\ge0}$ (let $\mathcal{E}$ be the Borel
$\sigma$-algebra of $\mathcal{U}$). The Poisson matching lemma can
be equivalently stated as: for any probability measures $\nu,\xi\ll\mu$,
and $\kappa:\mathcal{U}\times\mathcal{F}\to[0,1]$ a regular conditional
probability distribution (RCPD) of $\{\bar{U}_{i},T_{i}\}_{i}$ conditioned
on $\tilde{U}_{\nu}(\{\bar{U}_{i},T_{i}\}_{i})$ (i.e., $\kappa$
is a probability kernel, and $P_{\{\bar{U}_{i},T_{i}\}_{i}}(A\cap\tilde{U}_{\nu}^{-1}(B))=\int_{B}\kappa(u,A)\nu(du)$
for any $A\in\mathcal{F}$, $B\in\mathcal{E}$ , where $\tilde{U}_{\nu}^{-1}(B)$
denotes the preimage of $B$ under $\tilde{U}_{\nu}:\Omega\to\mathcal{U}$,
note that $\tilde{U}_{\nu}(\{\bar{U}_{i},T_{i}\}_{i})\sim\nu$), then
we have
\begin{equation}
\int\mathbf{1}\{\tilde{U}_{\xi}(\{\bar{u}_{i},t_{i}\}_{i})\neq u\}\kappa(u,\,d\{\bar{u}_{i},t_{i}\}_{i})\le1-\left(1+\frac{d\nu}{d\xi}(u)\right)^{-1}\label{eq:phidiv_int}
\end{equation}
for $\nu$-almost all $u$.

Consider the conditional Poisson matching lemma. We have the following
for $P_{X,U,Y}$-almost all $(x,u,y)$:
\begin{align*}
 & \mathbf{P}\left\{ \left.\tilde{U}_{Q_{U|Y}(\cdot|Y)}\neq U\,\right|\,X=x,U=u,Y=y\right\} \\
 & =\int\mathbf{1}\{\tilde{U}_{Q_{U|Y}(\cdot|y)}(\{\bar{u}_{i},t_{i}\}_{i})\neq u\}P_{\{\bar{U}_{i},T_{i}\}_{i}|X,U,Y}(d\{\bar{u}_{i},t_{i}\}_{i}|x,u,y)\\
 & \stackrel{(a)}{=}\int\mathbf{1}\{\tilde{U}_{Q_{U|Y}(\cdot|y)}(\{\bar{u}_{i},t_{i}\}_{i})\neq u\}P_{\{\bar{U}_{i},T_{i}\}_{i}|X,U}(d\{\bar{u}_{i},t_{i}\}_{i}|x,u)\\
 & \stackrel{(b)}{\le}1-\left(1+\frac{dP_{U|X}(\cdot|x)}{dQ_{U|Y}(\cdot|y)}(u)\right)^{-1},
\end{align*}
where (a) holds for $P_{X,U,Y}$-almost all $(x,u,y)$ due to $Y\leftrightarrow(X,U)\leftrightarrow\{\bar{U}_{i},T_{i}\}_{i}$,
and (b) is by \eqref{eq:phidiv_int} with $(\nu,\xi,\kappa)\leftarrow(P_{U|X}(\cdot|x),\,Q_{U|Y}(\cdot|y),\,P_{\{\bar{U}_{i},T_{i}\}_{i}|X,U}(\cdot|x,\cdot))$,
which holds for $P_{U|X}(\cdot|x)$-almost all $u$, and hence holds
for $P_{X,U,Y}$-almost all $(x,u,y)$. We now check that $P_{\{\bar{U}_{i},T_{i}\}_{i}|X,U}(\cdot|x,\cdot)$
satisfies the RCPD condition for $P_{X}$-almost all $x$. Since $X{\perp\!\!\!\perp}\{\bar{U}_{i},T_{i}\}_{i}$,
we have $P_{\{\bar{U}_{i},T_{i}\}_{i}}(\cdot)=P_{\{\bar{U}_{i},T_{i}\}_{i}|X}(\cdot|x)$
for $P_{X}$-almost all $x$. Since $U=\tilde{U}_{P_{U|X}(\cdot|X)}(\{\bar{U}_{i},T_{i}\}_{i})$,
we have $P_{\{\bar{U}_{i},T_{i}\}_{i}|X,U}(\tilde{U}_{P_{U|X}(\cdot|x)}^{-1}(\{u\})\allowbreak\,|\,x,u)=1$
for $P_{X,U}$-almost all $(x,u)$. Hence the following conditions
are satisfied for $P_{X}$-almost all $x$:
\begin{equation}
P_{\{\bar{U}_{i},T_{i}\}_{i}}(\cdot)=P_{\{\bar{U}_{i},T_{i}\}_{i}|X}(\cdot|x),\label{eq:cond_phidiv_xindep}
\end{equation}
\begin{equation}
P_{\{\bar{U}_{i},T_{i}\}_{i}|X,U}(\tilde{U}_{P_{U|X}(\cdot|x)}^{-1}(\{u\})|x,u)=1\;\text{for}\,P_{U|X}(\cdot|x)\text{-almost all}\,u.\label{eq:cond_phidiv_u}
\end{equation}
For any $x$ satisfying \eqref{eq:cond_phidiv_xindep} and \eqref{eq:cond_phidiv_u},
we have the following: for all $A\in\mathcal{F}$, $B\in\mathcal{E}$,
\begin{align*}
 & P_{\{\bar{U}_{i},T_{i}\}_{i}}(A\cap\tilde{U}_{P_{U|X}(\cdot|x)}^{-1}(B))\\
 & \stackrel{(a)}{=}P_{\{\bar{U}_{i},T_{i}\}_{i}|X}(A\cap\tilde{U}_{P_{U|X}(\cdot|x)}^{-1}(B)|x)\\
 & =\int P_{\{\bar{U}_{i},T_{i}\}_{i}|X,U}(A\cap\tilde{U}_{P_{U|X}(\cdot|x)}^{-1}(B)\,|\,x,u)P_{U|X}(du|x)\\
 & \stackrel{(b)}{=}\int P_{\{\bar{U}_{i},T_{i}\}_{i}|X,U}(A\cap\tilde{U}_{P_{U|X}(\cdot|x)}^{-1}(B)\cap\tilde{U}_{P_{U|X}(\cdot|x)}^{-1}(\{u\})\,|\,x,u)P_{U|X}(du|x)\\
 & =\int\mathbf{1}\{u\in B\}P_{\{\bar{U}_{i},T_{i}\}_{i}|X,U}(A\cap\tilde{U}_{P_{U|X}(\cdot|x)}^{-1}(\{u\})\,|\,x,u)P_{U|X}(du|x)\\
 & \stackrel{(c)}{=}\int_{B}P_{\{\bar{U}_{i},T_{i}\}_{i}|X,U}(A|x,u)P_{U|X}(du|x),
\end{align*}
where (a) is by \eqref{eq:cond_phidiv_xindep}, and (b), (c) are by
\eqref{eq:cond_phidiv_u}.

\subsection{Proof of Theorem \ref{thm:channel2}\label{subsec:channel2}}

Let $\{\bar{X}_{i},T_{i}\}_{i\in\mathbb{N}}$ be the points of a Poisson
process with intensity measure $P_{X}\times\lambda_{\mathbb{R}_{\ge0}}$
independent of $M$. The encoding function is $m\mapsto\tilde{X}_{P_{X}}(m)$
(i.e., $X=\tilde{X}_{P_{X}}(M)$), and the decoding function is $y\mapsto\Upsilon_{P_{X|Y}(\cdot|y)\Vert P_{X}}(1)$.
We have $(M,X,Y)\sim P_{M}\times P_{X}P_{Y|X}$,
\begin{align*}
 & \mathbf{P}\{M\neq\Upsilon_{P_{X|Y}(\cdot|Y)\Vert P_{X}}(1)\}\\
 & \stackrel{(a)}{=}\mathbf{P}\{\Upsilon_{P_{X}\Vert P_{X|Y}(\cdot|Y)}(M)>1\}\\
 & =\mathbf{E}\left[\mathbf{P}\left\{ \left.\Upsilon_{P_{X}\Vert P_{X|Y}(\cdot|Y)}(M)>1\,\right|\,M,X,Y\right\} \right]\\
 & \stackrel{(b)}{\le}\mathbf{E}\left[1-\left(1-\min\left\{ \frac{dP_{X}}{dP_{X|Y}(\cdot|Y)}(X),\,1\right\} \right)^{M}\right]\\
 & =\mathbf{E}\left[1-\left(1-\min\left\{ 2^{-\iota_{X;Y}(X;Y)},\,1\right\} \right)^{M}\right]\\
 & \stackrel{(c)}{\le}\mathbf{E}\left[1-\left(1-\min\left\{ 2^{-\iota_{X;Y}(X;Y)},\,1\right\} \right)^{(\mathsf{L}+1)/2}\right],
\end{align*}
where (a) is by the definition of $\Upsilon$, (b) is by the conditional
generalized Poisson matching lemma on $(\emptyset,M,X,Y,P_{X|Y})$,
and (c) is by $M{\perp\!\!\!\perp}(X,Y)$ and Jensen's inequality.
Therefore there exists a fixed $\{\bar{x}_{i},t_{i}\}_{i\in\mathbb{N}}$
attaining the desired bound.
\begin{flushright}
$\blacksquare$
\par\end{flushright}

\medskip{}

A noteworthy property of this construction is that both the encoder
and the decoder do not require knowledge of $\mathsf{L}$. The code
can transmit any integer $m\in\mathbb{N}$ with error probability
$\mathbf{E}\left[1-(1-\min\{2^{-\iota_{X;Y}(X;Y)},\,1\})^{m}\right]$,
assuming unlimited common randomness $\{\bar{X}_{i},T_{i}\}_{i\in\mathbb{N}}$
between the encoder and the decoder.

\subsection{Dispersion of Joint Source-Channel Coding\label{subsec:second_jscc}}

We show a second order result for joint source-channel coding using
Theorem \ref{thm:jscc} that coincides with the optimal dispersion
in \cite{kostina2013joint}. Consider an i.i.d. source sequence $W^{k}$
of length $k$, separable distortion measure $\mathsf{d}(w^{k},\hat{z}^{k})=\frac{1}{k}\sum_{i=1}^{k}\mathsf{d}(w_{i},\hat{z}_{i})$,
and $n$ uses of the memoryless channel $P_{Y|X}$. Let $P_{Z|W}$
attain the infimum of the rate-distortion function
\[
R(\mathsf{D}):=\inf_{P_{Z|W}:\,\mathbf{E}[\mathsf{d}(W,Z)]\le\mathsf{D}}I(W;Z).
\]
The $\mathsf{D}$-tilted information \cite{kostina2012lossy} is defined
as
\[
\jmath_{W}(w,\mathsf{D}):=-\log\mathbf{E}\left[2^{\nu^{*}(\mathsf{D}-\mathsf{d}(w,Z))}\right],
\]
where $Z\sim P_{Z}$ (the unconditional $Z$-marginal of $P_{W}P_{Z|W}$),
and $\nu^{*}=-R'(\mathsf{D})$ (the derivative exists if the infimum
in $R(\mathsf{D})$ is achieved by a unique $P_{Z|W}$ \cite{kostina2012lossy}).
We invoke a lemma in \cite{kostina2012lossy}:
\begin{lem}
[\cite{kostina2012lossy}, Lemma 2]\label{lem:kostina_lemma}If
the following conditions hold: 
\begin{itemize}
\item $\inf\{\tilde{\mathsf{D}}\ge0:\,R(\tilde{\mathsf{D}})<\infty\}<\mathsf{D}<\inf_{z\in\mathcal{Z}}\mathbf{E}[\mathsf{d}(W,z)]$,
\item the infimum in $R(\mathsf{D})$ is achieved by a unique $P_{Z|W}$,
\item there exists a finite set $\tilde{\mathcal{Z}}\subseteq\mathcal{Z}$
such that $\mathbf{E}[\min_{z\in\tilde{\mathcal{Z}}}\mathsf{d}(W,z)]<\infty$,
and
\item $\mathbf{E}_{P_{W}\times P_{Z}}[(\mathsf{d}(W,Z))^{9}]<\infty$ (computed
assuming $W,Z$ independent),
\end{itemize}
then there exist constants $\alpha,\beta,\gamma,k_{0}>0$ such that
for $k\ge k_{0}$,
\[
\mathbf{P}\left\{ -\log P_{Z^{k}}(\mathcal{B}_{\mathsf{D}}(W^{k}))\le\sum_{i=1}^{k}\jmath_{W}(W_{i},\mathsf{D})+\alpha\log k+\beta\right\} \ge1-\frac{\gamma}{\sqrt{k}},
\]
where $W^{k}\stackrel{iid}{\sim}P_{W}$, and $P_{Z^{k}}=P_{Z}^{\otimes k}$.
\end{lem}
We now show a second order result.
\begin{prop}
Fix $P_{X}$, $0<\epsilon<1$, $n,k\in\mathbb{N}$. We have $P_{e}=\mathbf{P}\{\mathsf{d}(W^{k},\hat{Z}^{k})>\mathsf{D}\}\le\epsilon$
if the conditions in Lemma \ref{lem:kostina_lemma} are satisfied,
$k\ge k_{0}$, and
\[
nC-kR(\mathsf{D})\ge\sqrt{nV+k\mathcal{V}(\mathsf{D})}\mathcal{Q}^{-1}\left(\epsilon-\frac{\eta}{\sqrt{\min\{n,k\}}}\right)+\alpha\log k+\frac{1}{2}\log n+\beta,
\]
where $C:=I(X;Y)$, $V:=\mathrm{Var}[\iota_{X;Y}(X;Y)]$, $\mathcal{V}(\mathsf{D}):=\mathrm{Var}[\jmath_{W}(W,\mathsf{D})]$,
and $\eta>0$ is a constant that depends on $P_{X,Y}$ and the distribution
of $\jmath_{W}(W,\mathsf{D})$. 
\end{prop}
\begin{IEEEproof}
We have
\begin{align*}
P_{e} & =\mathbf{P}\{\mathsf{d}(W^{k},\hat{Z}^{k})>\mathsf{D}\}\\
 & \stackrel{(a)}{\le}\mathbf{P}\left\{ -\log P_{Z^{k}}(\mathcal{B}_{\mathsf{D}}(W^{k}))>\sum_{i=1}^{k}\jmath_{W}(W_{i},\mathsf{D})+\alpha\log k+\beta\right\} \\
 & \;\;\;+\mathbf{E}\left[\left(1+2^{-\sum_{i=1}^{k}\jmath_{W}(W_{i},\mathsf{D})-\alpha\log k-\beta}2^{\iota_{X^{n};Y^{n}}(X^{n};Y^{n})}\right)^{-1}\right]\\
 & \stackrel{(b)}{\le}\frac{\gamma}{\sqrt{k}}+\frac{1}{\sqrt{n}}+\mathbf{P}\left\{ 2^{\sum_{i=1}^{k}\jmath_{W}(W_{i},\mathsf{D})-\iota_{X^{n};Y^{n}}(X^{n};Y^{n})+\alpha\log k+\beta}>\frac{1}{\sqrt{n}}\right\} \\
 & =\frac{\gamma}{\sqrt{k}}+\frac{1}{\sqrt{n}}+\mathbf{P}\left\{ \sum_{i=1}^{n}(\iota_{X;Y}(X_{i};Y_{i})-C)-\sum_{i=1}^{k}(\jmath_{W}(W_{i},\mathsf{D})-R(\mathsf{D}))<-nC+kR(\mathsf{D})+\alpha\log k+\frac{1}{2}\log n+\beta\right\} \\
 & \le\frac{\gamma}{\sqrt{k}}+\frac{1}{\sqrt{n}}+\mathbf{P}\left\{ \sum_{i=1}^{n}(\iota_{X;Y}(X_{i};Y_{i})-C)-\sum_{i=1}^{k}(\jmath_{W}(W_{i},\mathsf{D})-R(\mathsf{D}))<-\sqrt{nV+k\mathcal{V}(\mathsf{D})}\mathcal{Q}^{-1}\left(\epsilon-\frac{\eta}{\sqrt{\min\{n,k\}}}\right)\right\} \\
 & \stackrel{(c)}{\le}\frac{\gamma}{\sqrt{k}}+\frac{1}{\sqrt{n}}+\epsilon-\frac{\eta}{\sqrt{\min\{n,k\}}}+\frac{\eta-\gamma-1}{\sqrt{\min\{n,k\}}}\\
 & \le\epsilon
\end{align*}
where (a) is by Theorem \ref{thm:jscc}, (b) is by Lemma \ref{lem:kostina_lemma},
and (c) is by the Berry-Esseen theorem \cite{berry1941accuracy,esseen1942liapunov,feller1971introduction}
if we let $\eta-\gamma-1$ be a constant given by the Berry-Esseen
theorem.
\end{IEEEproof}
This coincides with the optimal dispersion in \cite{kostina2013joint}.
Although this is not a self-contained proof (it requires the lemma
in \cite{kostina2012lossy} for the dispersion of lossy source coding),
it shows how we can obtain the achievability of the dispersion in
joint source-channel coding from a result on the dispersion of lossy
source coding with little additional effort, using the Poisson matching
lemma. This proof is considerably simpler than that in \cite{kostina2013joint}.

\subsection{Properties of $\phi(t)$}

Let $\phi:\mathbb{R}_{>0}\to\mathbb{R}_{>0}$, $\phi(t)=ct^{-1}(\log(t+2))^{-2}$,
where $c>0$ such that $\sum_{j=1}^{\infty}\phi(j)=1$. Note that
$(\phi(t))^{-1}$ is convex. It can be checked numerically that $1\le c\le2$.
We prove a useful inequality about $\phi(t)$.
\begin{prop}
\label{prop:phi_ineq}For any $s>0$, $t\ge1$, we have
\[
\min\{s(\phi(t))^{-1},\,1\}\le\min\left\{ st\left(\log(s^{-1}+1)+1\right)^{2},\,1\right\} .
\]
Moreover, if $st\le2^{-\alpha}$, $t-1\le2^{\beta}$, and $\tilde{\alpha}\ge\max\{\alpha,0\}$,
then
\[
\min\left\{ st\left(\log(s^{-1}+1)+1\right)^{2},\,1\right\} \le2^{-\alpha}\left(2(\tilde{\alpha}+\beta)^{2}+2\tilde{\alpha}^{2}+14\right).
\]
\end{prop}
\begin{IEEEproof}
Write $\phi^{-1}(t)$ for the inverse function of $\phi$. Since
\[
\phi\left(\frac{c}{t\left(\log\left(c/t+2\right)\right)^{2}}\right)=\frac{t\left(\log\left(c/t+2\right)\right)^{2}}{\left(\log\left(\frac{c}{t\left(\log\left(c/t+2\right)\right)^{2}}+2\right)\right)^{2}}\ge t,
\]
we have 
\[
\phi^{-1}(t)\ge\frac{c}{t\left(\log\left(c/t+2\right)\right)^{2}}.
\]
By the convexity of $(\phi(t))^{-1}$ ,
\begin{align*}
\min\left\{ \frac{s}{\phi(t)},\,1\right\}  & \le\min\left\{ \frac{t}{\phi^{-1}(s)},\,1\right\} \\
 & \le\min\left\{ tc^{-1}s\left(\log\left(c/s+2\right)\right)^{2},\,1\right\} \\
 & \le\min\left\{ st\left(\log\left(2/s+2\right)\right)^{2},\,1\right\} \\
 & =\min\left\{ st\left(\log(s^{-1}+1)+1\right)^{2},\,1\right\} .
\end{align*}
If $st\le2^{-\alpha}$, $t-1\le2^{\beta}$, and $\tilde{\alpha}\ge\max\{\alpha,0\}$,
\begin{align*}
 & \min\left\{ st\left(\log(s^{-1}+1)+1\right)^{2},\,1\right\} \\
 & =\min\left\{ st\left(\log(t/(st)+1)+1\right)^{2},\,1\right\} \\
 & \le2^{-\alpha}\left(\log((2^{\beta}+1)2^{\alpha}+1)+1\right)^{2}\\
 & \le2^{-\alpha}\left(\log(2^{\tilde{\alpha}+\beta}+2^{\tilde{\alpha}}+1)+1\right)^{2}\\
 & \le2^{-\alpha}\left(\max\{\tilde{\alpha}+\beta,\,\tilde{\alpha}\}+\log3+1\right)^{2}\\
 & \le2^{-\alpha}\left(2(\tilde{\alpha}+\beta)^{2}+2\tilde{\alpha}^{2}+14\right),
\end{align*}
where the last inequality follows from considering whether $\beta$
is positive or negative, and the inequality $(x+y)^{2}\le2x^{2}+2y^{2}$.
\end{IEEEproof}

\subsection{Proof of Theorem \ref{thm:bc_cm} for Broadcast Channel with Common
Message\label{subsec:pf_bc_cm}}

The parameters $\mathsf{K}_{1},\mathsf{K}_{2}$ correspond to rate
splitting. We can split $M_{1}\in[1:\mathsf{L}_{1}]$ into $M_{10}\in[1:\mathsf{K}_{1}]$
and $M_{11}\in[1:\lceil\mathsf{L}_{1}\mathsf{K}_{1}^{-1}\rceil]$,
and treat $M_{10}$ as part of $M_{0}$ to be decoded by both decoders.
Although $M_{10}$ and $M_{11}$ may not be uniformly distributed,
we can apply a random cyclic shift to $M_{1}$ such that $M_{1}\sim\mathrm{Unif}[1:\mathsf{K}_{1}\lceil\mathsf{L}_{1}\mathsf{K}_{1}^{-1}\rceil]$
(and hence $M_{10},M_{11}$ are also uniform), and condition on a
fixed shift at the end. Also $M_{2}$ can be split similarly. Therefore
we assume $\mathsf{K}_{1}=\mathsf{K}_{2}=1$ without loss of generality.

Let $\mathfrak{P}_{0}=\{(\bar{U}_{0,i},\bar{M}_{00,i}),T_{0,i}\}_{i\in\mathbb{N}}$,
$\mathfrak{P}_{1}=\{(\bar{U}_{1,i},\bar{M}_{01,i},\bar{M}_{1,i}),T_{1,i}\}_{i\in\mathbb{N}}$,
$\mathfrak{P}_{2}=\{(\bar{U}_{2,i},\bar{M}_{02,i},\bar{M}_{2,i}),T_{2,i}\}_{i\in\mathbb{N}}$
be three independent Poisson processes with intensity measures $P_{U_{0}}\times P_{M_{0}}\times\lambda_{\mathbb{R}_{\ge0}}$,
$P_{U_{1}}\times P_{M_{0}}\times P_{M_{1}}\times\lambda_{\mathbb{R}_{\ge0}}$
and $P_{U_{2}}\times P_{M_{0}}\times P_{M_{2}}\times\lambda_{\mathbb{R}_{\ge0}}$
respectively, independent of $M_{0},M_{1},M_{2}$.

The encoder would generate $X$ such that
\begin{align}
 & (M_{0},M_{1},M_{2},U_{0},J,\{\check{U}_{1j}\}_{j\in[1:\mathsf{J}]},U_{1},U_{2},X)\nonumber \\
 & \sim P_{M_{0}}\times P_{M_{1}}\times P_{M_{2}}\times P_{U_{0}}P_{J}P_{U_{1}|U_{0}}^{\otimes\mathsf{J}}\delta_{\check{U}_{1J}}P_{U_{2}|U_{0},U_{1}}\delta_{x(U_{0},U_{1},U_{2})},\label{eq:bc2_dist}
\end{align}
where $P_{J}=\mathrm{Unif}[1:\mathsf{J}]$, and $\{\check{U}_{1j}\}_{j\in[1:\mathsf{J}]}\in\mathcal{U}_{1}^{\mathsf{J}}$
is an intermediate list (which can be regarded as a sub-codebook).
The term $P_{U_{1}|U_{0}}^{\otimes\mathsf{J}}\delta_{\check{U}_{1J}}$
in \eqref{eq:bc2_dist} means that $\{\check{U}_{1j}\}_{j}$ are conditionally
i.i.d. $P_{U_{1}|U_{0}}$ given $U_{0}$, and $U_{1}=\check{U}_{1J}$.
To accomplish this, the encoder computes $U_{0}=(\tilde{U}_{0})_{P_{U_{0}}\times\delta_{M_{0}}}$,
$\check{U}_{1j}=(\tilde{U}_{1})_{P_{U_{1}|U_{0}}(\cdot|U_{0})\times\delta_{M_{0}}\times\delta_{M_{1}}}(j)$
for $j=1,\ldots,\mathsf{J}$, \\
$U_{2}=(\tilde{U}_{2})_{\mathsf{J}^{-1}\sum_{j=1}^{\mathsf{J}}P_{U_{2}|U_{0},U_{1}}(\cdot|U_{0},\check{U}_{1j})\times\delta_{M_{0}}\times\delta_{M_{2}}}$
(which Poisson process we are referring to can be deduced from whether
we are discussing $U_{0}$, $U_{1}$ or $U_{2}$), $(J,U_{1})|(U_{0},\{\check{U}_{1j}\}_{j},U_{2})\sim P_{J,U_{1}|U_{0},\{\check{U}_{1j}\}_{j},U_{2}}$
(where $P_{J,U_{1}|U_{0},\{\check{U}_{1j}\}_{j},U_{2}}$ is derived
from \eqref{eq:bc2_dist}), and outputs $X=x(U_{0},U_{1},U_{2})$.
It can be verified that \eqref{eq:bc2_dist} is satisfied.

For the decoding function at the decoder $a\in[1:2]$, let $(\check{U}_{0aj},\check{M}_{0aj})=(\tilde{U}_{0},\tilde{M}_{00})_{P_{U_{0}|Y_{a}}(\cdot|Y_{a})\times P_{M_{0}}}(j)$
for $j\in\mathbb{N}$, $(\hat{U}_{a},\hat{M}_{0a},\hat{M}_{a})=(\tilde{U}_{a},\tilde{M}_{0a},\tilde{M}_{a})_{\sum_{j=1}^{\infty}\phi(j)(P_{U_{a}|U_{0},Y_{a}}(\cdot|\check{U}_{0aj},Y_{a})\times\delta_{\check{M}_{0aj}})\times P_{M_{a}}}$
where $\phi(j)\propto j^{-1}(\log(j+2))^{-2}$ with $\sum_{j=1}^{\infty}\phi(j)=1$.

\begin{figure}
\begin{centering}
\includegraphics{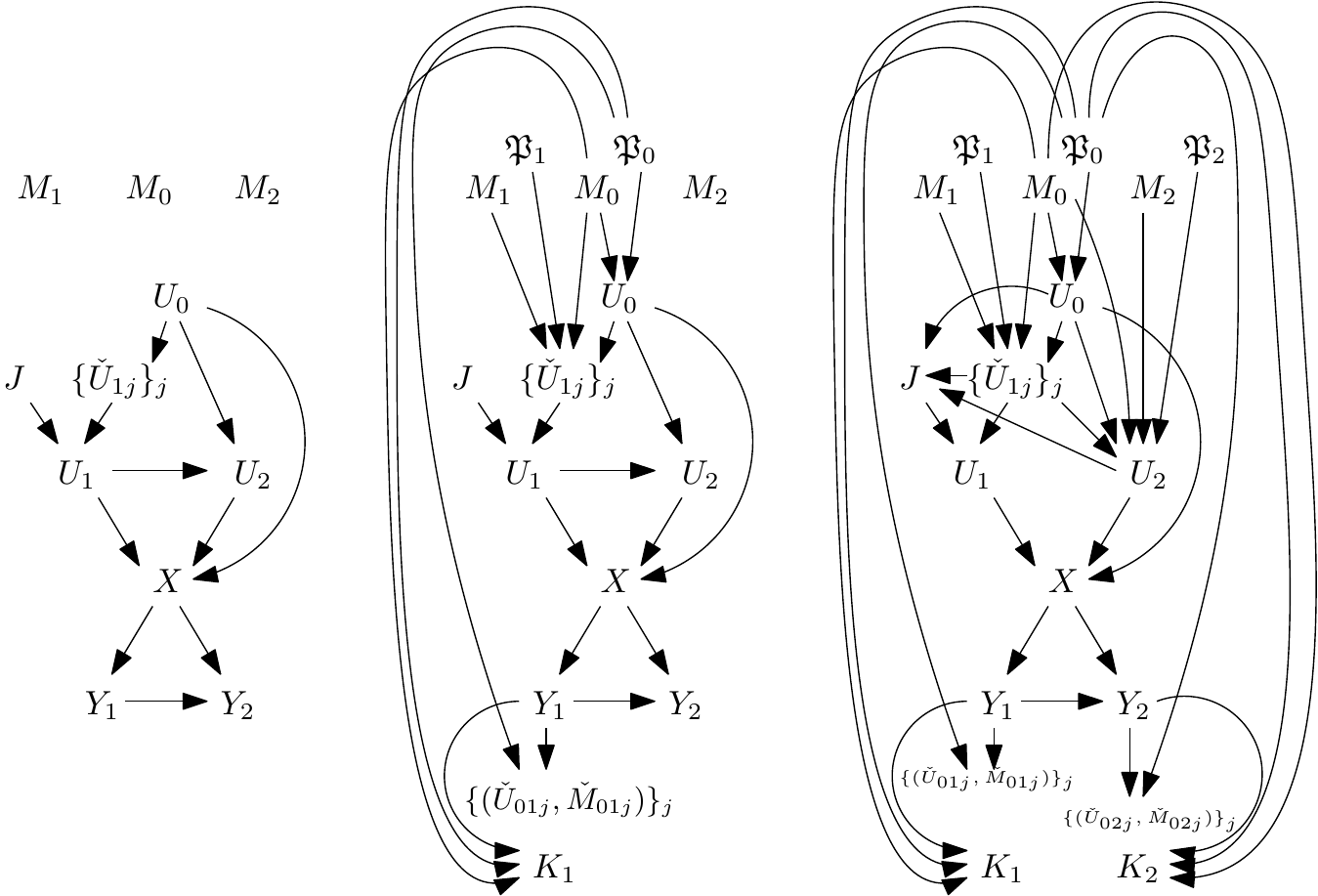}
\par\end{centering}
\caption{\label{fig:bc2_bayes}Left: The Bayesian network described in \eqref{eq:bc2_dist}.
Middle: The Bayesian network deduced from \eqref{eq:bc2_dist}, $U_{0}=(\tilde{U}_{0})_{P_{U_{0}}\times\delta_{M_{0}}}$,
$\check{U}_{1j}=(\tilde{U}_{1})_{P_{U_{1}|U_{0}}(\cdot|U_{0})\times\delta_{M_{0}}\times\delta_{M_{1}}}(j)$,
and $(\check{U}_{01j},\check{M}_{01j})=(\tilde{U}_{0},\tilde{M}_{00})_{P_{U_{0}|Y_{1}}(\cdot|Y_{1})\times P_{M_{0}}}(j)$.
Right: The Bayesian network describing the scheme. Note that all three
are valid Bayesian networks, and the desired conditional independence
relations can be deduced using d-separation.}
\end{figure}

Let $K_{a}=\Upsilon_{P_{U_{0}}\times\delta_{M_{0}}\Vert P_{U_{0}|Y_{a}}(\cdot|Y_{a})\times P_{M_{0}}}(1)$
(using the Poisson process $\mathfrak{P}_{0}$). By the conditional
generalized Poisson matching lemma on $(M_{0},\,1,\,(U_{0},M_{0}),\,Y_{a},\,P_{U_{0}|Y_{a}}\times P_{M_{0}})$,
almost surely,
\begin{align}
\mathbf{E}\left[\left.K_{a}\,\right|\,U_{0},Y_{a},M_{0}\right] & \le\frac{dP_{U_{0}}\times\delta_{M_{0}}}{dP_{U_{0}|Y_{a}}(\cdot|Y_{a})\times P_{M_{0}}}(U_{0},M_{0})+1\nonumber \\
 & =\mathsf{L}_{0}2^{-\iota_{U_{0};Y_{a}}(U_{0};Y_{a})}+1.\label{eq:bc2_Ka}
\end{align}
By \eqref{eq:bc2_dist}, $U_{0}=(\tilde{U}_{0})_{P_{U_{0}}\times\delta_{M_{0}}}$,
$\check{U}_{1j}=(\tilde{U}_{1})_{P_{U_{1}|U_{0}}(\cdot|U_{0})\times\delta_{M_{0}}\times\delta_{M_{1}}}(j)$,
and $(\check{U}_{01j},\check{M}_{01j})=(\tilde{U}_{0},\tilde{M}_{00})_{P_{U_{0}|Y_{1}}(\cdot|Y_{1})\times P_{M_{0}}}(j)$,
we have $(M_{0},M_{1},U_{0},J)\perp\!\!\!\perp\mathfrak{P}_{1}$ and
$(\{(\check{U}_{01j},\check{M}_{01j})\}_{j},Y_{1})\leftrightarrow(M_{0},M_{1},U_{0},J,U_{1})\leftrightarrow\mathfrak{P}_{1}$
(see Figure \eqref{fig:bc2_bayes} middle). Hence by the conditional
generalized Poisson matching lemma on $((M_{0},M_{1},U_{0}),\,J,\,(U_{1},M_{0},M_{1}),\,(\{(\check{U}_{01j},\check{M}_{01j})\}_{j},Y_{1}),\,\allowbreak\sum_{j=1}^{\infty}\phi(j)\allowbreak(P_{U_{1}|U_{0},Y_{1}}(\cdot|\check{U}_{01j},Y_{1})\times\delta_{\check{M}_{01j}})\times P_{M_{1}})$,
almost surely,
\begin{align*}
 & \mathbf{P}\biggl\{(\tilde{U}_{1},\tilde{M}_{01},\tilde{M}_{1})_{\sum_{j=1}^{\infty}\phi(j)(P_{U_{1}|U_{0},Y_{1}}(\cdot|\check{U}_{01j},Y_{1})\times\delta_{\check{M}_{01j}})\times P_{M_{1}}}\neq(U_{1},M_{0},M_{1})\,\biggl|\,U_{0},U_{1},U_{2},J,Y_{1},Y_{2},M_{0},M_{1}\biggr\}\\
 & \stackrel{(a)}{=}\mathbf{P}\biggl\{(\tilde{U}_{1},\tilde{M}_{01},\tilde{M}_{1})_{\sum_{j=1}^{\infty}\phi(j)(P_{U_{1}|U_{0},Y_{1}}(\cdot|\check{U}_{01j},Y_{1})\times\delta_{\check{M}_{01j}})\times P_{M_{1}}}\neq(U_{1},M_{0},M_{1})\,\biggl|\,U_{0},U_{1},J,Y_{1},M_{0},M_{1}\biggr\}\\
 & \stackrel{(b)}{\le}\mathbf{E}\left[\left.\min\left\{ \mathsf{J}\frac{dP_{U_{1}|U_{0}}(\cdot|U_{0})\times\delta_{M_{0}}\times\delta_{M_{1}}}{d(\sum_{j=1}^{\infty}\phi(j)(P_{U_{1}|U_{0},Y_{1}}(\cdot|\check{U}_{01j},Y_{1})\times\delta_{\check{M}_{01j}}))\times P_{M_{1}}}(U_{1},M_{0},M_{1}),\,1\right\} \,\right|U_{0},U_{1},J,Y_{1},M_{0},M_{1}\right]\\
 & \le\mathbf{E}\left[\left.\min\left\{ \frac{\mathsf{L}_{1}\mathsf{J}}{\phi(K_{1})}\frac{dP_{U_{1}|U_{0}}(\cdot|U_{0})\times\delta_{M_{0}}}{dP_{U_{1}|U_{0},Y_{1}}(\cdot|U_{0},Y_{1})\times\delta_{M_{0}}}(U_{1},M_{0}),\,1\right\} \,\right|U_{0},U_{1},J,Y_{1},M_{0},M_{1}\right]\\
 & =\mathbf{E}\left[\left.\min\left\{ \frac{\mathsf{L}_{1}\mathsf{J}}{\phi(K_{1})}2^{-\iota_{U_{1};Y_{1}|U_{0}}(U_{1};Y_{1}|U_{0})},\,1\right\} \,\right|U_{0},U_{1},J,Y_{1},M_{0},M_{1}\right]\\
 & \stackrel{(c)}{\le}\mathbf{E}\left[\left.K_{1}\mathsf{L}_{1}\mathsf{J}2^{-\iota_{U_{1};Y_{1}|U_{0}}(U_{1};Y_{1}|U_{0})}\left(\log(\mathsf{L}_{1}^{-1}\mathsf{J}^{-1}2^{\iota_{U_{1};Y_{1}|U_{0}}(U_{1};Y_{1}|U_{0})}+1)+1\right)^{2}\,\right|U_{0},U_{1},J,Y_{1},M_{0},M_{1}\right]\\
 & \stackrel{(d)}{\le}(\mathsf{L}_{0}2^{-\iota_{U_{0};Y_{1}}(U_{0};Y_{1})}+1)\mathsf{L}_{1}\mathsf{J}2^{-\iota_{U_{1};Y_{1}|U_{0}}(U_{1};Y_{1}|U_{0})}\left(\log(\mathsf{L}_{1}^{-1}\mathsf{J}^{-1}2^{\iota_{U_{1};Y_{1}|U_{0}}(U_{1};Y_{1}|U_{0})}+1)+1\right)^{2},
\end{align*}
where (a) is due to $(U_{2},Y_{2})\leftrightarrow(M_{0},M_{1},U_{0},J,U_{1},Y_{1})\leftrightarrow\mathfrak{P}_{1}$
(see Figure \ref{fig:bc2_bayes} middle), (b) is due to the aforementioned
application of the conditional generalized Poisson matching lemma,
(c) is by Proposition \ref{prop:phi_ineq}, and (d) is due to \eqref{eq:bc2_Ka}
and $K_{1}\leftrightarrow(U_{0},Y_{1},M_{0})\leftrightarrow(J,U_{1},M_{1})$
(see Figure \ref{fig:bc2_bayes} middle).

Also, since $(M_{0},M_{2},U_{0},\{\check{U}_{1j}\}_{j})\perp\!\!\!\perp\mathfrak{P}_{2}$
and $(\{(\check{U}_{02j},\check{M}_{02j})\}_{j},Y_{2})\leftrightarrow(M_{0},M_{2},U_{0},\{\check{U}_{1j}\}_{j},U_{2})\leftrightarrow\mathfrak{P}_{2}$
(see Figure \ref{fig:bc2_bayes} right), by the conditional Poisson
matching lemma on $((M_{0},M_{2},U_{0},\{\check{U}_{1j}\}_{j}),\,(U_{2},M_{0},M_{2}),\,(\{(\check{U}_{02j},\check{M}_{02j})\}_{j},Y_{2}),\,\allowbreak\sum_{j=1}^{\infty}\phi(j)\allowbreak(P_{U_{2}|U_{0},Y_{2}}\allowbreak(\cdot|\check{U}_{02j},\allowbreak Y_{2})\times\delta_{\check{M}_{02j}})\times P_{M_{2}})$,
almost surely,
\begin{align*}
 & \mathbf{P}\biggl\{(\tilde{U}_{2},\tilde{M}_{02},\tilde{M}_{2})_{\sum_{j=1}^{\infty}\phi(j)(P_{U_{2}|U_{0},Y_{2}}(\cdot|\check{U}_{02j},Y_{2})\times\delta_{\check{M}_{02j}})\times P_{M_{2}}}\neq(U_{2},M_{0},M_{2})\,\biggl|\,U_{0},U_{1},U_{2},Y_{1},Y_{2},M_{0},M_{2}\biggr\}\\
 & \stackrel{(a)}{=}\mathbf{P}\biggl\{(\tilde{U}_{2},\tilde{M}_{02},\tilde{M}_{2})_{\sum_{j=1}^{\infty}\phi(j)(P_{U_{2}|U_{0},Y_{2}}(\cdot|\check{U}_{02j},Y_{2})\times\delta_{\check{M}_{02j}})\times P_{M_{2}}}\neq(U_{2},M_{0},M_{2})\,\biggl|\,U_{0},U_{2},Y_{2},M_{0},M_{2}\biggr\}\\
 & \stackrel{(b)}{\le}\mathbf{E}\Biggl[\mathbf{E}\Biggl[\min\Biggl\{\frac{d(\mathsf{J}^{-1}\sum_{j=1}^{\mathsf{J}}P_{U_{2}|U_{0},U_{1}}(\cdot|U_{0},\check{U}_{1j}))\times\delta_{M_{0}}\times\delta_{M_{2}}}{d(\sum_{j=1}^{\infty}\phi(j)(P_{U_{2}|U_{0},Y_{2}}(\cdot|\check{U}_{02j},Y_{2})\times\delta_{\check{M}_{02j}}))\times P_{M_{2}}}(U_{2},M_{0},M_{2}),\,1\Biggr\}\\
 & \;\;\;\;\;\;\;\;\;\;\;\;\;\;\;\;\;\,\Biggl|\,\{\check{U}_{1j}\}_{j},U_{0},U_{2},Y_{2},M_{0},M_{2}\Biggr]\,\Biggl|\,U_{0},U_{2},Y_{2},M_{0},M_{2}\Biggr]\\
 & \le\mathbf{E}\left[\left.\min\left\{ \frac{\mathsf{L}_{2}}{\phi(K_{2})}\frac{d(\mathsf{J}^{-1}\sum_{j=1}^{\mathsf{J}}P_{U_{2}|U_{0},U_{1}}(\cdot|U_{0},\check{U}_{1j}))\times\delta_{M_{0}}}{dP_{U_{2}|U_{0},Y_{2}}(\cdot|U_{0},Y_{2})\times\delta_{M_{0}}}(U_{2},M_{0}),\,1\right\} \,\right|U_{0},U_{2},Y_{2},M_{0},M_{2}\right]\\
 & =\mathbf{E}\left[\left.\min\left\{ \frac{\mathsf{L}_{2}\mathsf{J}^{-1}}{\phi(K_{2})}\sum_{j=1}^{\mathsf{J}}2^{\iota_{U_{1};U_{2}|U_{0}}(\check{U}_{1j};U_{2}|U_{0})-\iota_{U_{2};Y_{2}|U_{0}}(U_{2};Y_{2}|U_{0})},\,1\right\} \,\right|U_{0},U_{2},Y_{2},M_{0},M_{2}\right]\\
 & \stackrel{(c)}{\le}\mathbf{E}\left[\left.\min\left\{ \frac{\mathsf{L}_{2}\mathsf{J}^{-1}}{\phi(K_{2})}2^{-\iota_{U_{2},Y_{2}|U_{0}}(U_{2};Y_{2}|U_{0})}(2^{\iota_{U_{1};U_{2}|U_{0}}(U_{1};U_{2}|U_{0})}+\mathsf{J}-1),\,1\right\} \,\right|U_{0},U_{2},Y_{2},M_{0},M_{2}\right]\\
 & \stackrel{(d)}{\le}\mathbf{E}\biggl[K_{2}\mathsf{L}_{2}\mathsf{J}^{-1}2^{-\iota_{U_{2},Y_{2}|U_{0}}(U_{2};Y_{2}|U_{0})}(2^{\iota_{U_{1};U_{2}|U_{0}}(U_{1};U_{2}|U_{0})}+\mathsf{J}-1)\\
 & \;\;\;\;\;\left(\log(\mathsf{L}_{2}^{-1}\mathsf{J}2^{\iota_{U_{2},Y_{2}|U_{0}}(U_{2};Y_{2}|U_{0})}(2^{\iota_{U_{1};U_{2}|U_{0}}(U_{1};U_{2}|U_{0})}+\mathsf{J}-1)^{-1}+1)+1\right)^{2}\,\biggl|\,U_{0},U_{2},Y_{2},M_{0},M_{2}\biggr]\\
 & \stackrel{(e)}{\le}(\mathsf{L}_{0}2^{-\iota_{U_{0};Y_{2}}(U_{0};Y_{2})}+1)\mathsf{L}_{2}\mathsf{J}^{-1}2^{-\iota_{U_{2},Y_{2}|U_{0}}(U_{2};Y_{2}|U_{0})}(2^{\iota_{U_{1};U_{2}|U_{0}}(U_{1};U_{2}|U_{0})}+\mathsf{J}-1)\\
 & \;\;\;\;\;\left(\log(\mathsf{L}_{2}^{-1}\mathsf{J}2^{\iota_{U_{2},Y_{2}|U_{0}}(U_{2};Y_{2}|U_{0})}(2^{\iota_{U_{1};U_{2}|U_{0}}(U_{1};U_{2}|U_{0})}+\mathsf{J}-1)^{-1}+1)+1\right)^{2},
\end{align*}
where (a) is due to $(U_{1},Y_{1})\leftrightarrow(U_{0},U_{2},Y_{2},M_{0},M_{2})\leftrightarrow\mathfrak{P}_{2}$
(see Figure \ref{fig:bc2_bayes} right), (b) is due to the aforementioned
application of the conditional Poisson matching lemma, (c) is by the
same arguments as in the proof of Theorem \ref{thm:bc_ncm}, (d) is
by Proposition \ref{prop:phi_ineq}, and (e) is due to \eqref{eq:bc2_Ka}
and $K_{2}\leftrightarrow(U_{0},Y_{2},M_{0})\leftrightarrow(U_{2},M_{2})$
(see Figure \ref{fig:bc2_bayes} right). Hence
\begin{align*}
 & \mathbf{P}\{(M_{0},M_{0},M_{1},M_{2})\neq(\hat{M}_{00},\hat{M}_{01},\hat{M}_{1},\hat{M}_{2})\}\\
 & \le\mathbf{E}\biggl[\min\biggl\{(\mathsf{L}_{0}2^{-\iota_{U_{0};Y_{1}}(U_{0};Y_{1})}+1)\mathsf{L}_{1}\mathsf{J}2^{-\iota_{U_{1};Y_{1}|U_{0}}(U_{1};Y_{1}|U_{0})}\left(\log(\mathsf{L}_{1}^{-1}\mathsf{J}^{-1}2^{\iota_{U_{1};Y_{1}|U_{0}}(U_{1};Y_{1}|U_{0})}+1)+1\right)^{2}\\
 & \;\;\;\;+(\mathsf{L}_{0}2^{-\iota_{U_{0};Y_{2}}(U_{0};Y_{2})}+1)\mathsf{L}_{2}\mathsf{J}^{-1}2^{-\iota_{U_{2},Y_{2}|U_{0}}(U_{2};Y_{2}|U_{0})}(2^{\iota_{U_{1};U_{2}|U_{0}}(U_{1};U_{2}|U_{0})}+\mathsf{J}-1)\\
 & \;\;\;\;\;\;\left(\log(\mathsf{L}_{2}^{-1}\mathsf{J}2^{\iota_{U_{2},Y_{2}|U_{0}}(U_{2};Y_{2}|U_{0})}(2^{\iota_{U_{1};U_{2}|U_{0}}(U_{1};U_{2}|U_{0})}+\mathsf{J}-1)^{-1}+1)+1\right)^{2},\,1\biggr\}\biggr]\\
 & \le\mathbf{E}\biggl[\min\biggl\{\mathsf{L}_{0}\mathsf{L}_{1}\mathsf{J}A2^{-\iota_{U_{0},U_{1};Y_{1}}(U_{0},U_{1};Y_{1})}+\mathsf{L}_{1}\mathsf{J}A2^{-\iota_{U_{1};Y_{1}|U_{0}}(U_{1};Y_{1}|U_{0})}\\
 & \;\;\;\;+\mathsf{L}_{0}\mathsf{L}_{2}\mathsf{J}^{-1}B2^{\iota_{U_{1};U_{2}|U_{0}}(U_{1};U_{2}|U_{0})-\iota_{U_{0},U_{2};Y_{2}}(U_{0},U_{2};Y_{2})}+\mathsf{L}_{0}\mathsf{L}_{2}(1-\mathsf{J}^{-1})B2^{-\iota_{U_{0},U_{2};Y_{2}}(U_{0},U_{2};Y_{2})}\\
 & \;\;\;\;+\mathsf{L}_{2}\mathsf{J}^{-1}B2^{\iota_{U_{1};U_{2}|U_{0}}(U_{1};U_{2}|U_{0})-\iota_{U_{2},Y_{2}|U_{0}}(U_{2};Y_{2}|U_{0})}+\mathsf{L}_{2}(1-\mathsf{J}^{-1})B2^{-\iota_{U_{2},Y_{2}|U_{0}}(U_{2};Y_{2}|U_{0})},\,1\biggr\}\biggr],
\end{align*}
where $A=(\log(\mathsf{L}_{1}^{-1}\mathsf{J}^{-1}2^{\iota_{U_{1};Y_{1}|U_{0}}(U_{1};Y_{1}|U_{0})}+1)+1)^{2}$,
$B=\bigl(\log((\mathsf{L}_{2}\mathsf{J}^{-1}2^{\iota_{U_{1};U_{2}|U_{0}}(U_{1};U_{2}|U_{0})-\iota_{U_{2},Y_{2}|U_{0}}(U_{2};Y_{2}|U_{0})}+\mathsf{L}_{2}(1-\mathsf{J}^{-1})2^{-\iota_{U_{2},Y_{2}|U_{0}}(U_{2};Y_{2}|U_{0})})^{-1}+1)+1\bigr)^{2}$.

For \eqref{eq:bc_pe2}, if the event in \eqref{eq:bc_pe2} does not
occur, by Proposition \ref{prop:phi_ineq},
\begin{align*}
 & (\mathsf{L}_{0}2^{-\iota_{U_{0};Y_{1}}(U_{0};Y_{1})}+1)\mathsf{L}_{1}\mathsf{J}2^{-\iota_{U_{1};Y_{1}|U_{0}}(U_{1};Y_{1}|U_{0})}\left(\log(\mathsf{L}_{1}^{-1}\mathsf{J}^{-1}2^{\iota_{U_{1};Y_{1}|U_{0}}(U_{1};Y_{1}|U_{0})}+1)+1\right)^{2}\\
 & \;\;+(\mathsf{L}_{0}2^{-\iota_{U_{0};Y_{2}}(U_{0};Y_{2})}+1)\mathsf{L}_{2}\mathsf{J}^{-1}2^{-\iota_{U_{2},Y_{2}|U_{0}}(U_{2};Y_{2}|U_{0})}(2^{\iota_{U_{1};U_{2}|U_{0}}(U_{1};U_{2}|U_{0})}+\mathsf{J}-1)\\
 & \;\;\;\;\;\left(\log(\mathsf{L}_{2}^{-1}\mathsf{J}2^{\iota_{U_{2},Y_{2}|U_{0}}(U_{2};Y_{2}|U_{0})}(2^{\iota_{U_{1};U_{2}|U_{0}}(U_{1};U_{2}|U_{0})}+\mathsf{J}-1)^{-1}+1)+1\right)^{2}\\
 & \le2^{1-\gamma}\left(2(\iota_{U_{1};Y_{1}|U_{0}}(U_{1};Y_{1}|U_{0}))^{2}+2\gamma^{2}+14\right)+2^{2-\gamma}\left(2(\iota_{U_{2},Y_{2}|U_{0}}(U_{2};Y_{2}|U_{0}))^{2}+2\gamma^{2}+14\right)\\
 & \le2^{-\gamma}\left(8(\iota_{U_{1};Y_{1}|U_{0}}(U_{1};Y_{1}|U_{0}))^{2}+8(\iota_{U_{2},Y_{2}|U_{0}}(U_{2};Y_{2}|U_{0}))^{2}+12\gamma^{2}+84\right).
\end{align*}

\bibliographystyle{IEEEtran}
\bibliography{ref}

\end{document}